\documentclass[12pt]{article}

\let\plural=\relax

\makeatletter
\def\newleaf{\newpage
\newcount\tmp
\tmp=\c@page
\divide\tmp by 2
\multiply\tmp by 2
\ifnum\c@page=\tmp
~\newpage
\fi
}
\makeatother

\expandafter\ifx\csname proofnewpage\endcsname\relax

\fi

\def\color[#1]#2{}

\long\def\nop#1{}

\def\comment{\edef\cps{\the\parskip} \parskip=0.5cm \begingroup \tt}

\begingroup
\makeatletter
\global\def\fakelabel#1#2{
\expandafter\ifx\csname fakelabelsome\endcsname\relax
\let\fakelabelsome\par
\AtEndDocument{\typeout{}}
\fi
\AtEndDocument{\typeout{NOTE: #1 is a fake label, marked #2}\typeout{}}
\@ifundefined {r@#1}
{\global\@namedef{r@#1}{#2}}
{}
}
\endgroup

\long\def\figurearrow#1#2{
\newbox\before
\setbox\before=\hbox{#1}
\newbox\after
\setbox\after=\hbox{#2}
\newdimen\vdim
\vdim=\ht\before
\ifdim\vdim<\the\ht\after
  \vdim=\ht\after
\fi
\hbox{
\vbox to \the\vdim{\vfill\box\before\vfill}
\vbox to \the\vdim{\vfill\hbox to 3cm{\hfill\Huge$\Rightarrow$\hfill}\vfill}
\vbox to \the\vdim{\vfill\box\after\vfill}
}
}

\expandafter\ifx\csname shortcite\endcsname\relax

\fi

\newbox\current

\long\def\plframebox#1{
\setbox\current\vbox{#1}		

\vbox to \ht\current {\hrule\vss
\hbox to \wd\current {%
\vrule \hss\box\current\hss \vrule}
\vss\hrule }
}

\long\def\eatpar#1{%
\ifx#1\par                      
\let\nextmove=\eatpar           
\else
\let\nextmove=#1
\fi
\noexpand\nextmove
}

\def\modifymargins#1#2{
\newdimen\addtoh
\newdimen\addtow
\addtoh=#1
\addtow=#2

\advance\topmargin by -\addtoh
\multiply\addtoh by 2
\advance\textheight by \addtoh

\advance\oddsidemargin by -\addtow
\advance\evensidemargin by -\addtow
\multiply\addtow by 2
\advance\textwidth by \addtow
}

\begingroup
\catcode`\~=11
\gdef\centertilde#1{\lower #1pt\hbox{~}}
\endgroup

\newcount\currenttime
\newcount\hour
\newcount\minute

\def\printtime{%
\currenttime=\time
\hour=\currenttime
\divide\hour by 60
\minute=-\hour
\multiply\minute by 60
\advance\minute by \currenttime
\the\hour:\ifnum\minute<10 0\fi\the\minute
}

\begingroup
\makeatletter
\global\let\@@date=\@date
\gdef\@date{\@@date\ --- \printtime}
\endgroup

\def\oggi{\number\day\space 
\ifcase\month\or
Gennaio\or Febbraio\or Marzo\or Aprile\or Maggio\or Giugno\or
Luglio\or Agosto\or Settembre\or Ottobre\or Novembre\or Dicembre\fi
\space \number\year}

\newcounter{rmexample}

\def\proof{\noindent {\sl Proof.\ \ }}

\def\qed{\hfill{\boxit{}}
  \ifdim\lastskip<\medskipamount \removelastskip\penalty55\medskip\fi}
\def\qedn#1{\hfill{\boxit{}$_#1$}
  \ifdim\lastskip<\medskipamount \removelastskip\penalty55\medskip\fi}
\long\def\boxit#1{\vbox{\hrule\hbox{\vrule\kern3pt
                  \vbox{\kern3pt#1\kern3pt}\kern3pt\vrule}\hrule}}

\def\var{V\!ar}

\def\true{{\sf true}}
\def\false{{\sf false}}

\def\p{{\rm P}}
\def\np{{\rm NP}}
\def\conp{{\rm coNP}}

\def\bh#1{\if#1{}{\rm BH}\else\mbox{BH$_{#1}$}\fi}
\def\S#1{\mbox{$\Sigma^p_{#1}$}}
\def\P#1{\mbox{$\Pi^p_{#1}$}}

\def\pspace{{\rm PSPACE}}

\let\cedilla=\c
\def\c{\mbox{$\leadsto$}}

\def\profont{\sf}

\def\x3c{{\profont x3c}}

\def\possnewtheorem#1#2{
\expandafter\ifx\csname #1\endcsname\relax
\newtheorem{#1}{#2}
\fi
}

\def\possnewtheoremthree#1[#2]#3{
\expandafter\ifx\csname #1\endcsname\relax
\newtheorem{#1}[#2]{#3}
\fi
}

\possnewtheorem{theorem}{Theorem}
\possnewtheorem{corollary}{Corollary}
\possnewtheorem{lemma}{Lemma}
\possnewtheoremthree{proposition}[theorem]{Proposition}
\possnewtheorem{definition}{Definition}
\possnewtheorem{question}{Question}
\possnewtheorem{example}{Example}
\possnewtheorem{nontheorem}{Counterexample}
\possnewtheorem{property}{Property}
\possnewtheorem{assumption}{Assumption}
\possnewtheorem{conjecture}{Conjecture}
\possnewtheorem{notation}{Notation}
\newenvironment{theorem*}[1]{{\noindent \bf Theorem~#1}\begin{it}}{\end{it}\

}

\def\after#1#2{#1~{\sf after}~#2}

\makeatletter

\long\def\state#1#2#3#4{#2#3%
\expandafter\gdef\csname st@ate#1\endcsname{#3#4}}

\def\restate#1{\csname st@ate#1\endcsname}

\long\def\statetheorem#1#2#3{
\begin{theorem}\label{#1}
\state{theorem#1}{}{#2}{}
\end{theorem}
\state{theoremproof#1}{}{}{#3}
}

\long\def\statelemma#1#2#3{
\begin{lemma}\label{#1}
\state{lemma#1}{}{#2}{}
\end{lemma}
\state{lemmaproof#1}{}{}{#3}
}

\newcount\restatec@unter

\def\re#1#2{
\expandafter\let\expandafter\counter\csname c@#1\endcsname
\restatec@unter=\counter
\expandafter\ifx\csname r@#2\endcsname\relax
\counter=0
\else
\counter=\ref{#2}
\fi
\advance\counter by -1
\begin{#1}
\restate{#1#2}
\end{#1}
\counter=\restatec@unter
\restate{#1proof#2}
}

\makeatother

\expandafter\ifx\csname ttytty\endcsname\relax\modifymargins{60pt}{40pt}
\else\modifymargins{60pt}{0pt}\def\_{\tt\char 95}
\raggedright\parindent 40pt\fi
\expandafter\ifx\csname ttytty\endcsname\relax
\def\ttytex#1{#1\nop}			
\else

\modifymargins{0pt}{-10pt}
\def\frac#1#2{#1/#2}
\def\_{\char 95}
\expandafter\ifx\csname makeatletter\endcsname\relax\makeatletter\fi
\def\@ttyfig#1#2{\def\specialtext{\special{txt:#2}}\specialtext\egroup}
\def\ttyfig{\relax\bgroup\catcode`\^^M\active\let^^M
\let\-\relax\catcode`\ \active\@ttyfig}
\let\ttytex\ttyfig
\expandafter\ifx\csname makeatother\endcsname\relax\makeatletter\fi
\fi

\possnewtheorem{algorithm}{Algorithm}
\possnewtheorem{condition}{Condition}
\expandafter\ifx\csname citeyear\endcsname\relax\let\citeyear=\cite\fi

\def\common{\equiv\equiv}
\def\forget{\mbox{forget}}

\def\hi{\mbox{\bf head\_implicates}}

\title{Common equivalence and size after forgetting}
\author{Paolo Liberatore\thanks{%
DIAG, Sapienza University of Rome.
{\tt liberato@diag.uniroma1.it}
}}
\date{}

\begin{document}

\maketitle

\let\oinput=\input
\def\xinput#1{}

\begin{abstract}

Forgetting variables from a propositional formula may increase its size.
Introducing new variables is a way to shorten it. Both operations can be
expressed in terms of common equivalence, a weakened version of equivalence. In
turn, common equivalence can be expressed in terms of forgetting. An algorithm
for forgetting and checking common equivalence in polynomial space is given for
the Horn case; it is polynomial-time for the subclass of single-head formulae.
Minimizing after forgetting is polynomial-time if the formula is also acyclic
and variables cannot be introduced, \np-hard when they can.

\end{abstract}


\section{Introduction}

Logical forgetting is removing variables from
consideration~\cite{delg-17,eite-kern-19}. Also called variable elimination, it
is done to
{} work with bounded memory~\cite{eite-kern-19},
{} simplify reasoning~\cite{delg-wang-15,erde-ferr-07,wang-etal-05},
{} clarify the relationship between variables~\cite{delg-17};
{} formalize the limited knowledge of agents~\cite{fagi-etal-95,raja-etal-14};
{} ensure privacy~\cite{gonc-etal-17};
{} merge information coming from different sources~\cite{wang-etal-14};
{} restore consistency~\cite{lang-marq-10}.

The first four aims are missed if the result is too large. The space needed to
store information increases instead of reducing. Reasoning from larger
knowledge bases is likely harder rather than easier. The relationships between
variables are probably obfuscated by an increase in size. A limit in knowledge
storage ability is never enforced by enlarging a formula. Regardless of the
aim, a large formula poses problems of storage, reasoning and ease of
interpretation.

That forgetting increases size is counterintuitive since forgetting is removing
facts or objects from consideration. Less information should take less memory.
Yet, less information may be more complicated to express. This is known to be
the case in various
logics~\cite{eite-wang-06,kone-etal-09,delg-wang-15,gonc-etal-16,libe-20}. To
complicate the matter, what results from forgetting may be large or small
depending on the original formula; and may be equivalent to small formulae or
not. For example, the classical syntactic definition of forget in propositional
logic always doubles the size of the formula for each forgotten variables, but
size may often be reduced.

Besides forgetting, this is the classical problem of logic
minimization~\cite{coud-94,coud-sasa-02,uman-etal-06}. It originates from
electronic circuit synthesis: given a Boolean function, design a circuit that
realizes it. The simpler the circuit, the better. Various solutions have been
developed like the Karnaugh maps~\cite{karn-53}, the Quine-McClunskey
method~\cite{mccl-56} and the Espresso algorithm~\cite{rude-sang-87}.

In spite of minimization, the result of forgetting may still be exponentially
large, or just too large for the intended application. What to do in such
cases? A way to make a formula smaller is to introduce new variables~%
\cite{bube-etal-10,bhat-etal-10,cado-etal-00,cost-etal-04,meng-wall-20}. As an
example, Mengel and Wallon~\citeyear{meng-wall-20} wrote: ``It is folklore that
adding auxiliary variables can decrease the size of an encoding: for example
the parity function has no subexponential CNF-representations but there is an
easy linear size encoding using auxiliary variables''. Bubeck and Kleine
B\"uning~\citeyear{bube-etal-10} wrote ``Using auxiliary variables to
introduce definitions is a popular and powerful technique in knowledge
representation which can lead to shorter and more natural encodings''.

This may look like a vicious circle: variables are first removed, then added
back. It is not. Variables that are unneeded or unwanted are forgotten; if the
result is too big, other variables are introduced to make it smaller. A
concrete example is a formula of 1000 variabiles, where only 20 are relevant to
a certain aim. Forgetting the remaining variables produces a formula of size
500000. Adding 14 other variables takes it down to size 200. The key is
``other'': these are other variables. The 980 original variables are removed
because they have to. They are not relevant to a certain context, they cause
inconsistencies, they have to be hidden for legal reasons. Since the result is
too large, 14 other variables are introduced to the sole aim of reducing size.
%
%
Adding elements when forgetting others had been considered in the very context
of forgetting, on abstract argumentation frameworks~\cite{baum-etal-20}, answer
set programming~\cite{eite-wang-08} and description logics~\cite{koop-schm-15}.

While introducing variables is mainly motivated by reducing size, it may give a
side benefit. If a new variable condenses a formula, a question arises: does
this happen by chance only? The variable has no meaning by itself, it just
reduces the size of the formula by pure luck. It so happens. But simpler
explanations are usually considered better than complicated ones. If the
shorter formula is a better representation of knowledge, it may be so because
it is closer to the situation it represents. The added variable represents a
real fact, rather than being the prop of a magic trick to reduce size. It was
missing from the original formula because the fact was hidden, not directly
observable. Reducing size had the indirect benefit of uncovering it. Formal
logic cannot tell its meaning, but tells that it may exist, and tells how it is
related to the other facts.

Forgetting and introducing variables have something in common: they both change
the alphabet of the formula while retaining some of its consequences. This is
formalized by restricted equivalence~\cite{flog-etal-93}, the equality of the
consequences on a given subset of variables. When this subset comprises all
variables, restricted equivalence is regular equivalence. When some variables
are missing, only the consequences that do not contain those variables matter.

\begin{description}

\item[Forgetting] is removing variables while retaining the consequences on the
others; the result is equivalent to the original when restricting to the other
variables.

\item[Introducing] is adding variables while retaining the consequences on the
original variables; the result is equivalent to the original when restricting
to the original variables.

\item[Forgetting and introducing] is removing and adding variables while
retaining the consequences on the original variables not to be forgotten; the
result is equivalent to the original when restricting to these variables.

\end{description}

Forgetting, introducing, forgetting and then introducing variables are all
formalized by restricted equivalence. A subcase of it, actually: the
restriction is on the variables that are shared between the input and the
output formula. This is common equivalence: equality of the consequences on
the common variables. Apart from the slight simplification of not explicitly
requiring a set of variables, common equivalence forbids reintroducing a
removed variable with a different meaning, which restricted equivalence allows.

Forget can be expressed in terms of restricted and common equivalence, but also
the other way around. Both forms of equivalence amount to forget the variables
that do not matter and then check regular equivalence. Theoretically, this can
always be done. Computationally, it may not: if the result of forgetting is
exponentially large, computing restricted or common equivalence this way
requires exponential space while both problems can be solved in polynomial
space. Size after forgetting matters again.

The exponentiality of forgetting is not just a possibility, nor it is due to a
specific method of forgetting. For certain formulae and variables to forget, it
is a certainty: the result of forgetting cannot be represented in polynomial
space. While reducing its size is important, it is not always possible. In such
cases, a forgetting algorithm takes exponential time just because its output is
always exponentially large. The time needed to just output it is exponential.
Yet, the required memory space may not. An algorithm for the Horn case is shown
that runs in polynomial space even when it produces an exponential output. A
consequence is that checking common equivalence takes polynomial space in the
Horn case.

This algorithm unearths a polynomial Horn subclass: it runs in polynomial time
when each variable is at most the head of a single clause.

Polynomial running time means polynomially-sized output. Fast enough, but not
always small enough. A polynomial output may be quadratic. But even if it is
only twice the size of the input, it is still a size increase. Forgetting fails
at making the formula smaller. But this is again the output of a specific
algorithm. Equivalent but smaller formulae may exist. Surprisingly, it depends
on whether they are required to be single-head or not. Either way, the minimal
formula can be found in polynomial time if the formula is
acyclic~\cite{hamm-koga-95}. Otherwise, a sufficient condition directs the
search for the clauses of the minimal formula.

Adding new variables allows reducing size. An algorithm for Horn formulae is
given. It is polynomial but unable to always find the minimal formula. Not a
fault of the algorithm, however: the problem is \np-complete. It is \np-hard
even in the single-head acyclic case. In the same conditions, the problem is
polynomial without new variables. The reason is that the new variables may
shorten the formula in many ways. Exploring them takes time.

Three implementations of the forgetting algorithm have been developed. They are
all based on the same algorithm, but they differ in how they realize
nondeterminism. The first is correct only in the single-head case, which does
not require nondeterminism. The second employs sets to represent the possible
outcomes of a nondeterministic choice; it is always correct but may take
exponential space. The third exploits multiple processes; it is always correct
and works in polynomial space. A script uses the third for checking common
equivalence. The algorithm for minimizing with new variables is implemented as
well.

Most of the examples formulae used in this article are made into test files for
these programs.

\iftrue

To finish this introduction, a summary of the contributions of this article is
given; it also outlines the organization of what follows.
%
%
First, variable forgetting, introducing and forgetting followed by introducing
are proved translatable into common equivalence, a specific case of restricted
equivalence, which is also shown to be translatable into forgetting. Some
results about common equivalence are proved, along with its \P{2}-completeness
in the general case and \conp-completeness in the Horn case. All of this is
Section~\ref{section-common}.
Second, forgetting is proved to exponentially increase size in some cases, even
if equivalent formulae are allowed and even when restricting to the definite
Horn case. These results are in Section~\ref{section-size}.
Third, an algorithm for forgetting and checking common equivalence in the Horn
case is presented. Contrary to previous methods for forgetting, it only
requires a polynomial amount of working memory. It is first defined and proved
correct in the definite Horn case in Section~\ref{section-definite} and then
applied to the general Horn case in Section~\ref{section-indefinite}.
Fourth, a subclass of Horn formulae that makes the algorithm polynomial in
time, in addition of working memory, is identified. The problem of minimizing
such formulae after forgetting is analyzed in Section~\ref{section-minimize}.
Fifth, an algorithm for reducing size when new variables can be introduced is
presented. The problem is shown \np-hard even in the simplest case considered
in this article. This is the content of Section~\ref{section-extended}.
The algorithms defined in this article are implemented in Python. Details are
in Section~\ref{section-python}.
Proofs of lemmas and theorems are in Appendix~\ref{section-proofs}.

\else

The next section defines variables forgetting and introduction in terms of
restricted and common equivalence and gives some general and complexity results
about common equivalence. Section~\ref{section-size} formally proves that
forget may exponentially increase size even when equivalent formulae are
allowed and when restricting to the definite Horn case.
Section~\ref{section-definite} presents the algorithms for forgetting and
checking common equivalence for definite Horn formulae. They are extended to
all Horn formulae in Section~\ref{section-indefinite}. The problem of
minimizing the formula after forgetting in the single-head case is analyzed in
Section~\ref{section-minimize}. Section~\ref{section-extended} presents the
algorithm for reducing size when variables can be introduced and shows that the
problem is \np-hard even in the acyclic single-head case.
Section~\ref{section-python} describes the implementations of the algorithms.

\fi

\section{Preliminaries}
\label{section-preliminaries}

Unless stated otherwise, logical formulae in this article are in Conjunctive
Normal Form (CNF): they are sets of clauses, each clause being a disjunction of
literal, where a literal is a propositional variable or its negation. Such a
set is equivalent to the conjunction of its elements. Writing formulae as sets
allows to compare them by containment: $A \subseteq B$ means that all clauses
in $A$ are also in $B$.

Some results are about Horn and definite Horn formulae. A clause is Horn if it
contains at most one positive (unnegated) literal. A formula is Horn if it
comprises Horn clauses only. A clause is definite Horn if it contains exactly
one positive literal. A formula is definite Horn if it comprises definite Horn
clauses only. Horn and definite Horn are also sets; therefore, they can be
compared by set containment.

Horn clauses are written as rules like $abc \rightarrow d$ instead of $\neg a
\vee \neg b \vee \neg c \vee d$. This format is also accepted by the Python
programs.

Two formulae are equisatisfiable if and only if either they are both
satisfiable or they both are not.

The size of a formula is the number of its literal occurrences. Equivalently,
it is the sum of the size of its clauses, where the size of a clause is the
number of literals it contains.

Forgetting is removing one or more variables from consideration. This is also
known as variable elimination~\cite{subb-prad-04} and uniform
interpolation~\cite{bilk-07}. While some authors also consider formula removal
a form of forgetting~\cite{eite-kern-19}, they are fundamentally different. The
first aims at reducing the alphabet while maintaining information as much as
possible. The second aims at making the removed formula no longer entailed; it
is a form of belief change~\cite{ferm-hans-18} rather than forgetting. The
first is disregarding something not of interest at the moment; the second is
removing the belief that information is true. One is focusing, the other is
changing.

In logics other than propositional logic, forgetting may be defined in
different ways depending on which properties it has to
achieve~\cite{gonc-etal-16}. This is not the case in propositional logic, where
variable forgetting is removing a variable while maintaining all consequences
that do not contain it. The only two variants are literal
forgetting~\cite{lang-etal-03} and forgetting with fixed
variables~\cite{moin-07}. Apart from these, forgetting some variables from a
formula is a new formula that entails exactly the same formulae on the other
variables~\cite{lang-etal-03}. Formally, $F'$ expresses forgetting the
variables $X$ from $F$ if and only if $F' \models C$ is equivalent to $F
\models C$ whenever $C$ is a formula that do not contain any variable in $X$.

\section{Common equivalence}
\label{section-common}

Given that size is important to many applications of forgetting, the question
is when forgetting actually reduces size or not. The traditional way to forget
a variable $x$ is by disjoining two copies of the formula, one for $x=\true$
and one for $x=\false$~\cite{bool-54}. The result is always double the size of
the original. For example, forgetting $x$ from $F$ produces $F'$:

\begin{eqnarray*}
F &=&
(x \vee y) \wedge (\neg x \vee \neg y) \wedge (a \vee b \vee c \vee d) \\
F' &=&
((\true \vee y) \wedge
 (\neg \true \vee \neg y) \wedge
 (a \vee b \vee c \vee d))
\vee \\
&&
((\false \vee y) \wedge
 (\neg \false \vee \neg y) \wedge
 (a \vee b \vee c \vee d))
\end{eqnarray*}

The resulting formula is larger than the original. Yet, applying some simple
rules such as $\false \vee A = A$ simplifies it to
{} $(\neg y \wedge (a \vee b \vee c \vee d))
{}  \vee
{}  (y \wedge (a \vee b \vee c \vee d))$,
which simplifies to
{} $(\neg y \vee y) \wedge (a \vee b \vee c \vee d))$
by factorization. Removing the tautology turns it into
{} $F'' = a \vee b \vee c \vee d$.
Summarizing: forgetting $x$ from $F$ produces a larger formula $F'$ equivalent
to the smaller formula $F''$.

Being equivalent, $F'$ and $F''$ have the same meaning: the same of the
original formula when $x$ is disregarded. The first formula $F'$ tells that
forgetting can double size, which is always the case. The second formula $F''$
tells that this can be avoided. This is what matters when evaluating size after
forgetting: how small the formula can be, not how large. Logical formulae can
always be inflated by adding useless parts such as $a \vee \neg a$. How large a
formula can be is trivial: any size. What counts is how small it can be.

The same problem without forgetting has been extensively
analyzed~\cite{hema-schn-11,cepe-kuce-08}. Equivalence complicates it, as a
formula may be equivalent to many others, some large and some small. The
question is not ``how large a formula is'' but ``how small a formula can be
made''. In the contex of forgetting, it is ``how small is a formula expressing
forgetting''. Which sizes are acceptable and which are not depends on the
application, so the maximal allowed size is part of the problem.

Technically, given a formula $F$, a set of variables and a bound $k$, the
question is whether some formula $F'$ of size $k$ or less expresses forgetting
the variables from $F$. This may or may not be the case. In the example above,
a formula of size 4 expresses forgetting $x$ from $F$. No formula of size 3 or
less do. No matter how
{} $F'' = a \vee b \vee c \vee d$
is manipulated, if equivalence is to be preserved the result is always a
formula that contains more than three literal occurrences.

A common technique to reduce size is to introduce new variables to represent
repeated
subformulae~\cite{cado-etal-99,cost-etal-04,giun-etal-06,meng-wall-20}. For
example, forgetting $x$ from
{} $F = \{ab \rightarrow x,
{}  xc \rightarrow d, xc \rightarrow e, xc \rightarrow f\}$
is expressed by
{} $F' = \{abc \rightarrow d, abc \rightarrow e, abc \rightarrow f\}$.
This formula contains twelve literal occurrences, the same as $F$. It is
minimal: it is equivalent to no smaller formula. Yet, its size can be reduced
by introducing a new variable $z$ to represent the repeated subformula $abc$.
The result is
{} $F'' =
{}  \{abc \rightarrow z, z \rightarrow d, z \rightarrow e, z \rightarrow f\}$,
which only contains ten literal occurrences, two less than the original.

The result of such an addition is not exactly equivalent to the original. The
original formula does not mention $z$; therefore, its value is unaffected by
the value of $z$. If a model satisfies the formula, it still does when changing
the value of $z$. This is not the case after adding the new variable $z$. For
example, the model that sets all variables to $\true$ satisfies the formula,
but no longer does when changing the value of $z$ to $\false$, since $a=\true$,
$b=\true$, $c=\true$ and $z=\false$ falsify the clause $abc \rightarrow z$.
Having different value on this model, $F$ and $F''$ are not equivalent.

Yet, $F''$ expresses the same information as $F'$ apart from $z$, which is just
a shorthand for $a \wedge b \wedge c$. They are equivalent when disregarding
$z$. For example, they entail the same consequences that do not contain $z$.
They are satisfied by the same partial models that do not evaluate $z$.
Excluding variables from the comparison is restricted
equivalence~\cite{flog-etal-93} or var-equivalence~\cite{lang-etal-03}.

\begin{definition}%
[Restricted equivalence~\cite{flog-etal-93} or
 Var equivalence~\cite{lang-etal-03}]

Two formulae $A$ and $B$ are restricted-equivalent or var-equivalent on the
variables $X$ if $A \models C$ holds if and only if $B \models C$ holds for
every formula $C$ over the alphabet $X$.

\end{definition}

Restricted equivalence formalizes the addition of new variables to the aim of
reducing size~\cite{bube-etal-10,bhat-etal-10}: the generated formulae are
equivalent to the original formula on the original variables. Given a formula
$A$ over variables $X$, the aim is to produce a smaller formula $B$ over
variables $X \cup Y$ that is restricted-equivalent to $A$ over the variables
$X$.

This is forgetting in reverse: instead of forgetting $Y$ from $B$ to produce
$A$, it adds $Y$ from $A$ to produce $B$. It can indeed be reformulated in
terms of forgetting: given $A$ over $X$, search for a formula $B$ of the given
size such that forgetting $Y$ from $B$ produces $A$.

In the other way around, restricted equivalence formalizes forgetting: the
result of forgetting $X$ from $A$ is a formula $B$ over the variables $\var(A)
\backslash X$ that is restricted-equivalent to $A$ over the variables $\var(A)
\backslash X$.

In the way around the other way around: $A$ and $B$ are restricted-equivalent
over $X$ if forgetting $\var(A) \backslash X$ from $A$ is equivalent to
forgetting $\var(B) \backslash Y$ from $B$~\cite{lang-etal-03}.

Restricted equivalence formalizes variable forgetting, variable introduction
and variable forgetting followed by variable introduction. All three forms of
change are restricted equivalence over the variables that are not forgotten and
not introduced.

The problem with restricted equivalence is that it is too powerful. It allows a
forgotten variable to be reintroduced with a different meaning. Forgetting $x$
from
{} $\{\neg x, abc \rightarrow d, abc \rightarrow e, abc \rightarrow f\}$
is expressed by
{} $\{abc \rightarrow d, abc \rightarrow e, abc \rightarrow f\}$,
which can be reduced size by variable introduction:
{} $\{abc \rightarrow x, x \rightarrow d, x \rightarrow e, x \rightarrow f\}$.
The variables that are neither forgotten nor introduced are $a,b,c,d,e,f$; the
two formulae are restricted equivalent over them. Restricted equivalence is
blind to $x$ being a fact that is false in the original formula and a shorthand
for $abc$ in the final. Such cases are avoided by comparing formulae on the
common variables instead of an arbitrary set of variables.

\begin{definition}

Two formulae $A$ and $B$ are common-equivalent, denoted $A \common B$, if $A
\models C$ if and only if $B \models C$ for every formula $C$ such that
$\var(C) \subseteq \var(A) \cap \var(B)$.

\end{definition}

Common equivalence is: same consequences on the common alphabet.

Forgetting variables $X$ from a formula $A$ results in a formula $B$ such that
$\var(B) = \var(A) \backslash X$ and $B \common A$. Adding new variables to a
formula $A$ produces a formula $B$ such that $\var(A) \subseteq \var(B)$ and $B
\common A$. Forgetting followed by adding is $\var(A) \backslash X \subseteq
\var(B)$ and $B \common A$. All three operations are defined in terms of common
equivalence and some simple condition over the variables.

A caveat on variable forgetting and introducing defined in terms of common
equivalence is that the formulae they produce are not always strictly minimal.
For example, forgetting $x$ from a formula built over the alphabet $\{x,y,z\}$
using this definition always produces a formula that contains $y$ and $z$; yet,
a formula that is minimal among the ones that contain $y$ and $z$ may be
equivalent to a smaller one that only contains $z$. Forcing the use of $y$ is
necessary to employ common equivalence, but may artificially increase size.
However, this presence is easily accomplished by subformulae such as $y \vee
\neg y$. They only increase size linearly in the number of variables.

Common equivalence is in line with the view of forgetting as language
reduction~\cite{delg-17}. It is not syntactical, but based on the consequences
on the common alphabet. Viewing the consequences of a formula as an explicit
representation of what the formula tells, $\common$ compares two formulae on
what they say about the things they both talk about.

\

Common equivalence can be defined in alternative ways based on consistency
rather than entailment.

\statetheorem{equisatisfiable}{

The condition $A \common B$ is equivalent to $A \cup S$ and $B \cup S$ being
equisatisfiable for every set of literals $S$ over $\var(A) \cap \var(B)$.

}{

\proof The definition of common equivalence is that $A \models C$ equates $B
\models C$ for every $C$ such that $\var(C) \subseteq \var(A) \cap \var(B)$.
This holds in particular if $C = \neg S$ where $S$ is a conjunction of
variables in $\var(A) \cap \var(B)$. The entailments $A \models \neg S$ and $B
\models \neg S$ are the same as the inconsistency of $A \cup S$ and $B \cup S$,
which therefore coincide.

The other direction is proved by assuming that $A \cup S$ is equisatisfiable
with $B \cup S$ for every conjunction of variables in $\var(A) \cap \var(B)$.
The claim is that $A \models C$ equates $B \models C$ whenever $\var(C)
\subseteq \var(A) \cap \var(B)$. Every formula is equivalent to one in CNF on
the same variables; let $C_1 \wedge \cdots \wedge C_m$ be a formula equivalent
to $C$, where each $C_i$ is a disjunction of literals. By assumption, $A \cup
\neg C_i$ is equisatisfiable with $B \cup \neg C_i$ since $\neg C_i$ is
equivalent to the conjunction of the variables in $C_i$ and $\var(C_i)
\subseteq \var(C) \subseteq \var(A) \cap \var(B)$. This proves that $A \models
C_i$ if and only if $B \models C_i$. Therefore, $A \models C$ if and only if $B
\models C$.~\qed

}

This condition can be further restricted: instead of checking consistency over
all sets of literals over the common alphabet, the ones that contain all common
variables are enough. In other words, the models over the common alphabet are
partial models of both formulae or none, if the formulae are common equivalent.

\statetheorem{equisatisfiable-complete}{

The condition $A \common B$ is equivalent to $A \cup S$ and $B \cup S$ being
equisatisfiable for every set of literals $S$ that contains exactly all
variables that are common to $A$ and $B$.

}{

\proof By Theorem~\ref{equisatisfiable}, if $A \common B$ then $A \cup S$ and
$B \cup S$ are equisatisfiable for every set of literals $S$ on the common
variables; this includes the sets that contain all common variables.

In the other direction, $A \cup S$ is satisfiable if and only if it has a model
$M$. Let $S'$ be the set of literals over $\var(A) \cap \var(B)$ that are
satisfied by $M$. Since both $A$ and $S'$ are satisfied by $M'$, the set $A
\cup S'$ is satisfiable. This proves that $A \cup S$ is satisfiable if and only
$A \cup S'$ is satisfiable for some set of literals over $\var(A) \cap
\var(B)$. As a result, if $A \cup S'$ and $B \cup S'$ are equisatisfiable for
every $S'$, then $A \cup S$ and $B \cup S$ are equisatisfiable as well. This
implies common equivalence by Theorem~\ref{equisatisfiable}.~\qed

}

A situation of particular interest is when one of the two formulae contains
only some variables of the other. It is the case when forgetting some
variables. It is also the case when introducing new variables. It is not when
first forgetting and then introducing variables.

\statelemma{contained}{

If $A \common B$ and $\var(B) \subseteq \var(A)$, then $A \models B$.

}{

\proof Since $\var(B) \subseteq \var(A)$, the common variables are $\var(A)
\cap \var(B) = \var(B)$. By common equivalence, $A \models C$ holds if and only
if $B \models C$ holds for every formula over the common variables. Formula $B$
is over the common alphabet in this case. As a result, $A \models B$ and $B
\models B$ are the same. Since the latter holds, the first follows.~\qed

}

The converse of this lemma does not hold. For example, $B = \{x\}$ does not
entail $A = \{x,y\}$ in spite of their common equivalence. Contrary to regular
equivalence, common equivalence is not the same as mutual implication. It only
contains implication in one direction, and only when the variables of a formula
are all in the other.

This particular case allows for a slight simplification of the definition.

\begin{theorem}
\label{contained-equisat}

If $\var(B) \subseteq \var(A)$, then $A \common B$ holds if and only if $A
\models B$ and the satisfiability of $B \cup S$ implies that of $A \cup S$ for
every consistent set of literals $S$ that contains exactly all variables in
$\var(B)$.

\end{theorem}


\

While equivalence is transitive, common equivalence is not. An example where
both $A \common B$ and $B \common C$ hold but $A \common C$ does not is:

\begin{eqnarray*}
A &=& x \wedge y \\
B &=& x \\
C &=& x \wedge \neg y
\end{eqnarray*}

Transitivity does not hold because $A$ and $C$ share the variable $y$ while
imposing different values on it, violating common equivalence; this variable is
not in $B$, and is therefore not shared between $B$ and $A$ and between $B$ and
$C$; the different values of $y$ in $A$ and $C$ do not prevent their common
equivalence to $B$.

This cannot happen if all variables of $A$ and $C$ are shared with $B$. This is
a general result: transitivity holds in these cases.

\statelemma{transitive}{

If $\var(A) \cap \var(C) \subseteq \var(B)$ then
$A \common B$ and $B \common C$ imply $A \common C$.

}{

\proof By Theorem~\ref{equisatisfiable}, the claim holds if $A \cup S$ and $C
\cup S$ are equisatisfiable for every set of literals $S$ over $\var(A) \cap
\var(C)$.

By assumption, $\var(A) \cap \var(C) \subseteq \var(B)$. Intersecting a set
with one of its supersets does not change it:
{} $\var(A) \cap \var(C) =
{}  \var(A) \cap \var(C) \cap \var(B)$.
Removing a set from an intersection may only enlarge the result:
{} $\var(A) \cap \var(C) \cap \var(B)$ is contained in both
{} $\var(A) \cap \var(B)$ and
{} $\var(C) \cap \var(B)$.
Therefore, the variables of $S$ are all contained in these two sets: the
variables shared between $A$ and $B$ and the variables shared between $B$ and
$C$.

By common equivalence, the consistency of $A \cup S$ is the same as that of $B
\cup S$, which is the same as that of $C \cup S$. This proves that $A \cup S$
and $C \cup S$ are equisatisfiable.~\qed

}

A following result requires a formula that is not in CNF. Every formula can be
turned into CNF, but this may exponentially increase its size. This can be
avoided by adding new variables. Does such an addition affect common
equivalence? The following lemma answers: it does not. If a formula is the
result of adding new variables to another, all properties related to common
equivalence are preserved.

\statelemma{new-variables}{

If $A \common A'$,
$\var(A) \subseteq \var(A')$ and
$(\var(A') \backslash \var(A)) \cap \var(B) = \emptyset$,
then $A \common B$ if and only if $A' \common B$.

}{

\proof The claim is proved by applying the limited transitivity of common
equivalence shown by Lemma~\ref{transitive}: if $A \common B$ and $B \common C$
then $A \common C$ if $\var(A) \cap \var(C) \subseteq \var(B)$. This lemma is
applied twice, the first time with $A', A, B$, the second with $A, A', B$.

\begin{itemize}

\item $A \common B$ is assumed and $A' \common B$ proved;

By standard set theory properties, it holds
{} $\var(A') \cap \var(B) =
{}  ((\var(A') \backslash \var(A)) \cup \var(A)) \cap \var(B) =
{}  (((\var(A') \backslash \var(A)) \cap \var(B)) \cup
{}  (\var(A) \cap \var(B)) =
{}  \emptyset \cup (var(A) \cap \var(B)) = \var(A) \cap \var(B)
{}  \subseteq \var(A)$.

Since $\var(A') \cap \var(B) \subseteq \var(A)$, common equivalence is
transitive in this case: $A' \common A$ and $A \common B$ imply $A' \common B$.

\item $A' \common B$ is assumed and $A \common B$ proved; since $\var(A)
\subseteq \var(A')$, it holds $\var(A) \cap \var(B) \subseteq \var(A') \cap
\var(A) \subseteq \var(A')$. Common equivalence is transitive also in this
case: $A \common A'$ and $A' \common B$ imply $A \common B$.

\end{itemize}

This proves that the assumptions of the lemma imply that $A \common B$ is the
same as $A' \common B$.~\qed

}

A formula that entails a literal is not equivalent to the formula with the
literal replaced by true. Yet, they only differ on that literal: the first
formula entails it, the second does not mention it; their consequences are
otherwise the same. This is exactly what common equivalence formalizes. It is
able to express that adding a literal to a formula and setting its value in the
formula are essentially the same. Regular equivalence it too picky.

\statelemma{replace}{

For every formula $F$ and variable $x$, the common equivalence
$F \cup \{\neg x\} \common F[\bot/x]$ holds.

}{

\proof Since $F[\bot/x]$ does not contain $x$, the variables shared between $F$
and $F[\bot/x]$ are a subset of $\var(F) \backslash \{x\}$. In other words, $x$
is not a shared variable. The claim holds if $F \cup \{\neg x\} \models C$ is
the same as $F[\bot/x] \models C$ for every formula $C$ over $\var(F)
\backslash \{x\}$.

If $F \cup \{\neg x\} \models C$ then
{} $(F \cup \{\neg x\})[\bot/x] \models C[\bot/x]$.
The formula in the left-hand side of this entailment can be rewritten
{} $F[\bot/x] \cup \{\neg x\}[\bot/x]$,
which is the same as
{} $F[\bot/x] \cup \{\neg \bot\}$,
or
{} $F[\bot/x]$.
The formula in the right-hand side $C[\bot/x]$ is the same as $C$ since $C$
does not contain $x$ by assumption. The entailment is therefore the same as
$F[\bot/x] \models C$.

The other direction is proved by expressing $F \cup \{\neg x\}$ by
the Shannon identity and rewriting it.

\begin{eqnarray*}
F \cup \{\neg x\}
& \equiv &
(     x \wedge (F \cup \{\neg x\}[\top/x]) \vee
(\neg x \wedge (F \cup \{\neg x\}[\bot/x])				\\
& \equiv &
(     x \wedge (F[\top/x] \cup \{\neg \top\}) \vee
(\neg x \wedge (F[\bot/x] \cup \{\neg \bot\})				\\
& \equiv &
(     x \wedge (F[\top/x] \cup \{\bot\}) \vee
(\neg x \wedge (F[\bot/x] \cup \{\top\})				\\
& \equiv &
(     x \wedge \bot) \vee 
(\neg x \wedge F[\bot/x])						\\
& \equiv &
\neg x \wedge F[\bot/x]
\end{eqnarray*}

Since $F[\bot/x] \models C$ implies $\neg x \wedge F[\bot/x] \models C$ by
monotonicity of entailment, it also implies $F \cup \{\neg x\} \models C$.~\qed

}

Regular equivalence is not only transitive but also monotonic: if two formulae
are equivalent, they remain equivalent after conjoining both with the same
formula. The same holds for common equivalence when conjoining with a formula
on the shared variables.

\statelemma{monotonic}{

If $A \common B$ then $A \cup C \common B \cup C$
if $\var(C) \subseteq \var(A) \cap \var(B)$.

}{

\proof Since the variables of $C$ are already shared between $A$ and $B$, the
addition of $C$ to $A$ and $B$ does not change their shared variables. In
formulae, $\var(A \cup C) \cap \var(A \cup B) = (\var(A) \cap \var(B)) \cup
(\var(C) \cap \var(B)) \cup (\var(A) \cap \var(C)) \cup (\var(C) \cap
\var(C))$. Factoring out $\var(C)$ from the last three sets of this union turns
it into $(\var(A) \cap \var(B)) \cup (\var(C) \cap \ldots)$. Since $\var(C)
\subseteq \var(A) \cap \var(B)$, the second part of this union is a subset of
the first, which is therefore the same as the whole: $\var(A) \cap \var(B)$.

By Theorem~\ref{equisatisfiable-complete}, the claim is the same as $A \cup C
\cup S$ being equisatisfiable with $B \cup C \cup S$ for every set of literals
$S$ that contains exactly all variables $\var(A) \cap \var(B)$. If $C \cup S$
is unsatisfiable, then both $A \cup C \cup S$ and $B \cup C \cup S$ are
unsatisfiable and the claim is proved. Otherwise, $C \cup S$ is equivalent to
$S$ since $S$ is a set of literals and the variables of $C$ are a subset of
those of $S$. Therefore, $A \cup C \cup S \equiv A \cup S$ and $B \cup C \cup S
\equiv B \cup S$. Since $A$ and $B$ are common equivalent, $A \cup S$ and $B
\cup S$ are either both satisfiable or both unsatisfiable.~\qed

}


\

Flogel et~al.~\cite{flog-etal-93} proved restricted equivalence to be
\conp-complete if the two formulae are Horn. Lang et~al.~\cite{lang-etal-03}
proved that var-equivalence is \P{2}-complete in the general case. Membership
to these classes also apply to common equivalence, which is restricted
equivalence or var equivalence on the variables that are shared among the two
formulae.

Hardness in the general case is proved by the following theorem. It holds even
when the alphabet of a formula is a subset of the other. This is the case with
variables forgetting alone and with variable introduction alone.

\statetheorem{common-hard}{

The problem of establishing whether $A \common B$ is \P{2}-complete.
Hardness holds even if $\var(B) \subseteq \var(A)$.

}{

\proof Common equivalence of $A$ and $B$ can be formulated as:

\[
A \common B \mbox{ iff }
\forall C ~.~ \var(C) \subseteq \var(A) \cap \var(B) \Rightarrow
(A \models C \mbox{ iff } B \models C)
\]

The problem is in \P{2} because it can be reduced to $\forall \exists QBF$.

\

Hardness is proved by reduction from the validity of a formula $\forall X
\exists Y . F$ where $F$ is in DNF. The formulae $A$ and $B$ that correspond to
this QBF are:

\begin{eqnarray*}
A &=& (a \vee F) \wedge (\neg a \vee X)			\\
B &=& \neg a \vee X
\end{eqnarray*}

The variables of these formulae are $\var(A) = \{a\} \cup X \cup Y$ and
$\var(B) = \{a\} \cup X$. The condition $\var(B) \subseteq \var(A)$ is met.

By Theorem~\ref{equisatisfiable-complete}, $A$ and $B$ are common equivalent if
and only if $A \cup S$ and $B \cup S$ are equisatisfiable for every set of
literals $S$ that contains all variables of $\var(A) \cap \var(B)$. In this
case, $\var(A) \cap \var(B) = (\{a\} \cup X \cup Y) \cap (\{a\} \cup X) = \{a\}
\cup X = \var(B)$.

If $S$ is not consistent with $B$ it is not consistent with $A$ either, since
$A$ entails $B$. Therefore, the claim holds if every set $S$ consistent with
$B$ is also consistent with $A$, if the QBF is valid.

Being a disjunction, $B$ is consistent with all sets of literals $S$ that
contain $\neg a$ and arbitrary literals over $X$ and with the single set $S$
that contains $a$ and all of $X$. The latter is consistent with $A$ because $a$
satisfies its first conjunct and $X$ its second.

What remains to be proved is that $A$ is consistent with all sets of literals
$S$ that contain $\neg a$ and arbitrary literals over all variables of $X$ if
and only if the QBF is valid. Since $S$ contains $\neg a$, the union $A \cup S$
simplifies to $((a \vee F) \wedge (\neg a \vee X)) \cup S = ((\false \vee F)
\wedge (\true)) \cup S = F \cup S$. This formula is satisfiable if and only if
a truth evaluation over $Y$ makes $F \cup S$ true. This happens for all sets
$S$ that contain literals over all variables $X$ if it happens for all truth
evaluations over $X$: for every $X$, there exists $Y$ that makes $F$ true.

This proves that $\forall X \exists Y . F$ is valid if and only if $A \common
B$. The formulae $A$ and $B$ are not in CNF, but can be turned so.
Lemma~\ref{new-variables} states that adding new variables to a formula does
not change its common equivalence with other formulae. First, $A$ is turned
into CNF without exponentially increasing its size by adding new
variables~\cite{lesc-conc-09}. Let $A'$ be the result of this transformation.
By Lemma~\ref{new-variables}, $A' \common B$ is the same as $A \common B$.
Since $\forall X \exists Y . F$ is valid if and only if $A \common B$, it is
also valid if and only if $A' \common B$. The same is done for $B$. This proves
the \P{2} hardness of the problem of common equivalence when the formulae are
in CNF.~\qed

}


\

Being a restriction of restricted equivalence, common equivalence is in \conp\ 
as well in the Horn case. In spite of being a restriction, it is also still
\conp-hard. It remains \conp-hard even when a formula is a subset of the other,
which implies that its variables are a subset of those of the other. This is
the case with variable forgetting: the result is a common-equivalent formula on
a subset of the variables. Therefore, establishing whether a formula is a valid
way of forgetting variables from another is \conp-hard in the Horn case. This
is the same problem as checking whether a formula is a valid way of introducing
variables in the other since forgetting is the opposite of introducing. Both
problems are therefore \conp-hard. Hardness holds even if $B$ is a subset of
the Horn clauses of $A$.

\statetheorem{horn-conp-complete}{

If $A$ and $B$ are Horn, establishing whether $A \common B$ is \conp-complete.
Hardness holds even if $B \subseteq A$.

}{

\proof Membership: $A$ and $B$ are not common-equivalent if:

\[
\exists S ~.~
\var(S) \subseteq \var(A) \cap \var(B)
\mbox{ and }
A \cup S \not\models \bot \mbox{ and } 
B \cup S \models \bot
\mbox{ or }
A \cup S \models \bot \mbox{ and } 
B \cup S \not\models \bot
\]

Since the size of $S$ is bounded by the number of variables, this condition is
in \np. This is the converse of common equivalence, which means that common
equivalence is in \conp.

Hardness is proved by showing that the satisfiability of a general formula $F =
\{f_1,\ldots,f_m\}$ over variables $X = \{x_1,\ldots,x_n\}$ is the same as the
non-common equivalence of two Horn formulae $A$ and $B$ with $\var(B) \subseteq
\var(A)$. The satisfiability problem is \np-hard because $F$ is not restricted
to the Horn form.

\begin{eqnarray*}
A &=&
\{\neg x_i \vee \neg n_i \mid x_i \in X\} \cup
\{\neg x_i \vee t_i, \neg n_i \vee t_i \mid x_i \in X\}		\\
&&
\{\neg x_i \vee c_j \mid x_i \in f_j\} \cup
\{\neg n_i \vee c_j \mid \neg x_i \in f_j\} \cup		\\
&&
\{
\neg t_1 \vee \cdots \vee \neg t_n \vee
\neg c_1 \vee \cdots \vee \neg c_m
\}								\\
B &=& \{\neg x_i \vee \neg n_i \mid x_i \in X\}
\end{eqnarray*}

\vbox{
\setlength{\unitlength}{5000sp}%
\begingroup\makeatletter\ifx\SetFigFont\undefined%
\fi\endgroup%
\begin{picture}(1734,2547)(5029,-5386)
\thinlines
{\color[rgb]{0,0,0}\put(5221,-3571){\framebox(360,180){}}
}%
{\color[rgb]{0,0,0}\put(5221,-3931){\framebox(360,180){}}
}%
{\color[rgb]{0,0,0}\put(5221,-4291){\framebox(360,180){}}
}%
{\color[rgb]{0,0,0}\put(6121,-3121){\framebox(540,180){}}
}%
{\color[rgb]{0,0,0}\put(6121,-3391){\framebox(540,180){}}
}%
{\color[rgb]{0,0,0}\put(6121,-3661){\framebox(540,180){}}
}%
{\color[rgb]{0,0,0}\put(6121,-3931){\framebox(540,180){}}
}%
{\color[rgb]{0,0,0}\put(5581,-3436){\line( 4, 3){540}}
}%
{\color[rgb]{0,0,0}\put(5581,-3481){\line( 3, 1){540}}
}%
{\color[rgb]{0,0,0}\put(5581,-3526){\line( 1, 0){540}}
}%
{\color[rgb]{0,0,0}\put(5581,-3841){\line( 1, 0){540}}
}%
{\color[rgb]{0,0,0}\put(5236,-4636){\oval(210,210)[bl]}
\put(5236,-3406){\oval(210,210)[tl]}
\put(5566,-4636){\oval(210,210)[br]}
\put(5566,-3406){\oval(210,210)[tr]}
\put(5236,-4741){\line( 1, 0){330}}
\put(5236,-3301){\line( 1, 0){330}}
\put(5131,-4636){\line( 0, 1){1230}}
\put(5671,-4636){\line( 0, 1){1230}}
}%
{\color[rgb]{0,0,0}\put(5221,-4651){\framebox(360,180){}}
}%
{\color[rgb]{0,0,0}\put(6121,-4561){\framebox(540,180){}}
}%
{\color[rgb]{0,0,0}\put(6121,-4831){\framebox(540,180){}}
}%
{\color[rgb]{0,0,0}\put(5581,-4516){\line( 6, 1){540}}
}%
{\color[rgb]{0,0,0}\put(5581,-4606){\line( 4,-1){540}}
}%
{\color[rgb]{0,0,0}\put(5146,-5086){\oval(210,210)[bl]}
\put(5146,-2956){\oval(210,210)[tl]}
\put(6646,-5086){\oval(210,210)[br]}
\put(6646,-2956){\oval(210,210)[tr]}
\put(5146,-5191){\line( 1, 0){1500}}
\put(5146,-2851){\line( 1, 0){1500}}
\put(5041,-5086){\line( 0, 1){2130}}
\put(6751,-5086){\line( 0, 1){2130}}
}%
\put(5401,-3526){\makebox(0,0)[b]{\smash{{\SetFigFont{12}{24.0}
{\rmdefault}{\mddefault}{\updefault}{\color[rgb]{0,0,0}$M_1$}%
}}}}
\put(5401,-3886){\makebox(0,0)[b]{\smash{{\SetFigFont{12}{24.0}
{\rmdefault}{\mddefault}{\updefault}{\color[rgb]{0,0,0}$M_2$}%
}}}}
\put(5401,-4606){\makebox(0,0)[b]{\smash{{\SetFigFont{12}{24.0}
{\rmdefault}{\mddefault}{\updefault}{\color[rgb]{0,0,0}$M_4$}%
}}}}
\put(5401,-4246){\makebox(0,0)[b]{\smash{{\SetFigFont{12}{24.0}
{\rmdefault}{\mddefault}{\updefault}{\color[rgb]{0,0,0}$M_3$}%
}}}}
\put(6391,-3346){\makebox(0,0)[b]{\smash{{\SetFigFont{12}{24.0}
{\rmdefault}{\mddefault}{\updefault}{\color[rgb]{0,0,0}$M_1^2$}%
}}}}
\put(6391,-3076){\makebox(0,0)[b]{\smash{{\SetFigFont{12}{24.0}
{\rmdefault}{\mddefault}{\updefault}{\color[rgb]{0,0,0}$M_1^1$}%
}}}}
\put(6391,-3616){\makebox(0,0)[b]{\smash{{\SetFigFont{12}{24.0}
{\rmdefault}{\mddefault}{\updefault}{\color[rgb]{0,0,0}$M_1^3$}%
}}}}
\put(6391,-3886){\makebox(0,0)[b]{\smash{{\SetFigFont{12}{24.0}
{\rmdefault}{\mddefault}{\updefault}{\color[rgb]{0,0,0}$M_2^1$}%
}}}}
\put(6391,-4516){\makebox(0,0)[b]{\smash{{\SetFigFont{12}{24.0}
{\rmdefault}{\mddefault}{\updefault}{\color[rgb]{0,0,0}$M_4^1$}%
}}}}
\put(6391,-4786){\makebox(0,0)[b]{\smash{{\SetFigFont{12}{24.0}
{\rmdefault}{\mddefault}{\updefault}{\color[rgb]{0,0,0}$M_4^2$}%
}}}}
\put(5401,-4921){\makebox(0,0)[b]{\smash{{\SetFigFont{12}{24.0}
{\rmdefault}{\mddefault}{\updefault}{\color[rgb]{0,0,0}$B$}%
}}}}
\put(5806,-5371){\makebox(0,0)[b]{\smash{{\SetFigFont{12}{24.0}
{\rmdefault}{\mddefault}{\updefault}{\color[rgb]{0,0,0}$A$}%
}}}}
\end{picture}%
}
\nop{
       M11
M1 --- M12
       M13
                             .
M2 --- M21
                             .
M3
                             .
M4 --- M41
       M42
                             .
--
B
                             .
----------
    A
}

This figure shows how the overall reduction works. Common equivalence requires
every model of $B$ to be extendable to form a model of $A$ with the addition of
the values for the variables not in $B$. For example, $M_1$ can be extended by
the addition of $M_1^1$, $M_1^2$ and $M_1^3$ while $M_3$ cannot. If a model of
$B$ can always be extended this way regardless of $F$, it is irrelevant to
common equivalence; this way, some models can be disregarded as irrelevant to
common equivalence. The only models that matter are those that have an
extension or not depending on $F$. This particular reduction has the models
with $x_i = n_i = \false$ for some $i$ being always extendable, and therefore
irrelevant. Instead, the models where $n_i$ is opposite to $x_i$ for every $i$
can be extended to satisfy $A$ if and only if the values of $X$ falsify $F$.

\

That was an outline of the proof, which is now technically detailed. Since $B$
only comprises $\neg x_i \vee \neg n_i$, its models are exactly those setting
either $x_i$ or $n_i$ to false for each $i$. Common equivalence is achieved if
every such evaluation can be extended to form a model of $A$.

The models where $x_i = n_i = \false$ for some $i$ can be extended by adding
$t_i = \false$ and all other $t_i$ and $c_j$ to true.

The other models have $n_i$ opposite to $x_i$ for every $i$. As a result, the
clauses $\neg n_i \vee c_j${\plural} are the same as $x_i \vee c_j$.

If the truth evaluation over $X$ falsifies $F$, at least a clause $f_j$ is
false: all its literals are false. If $x_i \in f_j$ the variable $x_i$ is
false, and the clause $\neg x_i \vee c_j$ is satisfiable with $c_j = \false$;
if $\neg x_i \in f_j$ then $x_i$ is true and $x_i \vee c_j$ is satisfiable with
$c_j = \false$. By setting all $t_i = \false$ and all other $c_j = \false$, all
clauses of $A$ are satisfied.

If the truth evaluation over $X$ satisfies $F$, for each clause $f_j$ at least
one of its literals is true. If $A$ contains $\neg x_i \vee c_j$ then $x_i$ is
true, and if $A$ contains $x_i \vee c_j$ then $x_i$ is false. The only way to
satisfy these clauses is to set $c_j = \true$. Since $A$ also contains $\neg
x_i \vee t_i$ and $x_i \vee t_i$, also $t_i = \true$ is necessary. The last
clause
{} $\neg t_1 \vee \cdots \vee \neg t_n \vee
{}  \neg c_1 \vee \cdots \vee \neg c_m$
is therefore falsified.

This means that equisatisfiability is lost only if the truth evaluation over
$X$ falsifies $F$. Globally, $A \common B$ only if $F$ is unsatisfiable. This
proves that common equivalence in the Horn case is \conp-hard.~\qed

}

The problem remains \conp-hard even if the formulae are definite Horn. Also in
this case, hardness holds even if $B$ is a subset of the Horn clauses of $A$.

\statetheorem{definite-horn-conp-complete}{

Checking $A \common B$ is \conp-hard even if $A$ and $B$ are definite Horn and
$B \subseteq A$.

}{

\proof The reduction is a variant of that in Theorem~\ref{horn-conp-complete},
where a new variable $z$ is added to all negative clauses of $A$ and $B$,
making them definite.

\begin{eqnarray*}
A &=&
\{z \vee \neg x_i \vee \neg n_i \mid x_i \in X\} \cup
\{\neg x_i \vee t_i, \neg n_i \vee t_i \mid x_i \in X\}		\\
&&
\{\neg x_i \vee c_j \mid x_i \in f_j\} \cup
\{\neg n_i \vee c_j \mid \neg x_i \in f_j\} \cup		\\
&&
\{
z \vee
\neg t_1 \vee \cdots \vee \neg t_n \vee
\neg c_1 \vee \cdots \vee \neg c_m
\}								\\
B &=& \{z \vee \neg x_i \vee \neg n_i \mid x_i \in X\}
\end{eqnarray*}

The models that assign false to $z$ satisfy these formulae if and only if they
satisfy the formulae in the proof of Theorem~\ref{horn-conp-complete}, since
the added literal $z$ is false.

The claim is proved by showing that all models of $B$ with $z=\true$ can be
extended to form a model of $A$. Such a model contains $z=\true$ and an
arbitrary evaluation for the variables $x_i$ and $n_i$. The value of $z$
satisfies all clauses containing it. The only remaining clauses are $\neg x_i
\vee t_i$, $\neg n_i \vee t_i$, $\neg x_i \vee c_j$ and $\neg n_i \vee c_j$.
All of them are satisfied by setting all $t_i$ and $c_j$ to true.~\qed

}


Complexity of common equivalence is established when the problem is formalized
as a decision: are two Horn formulae common-equivalent? The two formulae are
both assumed Horn, excluding the case when forgetting turns a Horn formula into
a non-Horn formula. Or when variable introduction makes a Horn formula
non-Horn.

These cases are both possible, in the sense that a specific procedure for
forgetting may generate non-Horn clauses. As an example, the non-Horn formula
$F' = \{x, y, x \vee y\}$ is a way of forgetting $z$ from the Horn formula $F =
\{x, y, z\}$. Not a natural way to do so, but still valid. Yet, a better way of
forgetting is $F'' = \{x, y\}$, which is Horn. Incidentally, $F''$ is also the
result of minimizing $F'$.

This is not specific to the example, tells the next theorem.

\statetheorem{horn-common-horn}{

If $A$ is Horn, $A \common B$ and $\var(B) \subseteq \var(A)$, then $B$ is
equivalent to a Horn formula.

}{

\proof A Horn formula is satisfied by the intersection of every pair of its
models, where the intersection is the model that assigns true to a variable if
and only if both models assigns it to true.

This holds for $A$ by assumption. It is proved for $B$ as a consequence. Let
$M_1$ and $M_2$ be two models over the alphabet of $B$. Let $S_1$ be the sets
of the literals over $\var(B)$ that are satisfies by $M_1$, and the same for
$M_2$. The union $B \cup S_1$ is satisfiable because $M_1$ satisfies $B$ by
assumption and $S_1$ by construction. The same holds for $B \cup S_2$. Let
$M_1$ be a model of $B \cup S_1$ and $M_2$ of $B \cup S_2$.

By Theorem~\ref{equisatisfiable}, the satisfiability of $B \cup S_1$ implies
that of $A \cup S_1$. Let $M_1'$ be a model of $A \cup S_1$. For the same
reason, $A \cup S_2$ has at least a model $M_2'$.

Since $S_1$ and $S_2$ contain a literal for each variable of $\var(B)$, they
are only satisfied by models that assign the same values of $M_1$ and $M_2$ to
all variables of $\var(B)$. As a result, $M_1'$ and $M_2'$ respectively
coincide with $M_1$ and $M_2$ on the variables of $B$.

Since $A$ is Horn and is satisfied by $M_1'$ and $M_2'$, it is also satisfied
by their intersection $M_3'$. Let $M_3$ be the restriction of $M_3'$ to the
variables of $B$, and $S_3$ be the only set of literals that contains all
variables of $B$ and is satisfied by $M_3$. The union $A \cup S_3$ is
satisfiable because $M_3'$ satisfies it. By Theorem~\ref{equisatisfiable}, $B
\cup S_3$ is also satisfiable. This implies $M_3 \models B$ since $M_3$ is the
only model of $S_3$ on the variables $\var(B)$.

This proves that if $M_1$ and $M_2$ are models of $B$, then $B$ is also
satisfied by their intersection. Therefore, $B$ is equivalent to a Horn
formula.~\qed

}

The theorem holds when the variables of $B$ are a subset of the variables of
$A$, not the opposite. For example, the Horn formula $A = \{x\}$ is
common-equivalent to $B = \{x, y \vee z\}$, which is not Horn and is not
equivalent to any Horn formula. In practice: forgetting variables may turn a
non-Horn formula into a Horn formula. Also: introducing variables may generate
a non-Horn formula.

\section{Size of forgetting}
\label{section-size}

Forgetting may not only fail at reducing size. It may increase it. Even
exponentially. Even with equivalence: the result of forgetting is only
equivalent to exponentially-sized formulae.

An example of size increase is forgetting $x$ from
{} $A =
{}  \{abc \rightarrow x, x \rightarrow l, x \rightarrow m, x \rightarrow n\}$.
This formula tells that $abc$ implies $x$ and $x$ implies $l$. Therefore, it
also tells that $abc$ implies $l$. This implication survives forgetting since
it does not involve $x$. The same holds for $abc$ implying $m$ and $n$.
Forgetting indeed produces
{} $B = \{abc \rightarrow l, abc \rightarrow m, abc \rightarrow n\}$,
which contains more literal occurrences than $A$. This example is in the test
file {\tt enlarge.py} of the forgetting programs described in
Section~\ref{section-python}.

This is not a proof that forgetting may increase size, since $B$ is not the
only way to forget $x$ from $A$. Every formula $B'$ that is equivalent to $B$
is a way to do that. Some of them could be smaller than $B$. The following
lemma proves that this is not the case.

\statelemma{no-smaller-cnf} {

No CNF formula over variables $\{a,b,c,l,m,n\}$ is equivalent to
{} $B = \{abc \rightarrow l, abc \rightarrow m, abc \rightarrow n\}$
and shorter than it.

}{

\proof Let $C$ be a formula equivalent to $B$. This implies $B \models C$.
Therefore, every clause of $C$ is entailed by $B$. All clauses entailed by $B$
are supersets of clauses resulting from resolving some clauses of $B$
\cite{pelt-16}. In other words, every clause of $C$ can be obtained by
resolving clauses of $B$ and then adding literals. But the clauses of $B$ do
not resolve since $a$, $b$ and $c$ only occur negative and $l$, $m$ and $n$
only positive. Therefore, $C$ may only contain supersets of clauses of $B$.

Since the alphabet of $B$ is $\{a,b,c,l,m,n\}$, the only possible supersets of
$abc \rightarrow l$ are
{} $abc \rightarrow l \vee m$,
{} $abc \rightarrow l \vee n$,
{} $abc \rightarrow l \vee m \vee n$,
{} $abcm \rightarrow l$,
{} $abcn \rightarrow l$ and
{} $abcnm \rightarrow l$.

If $C$ only contains some of the first three clauses, it is satisfied by the
model setting $\{a,b,c,m,n\}$ to true and $l$ to false. Since $B$ is falsified
by this model, it is not equivalent to $C$. This proves that every formula $C$
that is equivalent to $B$ contains either the original clause $abc \rightarrow
l$ or at least one of the last three supersets of it:
{} $abcm \rightarrow l$,
{} $abcn \rightarrow l$ and
{} $abcnm \rightarrow l$.

The same holds for the other two clauses of $B$. None of the three supersets of
$abc \rightarrow l$ is also a superset of these other two clauses $abc
\rightarrow m$ and $abc \rightarrow n$, which contain either $m$ or $n$
positive while $abc \rightarrow l$ and its three superset do not. This proves
that the supersets of $abc \rightarrow l$ and the supersets of the other two
clauses of $B$ are all different. Therefore, $C$ contains at least three
clauses, each one being the same size or larger than a clause of $B$.~\qed

}

This proves that $B$ is minimal in its own alphabet. The aim is proving that
forgetting $x$ from $A$ always produces a formula larger than $A$. What is
missing is proving that $B$ is actually the result of forgetting $x$ from $A$.
Formally: $\var(B) = \var(A) \backslash \{x\}$ and $B \common A$. The first
holds because
{} $\var(A) \backslash \{x\} =
{}	\{a,b,c,x,l,m,n\} \backslash \{x\} =
{}	\{a,b,c,l,m,n\} = \var(B)$.
The second is proved by the following lemma.

\statelemma{a-common-b}{

Formulae $A$ and $B$ are common equivalent:

\begin{eqnarray*}
A &=& \{
	abc \rightarrow x,
	x \rightarrow l,
	x \rightarrow m,
	x \rightarrow n
\}
\\
B &=& \{
	abc \rightarrow l,
	abc \rightarrow m,
	abc \rightarrow n
\}
\end{eqnarray*}

}{

\proof The common alphabet of $A$ and $B$ is $\var(A) \cap \var(B) =
\{a,b,c,l,m,n\}$.

The clauses entailed by these formulae are obtained by resolution and literal
adding. Since the clauses of $B$ do not resolve, the consequences of $B$ are
the supersets of the clauses of $B$. The clauses of $A$ only resolve on
variable $x$. The result are exactly the clauses of $B$.~\qed

}

Since $\var(B) = \var(A) \backslash \{x\}$, proving $B \common A$ shows that
$B$ is the result of forgetting $x$ from $A$. Since $B$ is minimal and is
larger than $A$, the claim that forgetting $x$ from $A$ always increases size
is proved.

\begin{theorem}

There exists a definite Horn formula $A$ and a variable $x$ such that for every
CNF formula $B$ if $\var(B) = \var(A) \backslash \{x\}$ and $B \common A$ then
$B$ is larger than $A$.

\end{theorem}

The central point of this theorem is ``for every formula $B$''. No way of
forgetting $x$ from $A$ reduces size or keeps it the same.

Looking at these results in the opposite direction, $B$ may be taken as the
original formula and $A$ as a way of reducing its size by adding the new
variable $x$. This size reduction is only due to the addition of $x$, since $B$
is minimal in its own alphabet. In other words, variable introduction may
reduce the size of otherwise minimal formulae. Because of minimality, such a
reduction is unachievable on the original alphabet.

The analogous result for forgetting is straightforward: $A = \{x,y\}$ is
minimal, but forgetting $x$ reduces it to $\{y\}$, which is smaller. Forgetting
unavoidably enlarges certain formulae, but also shrinks others, even some that
are minimal on their own alphabets.

The size increase or decrease is not limited to a handful of literal
occurrences. In some cases, it may be exponential.

\statelemma{exponential}{

There exists a definite Horn formula $A$ and a set of variables $X$ such that
for every CNF formula $B$ if $\var(B) = \var(A) \backslash X$ and $A \common B$
then $B$ is exponentially larger than $A$.

}{

\proof The proof is given for a set $X = \{a,b,c\}$ of three variables, but
extends to $n$ variables. The formula is the following, also in the test file
{\tt exponential.py} for the programs described in
Section~\ref{section-python}, with variables renamed:

\[
A = \{
	a_1 \rightarrow a, a_2 \rightarrow a,
	b_1 \rightarrow b, b_2 \rightarrow b,
	c_1 \rightarrow c, c_2 \rightarrow c,
	abc \rightarrow x
\}
\]

This formula entails all clauses $a_ib_jc_k \rightarrow x$ for every three
indexes $i,j,k$ in $\{1,2\}$. The number of such clauses is exponential in the
size of $A$. They only contain variables in $\var(A) \backslash X$; therefore,
if $B$ is common equivalent to $A$ and $\var(B) = \var(A) \backslash X$, it
entails all of them.

A clause whose body is a proper subset of $\{a_i,b_j,c_k\}$ is not entailed by
$B$ because it is not by $A$. For example, $A$ is consistent with the model
where $a_i$ and $b_j$ are true and all other variables are false except $a$ and
$b$; therefore, $A$ entails no clause $a_ib_j \rightarrow y$ for any other
variable $y$. Neither does $B$ because of common equivalence.

%

Since $B$ does not entail any clause whose body is a proper subset of
$\{a_i,b_j,c_k\}$, it does not contain any of them as well. Contrary to the
claim, $B$ is assumed not to contain any clause of body $a_ib_jc_k$ either. Let
$M$ be the truth assignment that sets $a_i$, $b_j$ and $c_k$ to true and all
other variables to false. Let $C$ be the body of an arbitrary clause of $B$.
Since $C$ is not a subset (proper or not) of $\{a_i,b_j,c_k\}$, it contains at
least a variable not in $\{a_i,b_j,c_k\}$. As a result, it is falsified by
$M$. The clause of body $C$ is therefore satisfied by $M$ regardless of its
head. This holds for every clause of $B$, making this formula satisfied by $M$.
Since $M$ satisfies $B$ but not $a_ib_jc_k \rightarrow x \in A$, it disproves
the common equivalence of $B$ and $A$.

The assumption leading to contradiction was that $B$ does not contain any
clause of body $a_ib_jc_k$. Its opposite is that $B$ contains some. This holds
for every three indexes $i$, $j$ and $k$, proving that $B$ contains an
exponential number of clauses.~\qed

}

This lemma proves that forgetting $X$ from $A$ always produces an exponential
increase in size. At the same time, it proves that $B$ can be exponentially
compacted by introducing new variables.

The take-away of this section is that forget may increase size, even
exponentially. This is sometimes due to the way forgetting is done but is
sometimes unavoidable, and happens regardless of how forgetting is done. This
is relevant to the following sections, which show algorithms for forgetting.

\section{Algorithm for definite Horn formulae}
\label{section-definite}

Exponentially large output requires exponential time. Yet, producing it may
only require a polynomial amount of memory. For example, the list of numbers
from $0$ to $2^n$ is exponentially longer than $n$, but can be printed by a
loop on a single variable of $n$ bits. Similarly, a formula expressing
forgetting may be exponentially long but may be generated by a program that
only needs a polynomial amount of memory.

Forgetting by $F[\true/x] \vee F[\false/x]$ doubles the size of the formula.
The result of forgetting a set of variables is exponential in their number. An
alternative is to resolve out each variable to forget: each clause that contain
$x$ is resolved with each that contain $\neg x$; the results are kept, the
original removed~\cite{delg-wass-13,delg-17}. For example, forgetting $x$ from
{} $\{x \vee a \vee b, \neg x \vee \neg c, \neg x \vee d \vee \neg f\}$
results in
{} $\{a \vee b \vee \neg c, a \vee b \vee d \vee \neg f\}$.
On Horn formulae, this is equivalent to unfolding $x$~\cite{wang-etal-05}:
replace every negative occurrence of $x$ with the negative literals of a clause
where $x$ occurs positive. For example, forgetting $x$ combines $xab
\rightarrow c$ with $d \rightarrow x$ and $ef \rightarrow x$ to produce $dab
\rightarrow c$ and $efab \rightarrow c$.

Unfolding a variable at time may generate large intermediate formulae even if
the final result is small. An example is forgetting all variables but $a$ and
$b$ from the following formula.

\begin{eqnarray*}
F &=&
\{
a \rightarrow c_1,
a \rightarrow c_2,
c_1 \rightarrow d_1,
c_1 \rightarrow d_2,
c_2 \rightarrow d_3,
c_2 \rightarrow d_4, \\
&&
d_1 \rightarrow e_1,
d_2 \rightarrow e_1,
d_3 \rightarrow e_2,
d_4 \rightarrow e_2,
e_1e_2 \rightarrow b
\}
\end{eqnarray*}

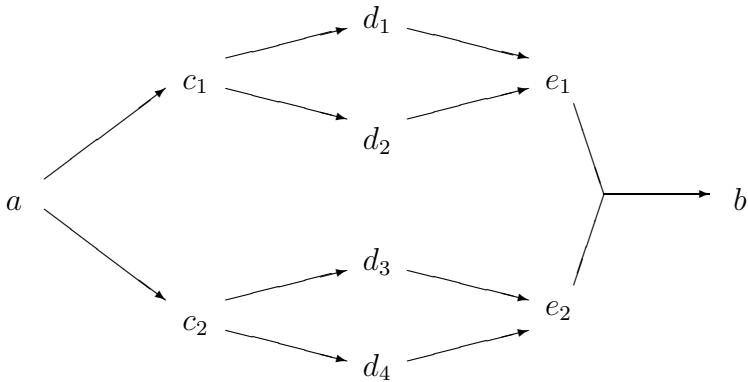
\vbox{
\setlength{\unitlength}{5000sp}%
\begingroup\makeatletter\ifx\SetFigFont\undefined%
\gdef\SetFigFont#1#2#3#4#5{%
  \reset@font\fontsize{#1}{#2pt}%
  \fontfamily{#3}\fontseries{#4}\fontshape{#5}%
  \selectfont}%
\fi\endgroup%
\begin{picture}(3630,1977)(3736,-3703)
\thinlines
{\color[rgb]{0,0,0}\put(6526,-2311){\line( 1,-3){150}}
\put(6676,-2761){\line(-1,-3){150}}
}%
{\color[rgb]{0,0,0}\put(6676,-2761){\vector( 1, 0){525}}
}%
\put(3751,-2836){\makebox(0,0)[b]{\smash{{\SetFigFont{12}{24.0}
{\rmdefault}{\mddefault}{\updefault}{\color[rgb]{0,0,0}$a$}%
}}}}
\put(7351,-2836){\makebox(0,0)[b]{\smash{{\SetFigFont{12}{24.0}
{\rmdefault}{\mddefault}{\updefault}{\color[rgb]{0,0,0}$b$}%
}}}}
{\color[rgb]{0,0,0}\put(5701,-3136){\vector( 4,-1){600}}
}%
{\color[rgb]{0,0,0}\put(5701,-3586){\vector( 4, 1){600}}
}%
\put(6451,-3361){\makebox(0,0)[b]{\smash{{\SetFigFont{12}{24.0}
{\rmdefault}{\mddefault}{\updefault}{\color[rgb]{0,0,0}$e_2$}%
}}}}
{\color[rgb]{0,0,0}\put(5701,-1936){\vector( 4,-1){600}}
}%
{\color[rgb]{0,0,0}\put(5701,-2386){\vector( 4, 1){600}}
}%
\put(6451,-2236){\makebox(0,0)[b]{\smash{{\SetFigFont{12}{24.0}
{\rmdefault}{\mddefault}{\updefault}{\color[rgb]{0,0,0}$e_1$}%
}}}}
{\color[rgb]{0,0,0}\put(4801,-2086){\vector( 4, 1){600}}
}%
\put(5551,-1936){\makebox(0,0)[b]{\smash{{\SetFigFont{12}{24.0}
{\rmdefault}{\mddefault}{\updefault}{\color[rgb]{0,0,0}$d_1$}%
}}}}
{\color[rgb]{0,0,0}\put(4801,-2236){\vector( 4,-1){600}}
}%
{\color[rgb]{0,0,0}\put(4801,-3286){\vector( 4, 1){600}}
}%
{\color[rgb]{0,0,0}\put(4801,-3436){\vector( 4,-1){600}}
}%
\put(5551,-2536){\makebox(0,0)[b]{\smash{{\SetFigFont{12}{24.0}
{\rmdefault}{\mddefault}{\updefault}{\color[rgb]{0,0,0}$d_2$}%
}}}}
\put(5551,-3136){\makebox(0,0)[b]{\smash{{\SetFigFont{12}{24.0}
{\rmdefault}{\mddefault}{\updefault}{\color[rgb]{0,0,0}$d_3$}%
}}}}
\put(5551,-3661){\makebox(0,0)[b]{\smash{{\SetFigFont{12}{24.0}
{\rmdefault}{\mddefault}{\updefault}{\color[rgb]{0,0,0}$d_4$}%
}}}}
{\color[rgb]{0,0,0}\put(3901,-2686){\vector( 4, 3){600}}
}%
{\color[rgb]{0,0,0}\put(3901,-2836){\vector( 4,-3){600}}
}%
\put(4651,-2236){\makebox(0,0)[b]{\smash{{\SetFigFont{12}{24.0}
{\rmdefault}{\mddefault}{\updefault}{\color[rgb]{0,0,0}$c_1$}%
}}}}
\put(4651,-3436){\makebox(0,0)[b]{\smash{{\SetFigFont{12}{24.0}
{\rmdefault}{\mddefault}{\updefault}{\color[rgb]{0,0,0}$c_2$}%
}}}}
\end{picture}%
}
\nop{
         ---> d1 --->
+---> c1              e1 ---+
|        ---> d2 --->       |
a                           +---> b
|        ---> d3 --->       |
+---> c2              e2 ---+
         ---> d4 --->
}

Unfolding $e_1$ turns
{} $e_1e_2 \rightarrow b$ into
{} $d_1e_2 \rightarrow b$ and
{} $d_2e_2 \rightarrow b$.
Unfolding $e_2$ turns the first clause into
{} $d_1d_3 \rightarrow b$ and $d_1d_4 \rightarrow b$
and the second into
{} $d_2d_3 \rightarrow b$ and $d_2d_4 \rightarrow b$.
Unfolding the variables $d_i$ turns all of them into $a \rightarrow b$.

Four clauses result from unfolding two variables. Increasing the number of the
variables $e_i$ from $2$ to $n$ and $d_i$ from $4$ to $2n$ makes the clauses
generated in the process $2^n$, even if only one is eventually output.

Resolving out a variable at time is the same as unfolding on Horn clauses;
therefore, it produces the same clauses.

The culprit is not resolution or unfolding, but the strategy of forgetting one
variable at time: forgetting $e_1$ turns $e_1e_2 \rightarrow b$ into two
clauses; forgetting $e_2$ makes them four. Forgetting a variable at time is
common for example in Answer Set Programming: most algorithms forget a single
variable, which implies that forgetting a set is done one variable at time~%
\cite{eite-wang-06,bert-etal-19,knor-alfe-14,zhan-foo-06}. Incidentally,
forgetting in Answer Set Programming is significantly different from forgetting
in propositional logic, as discussed in Section~\ref{section-conclusions}.


Exponential memory is not required in this example. The trick is to only do one
substitution at time in one clause at time. The other clauses and the other
substitutions are left waiting until the current clause is over. Starting with
{} $e_1e_2 \rightarrow b$,
its body variable $e_1$ can be replaced by $d_1$ and by $d_2$, but only the
first substitution is carried over; the other is left waiting. The result is
the single clause
{} $d_1e_2 \rightarrow b$.

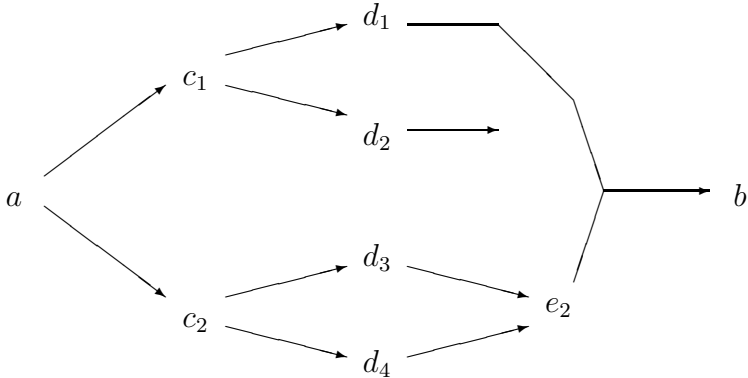
\vbox{
\setlength{\unitlength}{5000sp}%
\begingroup\makeatletter\ifx\SetFigFont\undefined%
\gdef\SetFigFont#1#2#3#4#5{%
  \reset@font\fontsize{#1}{#2pt}%
  \fontfamily{#3}\fontseries{#4}\fontshape{#5}%
  \selectfont}%
\fi\endgroup%
\begin{picture}(3630,1977)(3736,-3703)
\thinlines
{\color[rgb]{0,0,0}\put(6526,-2311){\line( 1,-3){150}}
\put(6676,-2761){\line(-1,-3){150}}
}%
{\color[rgb]{0,0,0}\put(6676,-2761){\vector( 1, 0){525}}
}%
\put(3751,-2836){\makebox(0,0)[b]{\smash{{\SetFigFont{12}{24.0}
{\rmdefault}{\mddefault}{\updefault}{\color[rgb]{0,0,0}$a$}%
}}}}
\put(7351,-2836){\makebox(0,0)[b]{\smash{{\SetFigFont{12}{24.0}
{\rmdefault}{\mddefault}{\updefault}{\color[rgb]{0,0,0}$b$}%
}}}}
{\color[rgb]{0,0,0}\put(5701,-3136){\vector( 4,-1){600}}
}%
{\color[rgb]{0,0,0}\put(5701,-3586){\vector( 4, 1){600}}
}%
\put(6451,-3361){\makebox(0,0)[b]{\smash{{\SetFigFont{12}{24.0}
{\rmdefault}{\mddefault}{\updefault}{\color[rgb]{0,0,0}$e_2$}%
}}}}
{\color[rgb]{0,0,0}\put(4801,-2086){\vector( 4, 1){600}}
}%
\put(5551,-1936){\makebox(0,0)[b]{\smash{{\SetFigFont{12}{24.0}
{\rmdefault}{\mddefault}{\updefault}{\color[rgb]{0,0,0}$d_1$}%
}}}}
{\color[rgb]{0,0,0}\put(4801,-2236){\vector( 4,-1){600}}
}%
{\color[rgb]{0,0,0}\put(4801,-3286){\vector( 4, 1){600}}
}%
{\color[rgb]{0,0,0}\put(4801,-3436){\vector( 4,-1){600}}
}%
\put(5551,-2536){\makebox(0,0)[b]{\smash{{\SetFigFont{12}{24.0}
{\rmdefault}{\mddefault}{\updefault}{\color[rgb]{0,0,0}$d_2$}%
}}}}
\put(5551,-3136){\makebox(0,0)[b]{\smash{{\SetFigFont{12}{24.0}
{\rmdefault}{\mddefault}{\updefault}{\color[rgb]{0,0,0}$d_3$}%
}}}}
\put(5551,-3661){\makebox(0,0)[b]{\smash{{\SetFigFont{12}{24.0}
{\rmdefault}{\mddefault}{\updefault}{\color[rgb]{0,0,0}$d_4$}%
}}}}
{\color[rgb]{0,0,0}\put(3901,-2686){\vector( 4, 3){600}}
}%
{\color[rgb]{0,0,0}\put(3901,-2836){\vector( 4,-3){600}}
}%
\put(4651,-2236){\makebox(0,0)[b]{\smash{{\SetFigFont{12}{24.0}
{\rmdefault}{\mddefault}{\updefault}{\color[rgb]{0,0,0}$c_1$}%
}}}}
\put(4651,-3436){\makebox(0,0)[b]{\smash{{\SetFigFont{12}{24.0}
{\rmdefault}{\mddefault}{\updefault}{\color[rgb]{0,0,0}$c_2$}%
}}}}
{\color[rgb]{0,0,0}\put(5701,-2461){\vector( 1, 0){450}}
}%
{\color[rgb]{0,0,0}\put(5701,-1936){\line( 1, 0){450}}
\put(6151,-1936){\line( 1,-1){375}}
}%
\end{picture}%
}
\nop{
         ---> d1 -----------+
+---> c1                    |
|        ---> d2 --->       |
a                           +---> b
|        ---> d3 --->       |
+---> c2              e2 ---+
         ---> d4 --->
}

The first variable $d_1$ of
{} $d_1e_2 \rightarrow b$
can be replaced by $c_1$ only, producing
{} $c_1e_2 \rightarrow b$.

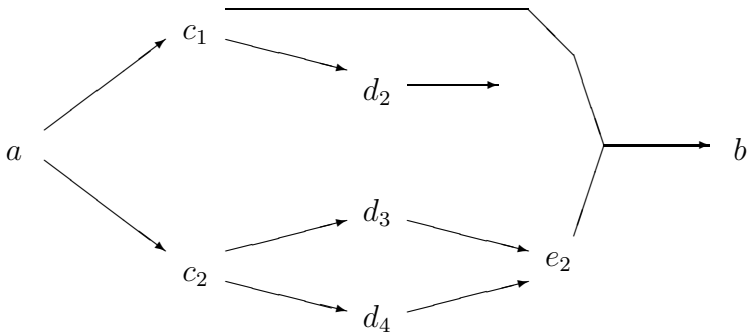
\vbox{
\setlength{\unitlength}{5000sp}%
\begingroup\makeatletter\ifx\SetFigFont\undefined%
\gdef\SetFigFont#1#2#3#4#5{%
  \reset@font\fontsize{#1}{#2pt}%
  \fontfamily{#3}\fontseries{#4}\fontshape{#5}%
  \selectfont}%
\fi\endgroup%
\begin{picture}(3630,1977)(3736,-3703)
\thinlines
{\color[rgb]{0,0,0}\put(6526,-2311){\line( 1,-3){150}}
\put(6676,-2761){\line(-1,-3){150}}
}%
{\color[rgb]{0,0,0}\put(6676,-2761){\vector( 1, 0){525}}
}%
\put(3751,-2836){\makebox(0,0)[b]{\smash{{\SetFigFont{12}{24.0}
{\rmdefault}{\mddefault}{\updefault}{\color[rgb]{0,0,0}$a$}%
}}}}
\put(7351,-2836){\makebox(0,0)[b]{\smash{{\SetFigFont{12}{24.0}
{\rmdefault}{\mddefault}{\updefault}{\color[rgb]{0,0,0}$b$}%
}}}}
{\color[rgb]{0,0,0}\put(5701,-3136){\vector( 4,-1){600}}
}%
{\color[rgb]{0,0,0}\put(5701,-3586){\vector( 4, 1){600}}
}%
\put(6451,-3361){\makebox(0,0)[b]{\smash{{\SetFigFont{12}{24.0}
{\rmdefault}{\mddefault}{\updefault}{\color[rgb]{0,0,0}$e_2$}%
}}}}
{\color[rgb]{0,0,0}\put(4801,-2236){\vector( 4,-1){600}}
}%
{\color[rgb]{0,0,0}\put(4801,-3286){\vector( 4, 1){600}}
}%
{\color[rgb]{0,0,0}\put(4801,-3436){\vector( 4,-1){600}}
}%
\put(5551,-2536){\makebox(0,0)[b]{\smash{{\SetFigFont{12}{24.0}
{\rmdefault}{\mddefault}{\updefault}{\color[rgb]{0,0,0}$d_2$}%
}}}}
\put(5551,-3136){\makebox(0,0)[b]{\smash{{\SetFigFont{12}{24.0}
{\rmdefault}{\mddefault}{\updefault}{\color[rgb]{0,0,0}$d_3$}%
}}}}
\put(5551,-3661){\makebox(0,0)[b]{\smash{{\SetFigFont{12}{24.0}
{\rmdefault}{\mddefault}{\updefault}{\color[rgb]{0,0,0}$d_4$}%
}}}}
{\color[rgb]{0,0,0}\put(3901,-2686){\vector( 4, 3){600}}
}%
{\color[rgb]{0,0,0}\put(3901,-2836){\vector( 4,-3){600}}
}%
\put(4651,-2236){\makebox(0,0)[b]{\smash{{\SetFigFont{12}{24.0}
{\rmdefault}{\mddefault}{\updefault}{\color[rgb]{0,0,0}$c_1$}%
}}}}
\put(4651,-3436){\makebox(0,0)[b]{\smash{{\SetFigFont{12}{24.0}
{\rmdefault}{\mddefault}{\updefault}{\color[rgb]{0,0,0}$c_2$}%
}}}}
{\color[rgb]{0,0,0}\put(5701,-2461){\vector( 1, 0){450}}
}%
{\color[rgb]{0,0,0}\put(4801,-2086){\line( 1, 0){1500}}
\put(6301,-2086){\line( 1,-1){225}}
}%
\end{picture}%
}
\nop{
+---> c1 -------------------+
|        ---> d2 --->       |
a                           +---> b
|        ---> d3 --->       |
+---> c2              e2 ---+
         ---> d4 --->
}

The first variable $c_1$ of
{} $c_1e_2 \rightarrow b$
can only be replaced by $a$, producing
{} $ae_2 \rightarrow b$.

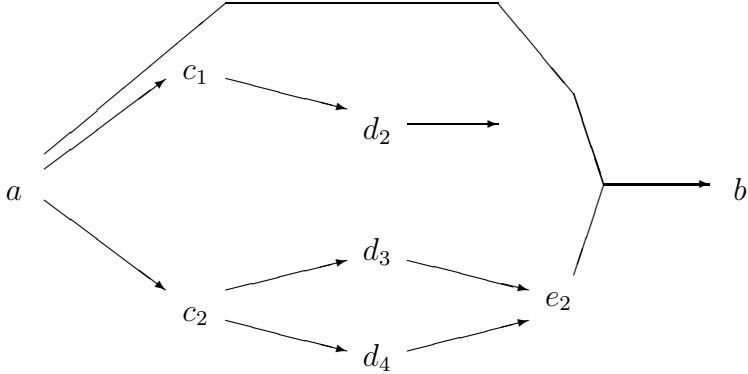
\vbox{
\setlength{\unitlength}{5000sp}%
\begingroup\makeatletter\ifx\SetFigFont\undefined%
\gdef\SetFigFont#1#2#3#4#5{%
  \reset@font\fontsize{#1}{#2pt}%
  \fontfamily{#3}\fontseries{#4}\fontshape{#5}%
  \selectfont}%
\fi\endgroup%
\begin{picture}(3630,1977)(3736,-3703)
\thinlines
{\color[rgb]{0,0,0}\put(6526,-2311){\line( 1,-3){150}}
\put(6676,-2761){\line(-1,-3){150}}
}%
{\color[rgb]{0,0,0}\put(6676,-2761){\vector( 1, 0){525}}
}%
\put(3751,-2836){\makebox(0,0)[b]{\smash{{\SetFigFont{12}{24.0}
{\rmdefault}{\mddefault}{\updefault}{\color[rgb]{0,0,0}$a$}%
}}}}
\put(7351,-2836){\makebox(0,0)[b]{\smash{{\SetFigFont{12}{24.0}
{\rmdefault}{\mddefault}{\updefault}{\color[rgb]{0,0,0}$b$}%
}}}}
{\color[rgb]{0,0,0}\put(5701,-3136){\vector( 4,-1){600}}
}%
{\color[rgb]{0,0,0}\put(5701,-3586){\vector( 4, 1){600}}
}%
\put(6451,-3361){\makebox(0,0)[b]{\smash{{\SetFigFont{12}{24.0}
{\rmdefault}{\mddefault}{\updefault}{\color[rgb]{0,0,0}$e_2$}%
}}}}
{\color[rgb]{0,0,0}\put(4801,-2236){\vector( 4,-1){600}}
}%
{\color[rgb]{0,0,0}\put(4801,-3286){\vector( 4, 1){600}}
}%
{\color[rgb]{0,0,0}\put(4801,-3436){\vector( 4,-1){600}}
}%
\put(5551,-2536){\makebox(0,0)[b]{\smash{{\SetFigFont{12}{24.0}
{\rmdefault}{\mddefault}{\updefault}{\color[rgb]{0,0,0}$d_2$}%
}}}}
\put(5551,-3136){\makebox(0,0)[b]{\smash{{\SetFigFont{12}{24.0}
{\rmdefault}{\mddefault}{\updefault}{\color[rgb]{0,0,0}$d_3$}%
}}}}
\put(5551,-3661){\makebox(0,0)[b]{\smash{{\SetFigFont{12}{24.0}
{\rmdefault}{\mddefault}{\updefault}{\color[rgb]{0,0,0}$d_4$}%
}}}}
{\color[rgb]{0,0,0}\put(3901,-2686){\vector( 4, 3){600}}
}%
{\color[rgb]{0,0,0}\put(3901,-2836){\vector( 4,-3){600}}
}%
\put(4651,-2236){\makebox(0,0)[b]{\smash{{\SetFigFont{12}{24.0}
{\rmdefault}{\mddefault}{\updefault}{\color[rgb]{0,0,0}$c_1$}%
}}}}
\put(4651,-3436){\makebox(0,0)[b]{\smash{{\SetFigFont{12}{24.0}
{\rmdefault}{\mddefault}{\updefault}{\color[rgb]{0,0,0}$c_2$}%
}}}}
{\color[rgb]{0,0,0}\put(5701,-2461){\vector( 1, 0){450}}
}%
{\color[rgb]{0,0,0}\put(3901,-2611){\line( 6, 5){900}}
\put(4801,-1861){\line( 1, 0){1350}}
\put(6151,-1861){\line( 5,-6){375}}
\put(6526,-2311){\line( 1,-3){150}}
}%
\end{picture}%
}
\nop{
+---------------------------+
|+--> c1                    |
||       ---> d2 --->       |
a                           +---> b
|        ---> d3 --->       |
+---> c2              e2 ---+
         ---> d4 --->
}

Since $a$ is not a variable to forget, it is not replaced. The only variable to
forget in
{} $ae_2 \rightarrow b$
is $e_2$. It can be replaced by $d_3$ and by $d_4$, but only the first is
carried over; the other is left waiting. What results from replacing $e_2$ with
$d_3$ is
{} $ad_3 \rightarrow b$.

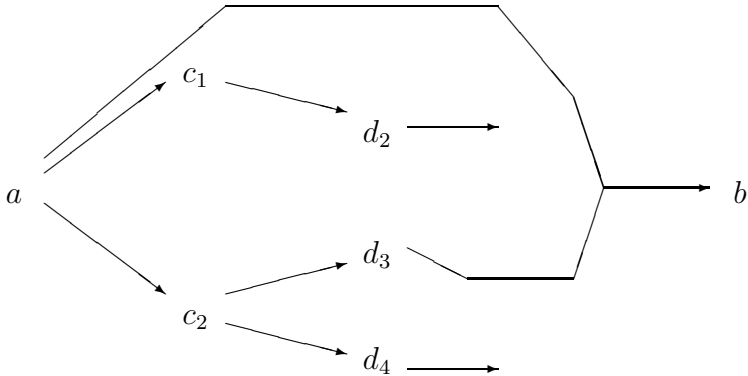
\vbox{
\setlength{\unitlength}{5000sp}%
\begingroup\makeatletter\ifx\SetFigFont\undefined%
\gdef\SetFigFont#1#2#3#4#5{%
  \reset@font\fontsize{#1}{#2pt}%
  \fontfamily{#3}\fontseries{#4}\fontshape{#5}%
  \selectfont}%
\fi\endgroup%
\begin{picture}(3630,1977)(3736,-3703)
\thinlines
{\color[rgb]{0,0,0}\put(6526,-2311){\line( 1,-3){150}}
\put(6676,-2761){\line(-1,-3){150}}
}%
{\color[rgb]{0,0,0}\put(6676,-2761){\vector( 1, 0){525}}
}%
\put(3751,-2836){\makebox(0,0)[b]{\smash{{\SetFigFont{12}{24.0}
{\rmdefault}{\mddefault}{\updefault}{\color[rgb]{0,0,0}$a$}%
}}}}
\put(7351,-2836){\makebox(0,0)[b]{\smash{{\SetFigFont{12}{24.0}
{\rmdefault}{\mddefault}{\updefault}{\color[rgb]{0,0,0}$b$}%
}}}}
{\color[rgb]{0,0,0}\put(4801,-2236){\vector( 4,-1){600}}
}%
{\color[rgb]{0,0,0}\put(4801,-3286){\vector( 4, 1){600}}
}%
{\color[rgb]{0,0,0}\put(4801,-3436){\vector( 4,-1){600}}
}%
\put(5551,-2536){\makebox(0,0)[b]{\smash{{\SetFigFont{12}{24.0}
{\rmdefault}{\mddefault}{\updefault}{\color[rgb]{0,0,0}$d_2$}%
}}}}
\put(5551,-3136){\makebox(0,0)[b]{\smash{{\SetFigFont{12}{24.0}
{\rmdefault}{\mddefault}{\updefault}{\color[rgb]{0,0,0}$d_3$}%
}}}}
\put(5551,-3661){\makebox(0,0)[b]{\smash{{\SetFigFont{12}{24.0}
{\rmdefault}{\mddefault}{\updefault}{\color[rgb]{0,0,0}$d_4$}%
}}}}
{\color[rgb]{0,0,0}\put(3901,-2686){\vector( 4, 3){600}}
}%
{\color[rgb]{0,0,0}\put(3901,-2836){\vector( 4,-3){600}}
}%
\put(4651,-2236){\makebox(0,0)[b]{\smash{{\SetFigFont{12}{24.0}
{\rmdefault}{\mddefault}{\updefault}{\color[rgb]{0,0,0}$c_1$}%
}}}}
\put(4651,-3436){\makebox(0,0)[b]{\smash{{\SetFigFont{12}{24.0}
{\rmdefault}{\mddefault}{\updefault}{\color[rgb]{0,0,0}$c_2$}%
}}}}
{\color[rgb]{0,0,0}\put(5701,-2461){\vector( 1, 0){450}}
}%
{\color[rgb]{0,0,0}\put(3901,-2611){\line( 6, 5){900}}
\put(4801,-1861){\line( 1, 0){1350}}
\put(6151,-1861){\line( 5,-6){375}}
\put(6526,-2311){\line( 1,-3){150}}
}%
{\color[rgb]{0,0,0}\put(5701,-3061){\line( 2,-1){300}}
\put(6001,-3211){\line( 1, 0){525}}
}%
{\color[rgb]{0,0,0}\put(5701,-3661){\vector( 1, 0){450}}
}%
\end{picture}%
}
\nop{
+---------------------------+
|+--> c1 ---> d2 --->       |
||
a                           +---> b
|        ---> d3 -----------+
+---> c2 
         ---> d4 --->
}

Replacing $d_3$ with $c_2$ in
{} $ad_3 \rightarrow b$
produces
{} $ac_2 \rightarrow b$,
where replacing $c_2$ with $a$ produces
{} $a \rightarrow b$.
This clause does not contain variables to forget. It is therefore output and
not further processed.

One of the substitutions on hold is now restarted. The last was replacing $e_2$
with $d_4$ in
{} $ae_2 \rightarrow b$.
Its effect is
{} $ad_4 \rightarrow b$.
The variable $d_4$ is replaced with $c_2$, which is then replaced with $a$. The
result is $a \rightarrow b$ again. The other replacement still waiting is
replacing $e_1$ with $d_2$ in $e_1e_2 \rightarrow b$, which produces the same
result.

Whenever a variable can be replaced in two or more ways, only one is done. The
others are stopped. Contrary to unfolding a variable at time, a line of
replacements is followed until no longer possible; only then the alternatives
are considered. Replaced variables are not replaced again to avoid looping.

How much memory is required? The current clause is not larger than the number
of variables, and this is linear space. Every variable is only replaced once,
meaning that the replacements done on the current clause are linear. Each
replacement may have alternatives: every alternative is another clause to
replace one variable in a clause. The number of alternatives is at most the
number of clauses at each step. The memory required to store all of them is
quadratic at most. Quadratic is polynomial.

While the required memory is polynomial, the produced output may be
exponential. This is unavoidable in general. Also, while the algorithm works in
polynomial space, it may take time exponential in the size of the output. This
is because the same output clause may be obtained in several ways.

\begin{eqnarray*}
F &=& \{
k \rightarrow a, k \rightarrow b,
l \rightarrow d, l \rightarrow e,
m \rightarrow g, m \rightarrow h,
\\
&&
a \rightarrow c, b \rightarrow c,
d \rightarrow f, e \rightarrow f,
g \rightarrow i, h \rightarrow i,
cfi \rightarrow j
\}
\end{eqnarray*}

Forgetting $\{a,b,c,d,e,f,i\}$ from $F$ results in the single clause $klm
\rightarrow j$. Yet, this clause is generated in eight different ways. Starting
from $cfi \rightarrow j$, the first premise $c$ can be replaced by either $a$
or $b$, the second $f$ by either $d$ or $e$, and the third $i$ by either $g$ or
$h$, for a total of eight nondeterministic branches. All of them eventually
produce $klm \rightarrow j$, but none could be cut short before realizing that
the generated clause is already generated. This example is in the test file
{\tt branches.py} for the programs described in Section~\ref{section-python}.

\subsection{Summary}
\label{section-definite-summary}

The proof of correctness of the method is long. A summary is given here.

The first step is a basic property of entailment in Horn logic:
Lemma~\ref{set-implies-set} in Section~\ref{section-definite-set-implies-set}
shows that if a formula entails a non-tautological Horn clause, it contains a
clause with the same head and with the body entailed by that of the entailed
clause.

The core of the method is a nondeterministic procedure $\mbox{\bf
body\_replace}(F,R,D)$, presented and analyzed in
Section~\ref{section-definite-body-replace}. It recursively replaces some
variables of $R$ with their premises in $F$, where $D${\plural} are the
variables already replaced. This procedure has two return values. The first is
a possible result of replacing variables in $R$. The second is the set of
variables that have been replaced. Once a variable is replaced with others, it
is not replaced again but just deleted; this is essential to avoid looping, as
otherwise the algorithm would replace variables in cycles of clauses forever.
Calling
{} $\mbox{\bf body\_replace}(F,P,\emptyset)$
with certain nondeterministic choices produces a piece of the result of
forgetting. Namely, if $P$ is the head of a clause $P \rightarrow x$ of $F$,
the first return value $P'$ is the body of a clause $P' \rightarrow x$ in the
result of forgetting. Collecting all possible results makes the result of
forgetting.

Most of the work of forgetting is done by
{} $\mbox{\bf body\_replace}(F,P,\emptyset)$.
Section~\ref{section-definite-head-implicates} presents a procedure $\mbox{\bf
head\_implicates}(F)$ that calls it on all bodies of the clauses of $F$. With
certain nondeterministic choices, it returns the result of forgetting a set of
variables from $F$.

This procedure is in turn called by $\mbox{\bf common\_equivalent}(A,B)$ to
generate all clauses of forget, checking whether each is entailed by the other
formula. It only takes polynomial space, which is not obvious since the result
of forgetting may be exponentially large. This is the content of
Section~\ref{section-definite-common}.

Finally, $\mbox{\bf forget}(F,X)$ forgets variables $X$ from formula $F$. This
procedure is described in Section~\ref{section-definite-forget}.

All these algorithms become deterministic if each variable is the head of at
most one clause. Since nondeterminism is what requires exponential time, this
restriction makes forgetting and checking common equivalence polynomial-time. A
formula satisfying this condition is called single-head.
Section~\ref{section-definite-singlehead} proves it makes $\mbox{\bf
common\_equivalent}(A,B)$ run in polynomial time.

\subsection{Set implies set}
\label{section-definite-set-implies-set}

The base of the algorithm for forgetting is the following lemma. It states that
every implication from a Horn formula requires its conclusion to be the head of
some clause whose body is entailed by its premises and the formula. This result
can be pushed a little further, because the latter implication only requires a
subset of the formula.

\begin{definition}

For every set of clauses $F$, the set $F^x$ contains all clauses of $F$ that
contain neither $x$ nor $\neg x$.

\end{definition}

In this article, clauses are assumed not to be tautologies: unless otherwise
noted, writing $P \rightarrow x$ implicitly presumes $x \not\in P$. This
condition is sometimes stated explicitly when important.

\statelemma{set-implies-set}{

If $F$ is a definite Horn formula, the following three conditions are
equivalent, where $P' \rightarrow x$ is not a tautology ($x \not \in P'$).

\begin{enumerate}

\item $F \models P' \rightarrow x$;

\item $F^x \cup P' \models P$ where $P \rightarrow x \in F$;

\item $F \cup P' \models P$ where $P \rightarrow x \in F$.

\end{enumerate}

}{

\proof The second condition implies the third by monotonicity of entailment.

The third entails the first because $P \rightarrow x \in F$ implies $F \models
P \rightarrow x$, which in turn implies $F \cup P \models x$. With $F \cup P'
\models P$, it implies $F \cup P' \models x$ by transitivity of entailment. By
the deduction theorem, $F \models P' \rightarrow x$ follows.

What remains to be proved is that the first condition implies the second.
{} The assumption is $F \models P' \rightarrow x$;
{} the claim is $F^x \cup P' \models P$ where $P \rightarrow x \in F$.

If $F$ does not contain a positive occurrence of $x$, it is satisfied by the
model that assigns all variables to true but $x$, since every clause of $F$
contains a positive variable that is not $x$. This model falsifies $P'
\rightarrow x$, contrary to the assumption. This proves that $F$ contains at
least a clause $P \rightarrow x$.

The claim is that
{} $F^x \cup P' \models P$ holds for some clause $P \rightarrow x \in F$.
Its contrary is that
{} $F^x \cup P' \not\models P$
holds for every clause $P \rightarrow x \in F$. This is proved impossible.

The condition
{} $F^x \cup P' \not\models P$
implies the existence of a model $M_P$ such that
{} $M_P \models F^x \cup P'$ and $M_P \not\models P$.
Let $M$ be the intersection of all these models $M_P$ for every $P \rightarrow
x \in F$: the model that evaluates to true exactly the variables that are true
in all these models $M_P$. Alternatively, it is the model that sets to false
exactly the variables that are false in at least one model $M_P$.

Since $F^x \cup P'$ is Horn and is satisfied by all models $M_P$, it is also
satisfied by their intersection $M$. Since $M_P$ does not satisfy $P$, which is
a set of positive literals, $M_P$ assigns at least a variable of $P$ to false;
by construction, that variable is also false in $M$. Therefore, $M \not\models
P$ for all $P \rightarrow x \in F$.

Let $M_{x=\false}$ be the model that assigns $x$ to false and all other
variables to the same value $M$ does. This model will be proved to satisfy $F
\cup P'$ but not $x$, contradicting the assumption $F \models P' \rightarrow
x$.

Since $M \models F^x$ and $F^x$ does not contain $x$, the condition
$M_{x=\false} \models F^x$ follows. Since $M \not\models P$ and $x \not\in P$,
it follows $M_{x=\false} \not\models P$, which implies $M_{x=\false} \models P
\rightarrow x$; this holds for every $P \rightarrow x \in F$. Since $x$ is
false in $M_{x=\false}$, this model also satisfies all clauses of $F$
containing $x$ in the body. This proves that $M_{x=\false}$ implies all clauses
of $F^x$, all clauses of $F$ containing $x$ in the head and all containing $x$
in the body. As a result, $M_{x=\false} \models F$. Since $x$ is not in $P'$
and $M \models P'$, also $M_{x=\false} \models P'$ holds. Since $M_{x=\false}$
sets $x$ to false, $F \cup P' \not\models x$ holds. By the deduction theorem,
$F \not\models P' \rightarrow x$, contrary to the assumption.~\qed

}

This lemma proves that all clauses $P' \rightarrow x$ entailed by $F$ are
either themselves in $F$, or are consequences of another clause $P \rightarrow
x \in F$ thanks to $F^x$ implying $P' \rightarrow P$. Seen from a different
angle, all derivations of $x$ from $P'$ use $P \rightarrow x \in F$ as the last
step, and only clauses of $F^x$ in the previous.

The condition $F^x \cup P' \models P$ can be rewritten as $F^x \models P'
\rightarrow y$ for every $y \in P$. This allows applying the lemma again, to
$P' \rightarrow y$: there exists $P'' \rightarrow y \in F$ such that $F^{xy}
\models P' \rightarrow P''$.

This recursion is not infinite: it ends when $P' = P$; not only the lemma does
not rule this case out, it is its base case. More generally, it does not forbid
$P$ and $P'$ to intersect. Rather the opposite: at some point back in the
recursion they must coincide. Indeed, $F$ becomes $F^x$, then $F^{xy}$ and so
on; infinite recursion is not possible because the formula becomes smaller at
every step; therefore, it ends up empty, and the lemma implies that the formula
contains at least a clause.

\subsection{body\_replace()}
\label{section-definite-body-replace}

The recursive application of Lemma~\ref{set-implies-set} allows for repeatedly
replacing unwanted variables with others implying them. When forgetting $y$,
every entailed clause $F \models P' \rightarrow x$ not containing $y$ must
survive. By Lemma~\ref{set-implies-set}, the entailment implies $F^x \models P'
\rightarrow P$ and $P \rightarrow x \in F$. If $y \not\in P$, clause $P
\rightarrow x$ survives the removal of $y$. If $y \in P$,
Lemma~\ref{set-implies-set} kicks in: $F^x \models P' \rightarrow y$ requires
some clauses $P'' \rightarrow y \in F$. Each can be combined with $P
\rightarrow x$ to obtain $((P'' \cup P) \backslash \{y\}) \rightarrow x$, which
does not contain $y$. This combination leaves $P' \rightarrow x$ entailed in
spite of the removal of $P \rightarrow x$. The same procedure allows forgetting
other variables at the same time.

\

\hrule
\begingroup
\obeylines
~\newline
\noindent \#\# %
recursively replace some variables in $R$ with their preconditions in $F$
\noindent \# %
input $F$: a formula
\noindent \# %
input $R$: a set of variables, some of which are replaced
\noindent \# %
input $D$: a set of variables to delete
\noindent \# %
first output: $R$ with some variables replaced
\noindent \# %
second output: the variables replaced in the process
~
\endgroup
\begingroup
\parskip=-10pt
\noindent
$\mbox{variables},\mbox{variables }
	\mbox{\bf body\_replace}(
		\mbox{formula } F,
		\mbox{ variables } R,
		\mbox{ variables } D)$

\begin{enumerate}
\parskip=-5pt

\item choose $R' \subseteq R \backslash D$

\item $S' = \emptyset$

\item $E' = \emptyset$

\item foreach $y \in R'$
\label{bodyreplace-loop}

\begin{enumerate}

\item if $y \in D \cup E'$ continue

\item choose $P \rightarrow y \in F$ else {\bf fail}

\item $S,E = \mbox{\bf body\_replace}(F^y, P, D \cup E')$
\label{bodyreplace-recourse}

\item $S' = S' \cup S$

\item $E' = E' \cup E \cup \{y\}$

\end{enumerate}

\item return $(R \backslash D \backslash R') \cup S', E'$

\end{enumerate}
\endgroup
\hrule

\

The choice of the initial values of $R$ and $D$ and the choice of $R'$ is left
open at this point because it simplifies some proofs, but the intention is that
initially $R$ is the body of some clause $R \rightarrow x$ where $x$ is a
variable not to be forgotten, $D$ is initially empty and $R'$ contains exactly
all variables of $R \backslash D$ to forget. In each call:

\begin{itemize}

\item $R$ is the body of some clause $R \rightarrow x$ where $x$ is a variable
not to forget;

\item $\mbox{\bf body\_replace}(F, R, D)$ tries to replace the variables of $R$
to forget with others that entail them and are not to forget;

\item the first return value is the set of replacing variables;

\item $D$ contains the variables that have already been replaced, so they can
be deleted;

\item the second return value contains the variables that have been replaced in
this call.

\end{itemize}

This procedure hinges around its second parameter $R$, a set of variables. It
replaces some of its elements with variables that entail them according to $F$;
the replacing variables are the first return value. In the base case, $y \in R$
is replaced by some $P$ with $P \rightarrow y \in F$, but then some elements of
$P$ may be recursively replaced in the same way. This is how forgetting
happens: if all variables to be forgotten are replaced by all sets of variables
that entail them, they disappear from the formula while leaving all other
consequences intact.

The last parameter $D$ is empty in the first call, and changes at each
recursive call. It is the set of variables already replaced. Every time a
variable $y \in R$ is replaced, it is added to $E'$, which is then passed to
every subsequent recursive call and eventually returned as the second return
value. This way, if $y$ has already been replaced it is removed instead of
replaced again; this is correct because its replacing variables are already in
$S'$, which is part of the first return value.

The two return values have the same meaning of $R$ and $D$ but after the
replacement: $(R \backslash D \backslash R') \cup S'$ is $R$ with some
variables replaced; $E'$ is the set of these replaced variables; to be precise,
these are only the variables replaced in this call and it subcalls, not in some
previous recursive call.

Termination is guaranteed by the use of $F^y$ in the recursive calls and by the
set of already-replaced variables $D$. The first makes the formula used in the
subcalls smaller and smaller; when it is empty, the nondeterministic choice of
a clause in the loop fails. The second makes replacing $y$ when $y \in D$ just
a matter of removing $y$ without a recursive call.

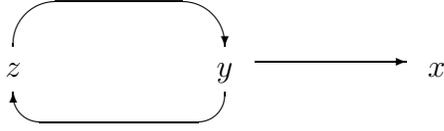
\vbox{
\setlength{\unitlength}{5000sp}%
\begingroup\makeatletter\ifx\SetFigFont\undefined%
\gdef\SetFigFont#1#2#3#4#5{%
  \reset@font\fontsize{#1}{#2pt}%
  \fontfamily{#3}\fontseries{#4}\fontshape{#5}%
  \selectfont}%
\fi\endgroup%
\begin{picture}(2130,618)(11686,-6220)
\thinlines
{\color[rgb]{0,0,0}\put(12751,-5836){\vector( 0,-1){0}}
\put(12226,-5836){\oval(1050,450)[tr]}
\put(12226,-5836){\oval(1050,450)[tl]}
}%
{\color[rgb]{0,0,0}\put(11701,-6061){\vector( 0, 1){0}}
\put(12226,-6061){\oval(1050,300)[bl]}
\put(12226,-6061){\oval(1050,300)[br]}
}%
{\color[rgb]{0,0,0}\put(12901,-5911){\vector( 1, 0){750}}
}%
\put(11701,-5986){\makebox(0,0)[b]{\smash{{\SetFigFont{12}{24.0}
{\rmdefault}{\mddefault}{\updefault}{\color[rgb]{0,0,0}$z$}%
}}}}
\put(12751,-5986){\makebox(0,0)[b]{\smash{{\SetFigFont{12}{24.0}
{\rmdefault}{\mddefault}{\updefault}{\color[rgb]{0,0,0}$y$}%
}}}}
\put(13801,-5986){\makebox(0,0)[b]{\smash{{\SetFigFont{12}{24.0}
{\rmdefault}{\mddefault}{\updefault}{\color[rgb]{0,0,0}$x$}%
}}}}
\end{picture}%
}
\nop{
 ---->
z     y ----> x
 <---
}

The first condition breaks loops like in
{} $\{y \rightarrow z, z \rightarrow y, y \rightarrow x\}$
when forgetting $y$ and $z$: after replacing $y$ by $z$, the clause $y
\rightarrow z$ is removed from the formula used in the subcall. This disallows
replacing $z$ by $y$, which would create an infinite chain of replacements.

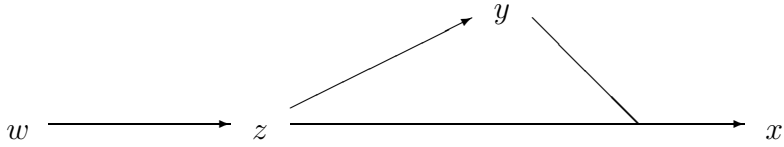
\vbox{
\setlength{\unitlength}{5000sp}%
\begingroup\makeatletter\ifx\SetFigFont\undefined%
\gdef\SetFigFont#1#2#3#4#5{%
  \reset@font\fontsize{#1}{#2pt}%
  \fontfamily{#3}\fontseries{#4}\fontshape{#5}%
  \selectfont}%
\fi\endgroup%
\begin{picture}(3780,750)(10186,-6301)
\thinlines
{\color[rgb]{0,0,0}\put(13276,-6211){\vector( 1, 0){525}}
}%
{\color[rgb]{0,0,0}\put(12751,-5686){\line( 1,-1){525}}
\put(13276,-6211){\line(-1, 0){1725}}
}%
{\color[rgb]{0,0,0}\put(11551,-6136){\vector( 2, 1){900}}
}%
{\color[rgb]{0,0,0}\put(10351,-6211){\vector( 1, 0){900}}
}%
\put(13951,-6286){\makebox(0,0)[b]{\smash{{\SetFigFont{12}{24.0}
{\rmdefault}{\mddefault}{\updefault}{\color[rgb]{0,0,0}$x$}%
}}}}
\put(12601,-5686){\makebox(0,0)[b]{\smash{{\SetFigFont{12}{24.0}
{\rmdefault}{\mddefault}{\updefault}{\color[rgb]{0,0,0}$y$}%
}}}}
\put(11401,-6286){\makebox(0,0)[b]{\smash{{\SetFigFont{12}{24.0}
{\rmdefault}{\mddefault}{\updefault}{\color[rgb]{0,0,0}$z$}%
}}}}
\put(10201,-6286){\makebox(0,0)[b]{\smash{{\SetFigFont{12}{24.0}
{\rmdefault}{\mddefault}{\updefault}{\color[rgb]{0,0,0}$w$}%
}}}}
\end{picture}%
}
\nop{
w ---> z ---> y ---+
       |           +---> x
       +-----------+
}

The second condition forbids multiple replacements, like in
{} $\{w \rightarrow z, z \rightarrow y, yz \rightarrow x\}$
when forgetting $y$ and $z$: after $y$ is replaced by $z$ and $z$ by $w$ in $yz
\rightarrow x$, the clause $yz \rightarrow x$ becomes $wz \rightarrow x$, and
the procedure moves to replacing $z$; this is recognized as unnecessary because
$z$ have already been replaced (by $w$), so it is simply deleted.

While termination is guaranteed, success is not. The set $R'$ may contain a
variable $y$ with no clause $P \rightarrow y$ in $F$. In such cases, $y$ cannot
be replaced by its preconditions in $F$. For example, when replacing $\{y,z\}$
in $F = \{w \rightarrow y, yz \rightarrow x\}$, the first variable $y$ can be
replaced by $w$, but the second variable $z$ cannot be replaced by anything,
since no clause has $z$ as its head. Running $\mbox{\bf body\_replace}(F,
\{y,z\}, \emptyset)$ fails if $R'$ is chosen to contain $z$. This is correct
because no subset of the other variables $\{x,w\}$ entail $z$ with $F$. Failure
indicates that no such replacement is possible.

The first step of proving that the procedure always terminates tells when no
recursive subcall is done. This is the base case of recursion.

\statelemma{r-prime-empty}{

A successful call to $\mbox{\bf body\_replace}()$ does not perform any
recursive subcall if and only if $R' = \emptyset$.

}{

\proof If $R' = \emptyset$ no loop iteration is executed; therefore, no
recursive subcall is performed.

The other direction is proved in reverse: $R' \not= \emptyset$ implies that
some recursive subcall is performed. Let the first variable chosen in the loop
be $y \in R'$. Since $R'$ is a subset of $R \backslash D$, it does not contain
any element of $D$. Initially, $E'$ is empty. As a result, $y \not\in D \cup
E'$. Therefore, the iteration is not cut short at the check $y \in D \cup E'$.
Since the call does not fail, the iteration is not cut short at the check $P
\rightarrow y \in F$ either. The next step is the recursive call.~\qed

}

The main component of the proofs about $\mbox{\bf body\_replace}()$ are its two
invariants. The first is recursive, a relation between parameters and return
values. The second is iterative, a relation between the local variables at each
iteration of the loop. All properties of the procedure are consequences of the
first, which requires the second to be proved. More precisely, each proves the
other. This is why they are in the same lemma.

\statelemma{loop-recursion-invariants}{

The following two invariants hold when running
$\mbox{\bf body\_replace}(F, R, D)$:

\begin{description}

\item[recursive invariant:] if it returns $S,E$, then
{} $F \cup S \cup D \models R \cup E$;

\item[loop invariant:] at the beginning and end of each iteration of its loop
(Step~\ref{bodyreplace-loop}), it holds
{} $F \cup S' \cup D \models E'$.

\end{description}

}{

\proof The two invariants are first shown to hold in their base cases. Then,
they are shown to each imply the other in an arbitrary call.

The base case for the loop invariant is the start of the first iteration of the
loop. Since $E'$ is initially empty, the condition $F \cup S' \cup D \models
E'$ holds.

The base case for the recursive invariant is a call where no recursive subcall
is done. This is only possible if $R' = \emptyset$ by
Lemma~\ref{r-prime-empty}. Since no loop iteration is executed, $S'$ and $E'$
retain their initial value $\emptyset$. The return values are
{} $(R \backslash D \backslash R') \cup S' = R \backslash D$
and $E' = \emptyset$. The recursive invariant is therefore
{} $F \cup (R \backslash D) \cup D \models R \cup \emptyset$,
which is equivalent to the trivially true proposition
{} $F \cup R \cup D \models R$.

The two invariants are now inductively proved in the general case.

The loop invariant is proved true at the end of each iteration of the loop. The
inductive assumptions are that it is true at the start of the iteration and
that the recursion invariant is true after each recursive subcall. Formally,
the two assumptions and the conclusion to be proved are:

\begin{eqnarray*}
\mbox{assumption (loop):}
	&& F \cup S' \cup D \models E'					\\
\mbox{assumption (recursion):}
	&& F^y \cup S \cup (D \cup E') \models P \cup E			\\
\mbox{conclusion to be proved (loop):}
	&& F \cup (S' \cup S) \cup D \models (E' \cup E \cup \{y\})
\end{eqnarray*}

The three parts of $E' \cup E \cup \{y\}$ are proved to follow from $F \cup S'
\cup S \cup D$ one at a time:

\begin{itemize}

\item since $F \cup S' \cup D \models E'$, it holds
$F \cup S' \cup S \cup D \models E'$;

\item The formula
{} $F \cup S' \cup S \cup D$
is a superset of
{} $F \cup S' \cup D$,
which entails $E'$ by the loop assumption. Therefore,
{} $F \cup S' \cup S \cup D \models E'$.
This implies the equivalence of
{} $F \cup S' \cup S \cup D$
and
{} $F \cup S' \cup S \cup D \cup E'$.
The latter formula is a superset of
{} $F^y \cup S \cup D \cup E'$,
which implies $E \cup P$ by the recursion assumption.
This proves the second part of the conclusion:
{} $F \cup S' \cup S \cup D \models E$;

\item the previous point also proves that
{} $F \cup S' \cup S \cup D \models P$;
since $P \rightarrow y \in F$, it follows $F \cup S' \cup S \cup D \models y$.

\end{itemize}

What remains to be proved is the recursion invariant in the general case. Since
the return values are $(R \backslash D \backslash R') \cup S'$ and $E'$, the
invariant to be proved is
{} $F \cup ((R \backslash D \backslash R') \cup S') \cup D \models R \cup E'$,
which can be rewritten as
{} $F \cup ((R \backslash R') \cup S') \cup D \models R \cup E'$.
The subset $R \backslash R'$ of the consequent $R$ is entailed because it is
also in the antecedent. Therefore, the invariant is equivalent to
{} $F \cup ((R \backslash R') \cup S') \cup D \models R' \cup E'$.
This is a consequence of
{} $F \cup S' \cup D \models R' \cup E'$,
which is now proved to hold. First $E'$ and then $R'$ are shown to be
consequences of $F \cup S' \cup D$.

\begin{itemize}

\item $F \cup S' \cup D \models E'$ is the loop invariant, which is assumed to
hold;

\item what remains to be proved is $F \cup S' \cup D \models R'$; this is
proved for every element of $R'$, that is, $F \cup S' \cup D \models y$ holds
for every $y \in R'$; the loop is run on every $y \in R'$; the iteration is cut
short if $y \in D \cup E'$; otherwise, iteration continues and $E' = E' \cup E
\cup \{y\}$ is executed; in both cases, $y \in D \cup E'$ holds at the end of
the iteration; since $D$ is never changed and $E'$ monotonically increases, $y
\in D \cup E'$ also holds at the end of the loop; if $y \in D$ then $F \cup S'
\cup D \models y$ is tautological; otherwise, $y \in E'$; the loop invariant
{} $F \cup S' \cup D \models E'$
implies
{} $F \cup S' \cup D \models y$.

\end{itemize}

Since the loop invariant and the recursion invariant are both true in their
base cases, and are both proved to hold in the induction cases, they are both
always true.~\qed

}

The aim of $\mbox{\bf body\_replace}(F, R, D)$ is to replace some variables of
$R$ with others implying them according to $F$. The recursive invariant after
{} $S,E = \mbox{\bf body\_replace}(F, R, D)$
is
{} $F \cup S \cup D \models R \cup E$.
The first return value entails the variables to be replaced according to $F$,
when $D = \emptyset$ as it should be in the first recursive call. Otherwise,
$D$ is a set of variables that have already been replaced; their replacements
are in the set $S$ for some call somewhere in the recursive tree. Since all
first return values are accumulated, they are returned by the first call.

All of this happens if the recursive call returns.

\

Forgetting maintains all consequences on the non-forgotten variables. Let $P'
\rightarrow x$ be such a consequence: $F \models P' \rightarrow x$. This
condition is the same as $F^x \models P' \rightarrow P$ and $P \rightarrow x$
by Lemma~\ref{set-implies-set}. While $x$ is not one of the variables to forget
by assumption, $P$ may contain some. If it does, they can be replaced by their
preconditions, as $\mbox{\bf body\_replace}(F, P, \emptyset)$ does. But
variables may have more than one set of possible preconditions, and some may
not be useful to ensure the entailment of $P' \rightarrow x$. Only the ones
entailed by $P'$ are. The others are not: if a replacing variable is not
entailed by $P'$, the result of replacement cannot be used to entail $P'
\rightarrow x$, still because of Lemma~\ref{set-implies-set}. The following
lemma confirms that the right choice of preconditions is always possible.

\statelemma{entailed-arg}{

If $F \cup P' \models P$, some nondeterministic choices of clauses in
$\mbox{\bf body\_replace}(F,P,\emptyset)$ and its subcalls ensure their
successful termination and the validity of the following invariants for every
recursive call
{} $S,E = \mbox{\bf body\_replace}(F,R,D)$
and every possible choices of $R'$ that do not include a variable in $P'$:

\begin{eqnarray*}
F \cup P'	& \models &	R			\\
F \cup P'	& \models &	S
\end{eqnarray*}

}{

\proof The first invariant is proved from calls to subcalls, the second in the
other direction.

The assumption $F \cup P' \models P$ of the lemma is the first invariant on the
first call $\mbox{\bf body\_replace}(F,P,\emptyset)$. This proves the first
invariant in the base case. Given that it is true for an arbitrary call, it is
proved true for all recursive subcalls. The inductive assumption is $F \cup P'
\models R$ at the beginning of the call $\mbox{\bf body\_replace}(F,R,D)$.
Since $R' \subseteq R$, it holds $F \cup P' \models y$ for every $y \in R'$. If
$y \in D \cup E'$, no recursive subcall is done. Otherwise, the clause $P
\rightarrow y$ is chosen according to Lemma~\ref{set-implies-set}: since $F
\cup P' \models y$ and $y$ is not in $P'$ because $R' \cap P' = \emptyset$,
there exists $P \rightarrow y \in F$ such that $F^y \cup P' \models P$. The
recursive subcall is $\mbox{\bf body\_replace}(F^y,P,D \cup E')$; the invariant
is therefore $F^y \cup P' \models P$, which has just been shown to hold. This
proves that the first invariant is true for all recursive calls.

This first invariant $F \cup P' \models R$ proves that the procedure always
terminates. Since $R'$ does not contain any element of $P'$, it holds $F \cup
P' \models y$ and $y \not\in P'$ for every $y \in R'$. Therefore, for every
such $y \in R'$ Lemma~\ref{set-implies-set} tells that $F$ contains a clause $P
\rightarrow y$. This means that failure never occurs. The contrary of the claim
is the existence of an infinite chain of recursive calls. Since the formula
$F^y$ used in the recursive subcall is a strict subset of the formula of the
call $F$, at some point the formula that is the argument of the call is empty.
The condition $P \rightarrow y \in F$ for every $y \in R'$ is possible when $F
= \emptyset$ only if $R' = \emptyset$. By Lemma~\ref{r-prime-empty}, no
recursive subcall is done, contrary to assumption.

The second invariant is proved to hold for the same choices of clauses. Since
the choices are the same, the first invariant holds for every call: $F \cup P'
\models R$. The base case for the second invariant are the calls when no
recursive subcall is done. This is only possible if $R' = \emptyset$ by
Lemma~\ref{r-prime-empty}. No loop iteration is performed; therefore, $S' =
\emptyset$. The first return value is
{} $(R \backslash D \backslash R') \cup S' = 
{}  (R \backslash D \backslash \emptyset) \cup \emptyset =
{}  R \backslash D$.
The second invariant is true if this set is entailed by $F \cup P'$. This holds
because of the first invariant, $F \cup P' \models R$. This proves the base
case for the second invariant.

For a general recursive call, the inductive assumption is that the second
invariant holds for the recursive subcalls. In formulae, $F^y \cup P' \models
S$ in every loop iteration. Since $S'$ accumulates the sets $S$, this implies
$F \cup P' \models S'$. Since $F \cup P' \models R$ by the first invariant, $F
\cup P' \models R \cup S'$. The return value
{} $(R \backslash D \backslash R') \cup S'$
is a subset of $R \cup S'$, and is therefore also entailed by $R \cup S'$.~\qed

}

Since the invariants hold for all calls, they hold for the first: $F \cup P'$
implies the first return value of $\mbox{\bf body\_replace}(F,P,\emptyset)$ for
appropriate nondeterministic choices. In the other way around, the
nondeterministic choices can be taken so to realize this implication.

The second requirement of forgetting is that all variables to be forgotten are
removed from the formula. Calling $\mbox{\bf body\_replace}(F,P,\emptyset)$
replaces the variables in $R'$ from $P$. Forgetting is obtained by choosing $R'$
as the variables to be forgotten in $P$ in the first call. In an arbitrary call
$\mbox{\bf body\_replace}(F,R,D)$, the variables in $D$ have already been
replaced, so they are not to be replaced again. Therefore, $R'$ comprises the
variables to be forgotten of $R \backslash D$.

The following lemmas concern $\mbox{\bf body\_replace}(F,R,D)$ when $R'$ is
always $R \backslash D \backslash V$ for some set of variables $V$. Forgetting
is achieved if $V$ is the set of variables to be retained.

\statelemma{alphabet}{

If $R'$ is always chosen equal to $R \backslash D \backslash V$ for a given set
of variables $V$, the first return value of
{} $\mbox{\bf body\_replace}(F, P, \emptyset)$
and every recursive subcall is a subset of $V$, if the call returns.

}{

\proof As proved by Lemma~\ref{r-prime-empty}, the base case of recursion is
$R' = \emptyset$. Since no loop iteration is performed, $S'$ retains its
initial value $\emptyset$. The return value
{} $(R \backslash D \backslash R') \cup S'$
is therefore the same as $R \backslash D$. Since
{} $R' = R \backslash D \backslash V$
and
{} $R' = \emptyset$,
the set
{} $R' = R \backslash D \backslash V$
is empty. This is only possible if $R \backslash D$ contains only elements of
$V$.

In the induction case, all return values $S$ are subsets of $V$. Since $S'$
accumulates the sets $S$, also $S' \subseteq V$ holds. Since $R'$ contains all
variables of $R \backslash D$ that are not in $V$, the set $R \backslash D
\backslash R'$ contains only variables in $V$. The return value
{} $(R \backslash D \backslash R') \cup S'$
is therefore a subset of $V$.~\qed

}

Termination is not guaranteed by this lemma; for example, a failure is
generated by
{} $\mbox{\bf body\_replace}(F, \{y\}, \emptyset)$
when $F = \{y \rightarrow x\}$ with $V=\{x\}$, since $R' = R \backslash
\emptyset \backslash \{x\} = \{y\}$, but no clause of $F$ has $y$ in the head.
The claim of the lemma only concerns the case of termination, as the words ``if
this call returns'' clarify.

The following lemma shows that Lemma~\ref{entailed-arg} holds even if $R'$ is
always chosen to be $R \backslash D \backslash V$. This is not obvious because
that lemma only states that its claim holds for some nondeterministic choices.

\statelemma{entailed}{

For some nondeterministic choices of clauses the call
{} $S,E=\mbox{\bf body\_replace}(F, P, \emptyset)$
returns and the first return value satisfies $F \cup P' \models S$, provided
that $F \cup P' \models P$, $P' \subseteq V$ and the nondeterministic choices
of variables are always
{} $R' = R \backslash D \backslash V$.

}{

\proof The choices $R' = R \backslash D \backslash V${\plural} satisfy the
condition $R' \cap P' = \emptyset$ of Lemma~\ref{entailed-arg} since it does
not contain any variable of $V$ while $P'$ only contains variable of $V$. The
other condition $F \cup P' \models P$ of that lemma is one of the assumptions
of this one. Lemma~\ref{entailed-arg} tells that a suitable sequence of
nondeterministic choices of clauses makes
{} $S,E=\mbox{\bf body\_replace}(F, P, \emptyset)$
returns and that the first return value satisfies $F \cup P' \models S$.~\qed

}

The conclusion of this string of lemmas is that
{} $S,E=\mbox{\bf body\_replace}(F,R,\emptyset)$
return a set $S \subseteq V$ such that
{} $F \models S \rightarrow R$ and
{} $F \models R \rightarrow P$
when it does certain nondeterministic choices.

\subsection{head\_implicates()}
\label{section-definite-head-implicates}

Forgetting requires all consequences $P' \rightarrow x$ on the variables not to
forget to be retained while all variables to forget disappear. By
Lemma~\ref{set-implies-set}, $F \models P' \rightarrow x$ is the same as $F^x
\models P' \rightarrow P$ and $P \rightarrow x$. A way to entail $P'
\rightarrow x$ from another formula $G$ is by ensuring $G \models P'
\rightarrow S$ and $S \rightarrow x \in G$. The lemmas in the previous section
tell how to obtain this condition from a common-equivalent formula $G$ over
the variables $V$ not to forget: by running
{} $S,E=\mbox{\bf body\_replace}(F,P',\emptyset)$
so that
{} $S \subseteq V$,
{} $F \models P' \rightarrow S$ and
{} $F \models S \rightarrow P$.
Since $P'$, $S$ and $x$ are common variables, $G$ may entail $P' \rightarrow S$
and it may contain $S \rightarrow x$. Therefore, $S \rightarrow x$ is a valid
choice for a clause in $G$, and allows entailing $P' \rightarrow x$ if $G
\models P' \rightarrow S$. The latter condition is achieved by ensuring $G
\models P' \rightarrow s$ for each $s \in S$ in a similar way. In other words,
if every $P \rightarrow x \in F$ is turned into $S \rightarrow x$ for all $S$
that are the first return value of
{} $\mbox{\bf body\_replace}(F,P',\emptyset)$,
all common consequences are retained while all variables to forget are removed.

This is what the following procedure does.

\

\hrule
\begingroup
\obeylines
~\newline
\noindent \#\# %
replace part of a body of $F$ with their preconditions
\noindent \# %
input $F$: a formula
\noindent \# %
output: a clause of $F$ with part of its body replaced

\endgroup
\begingroup
\noindent
clause $\mbox{\bf head\_implicates}(\mbox{formula } F)$
\parskip=-10pt

\begin{enumerate}
\parskip=-5pt

\item nondet $x \in \var(F)$

\item nondet $P \rightarrow x \in F$ else {\bf fail}

\item $S,E = \mbox{\bf body\_replace}(F^x, P, \emptyset)$

\item return $S \rightarrow x$

\end{enumerate}
\endgroup
\hrule

\

In order to derive all common consequences $P' \rightarrow x$ from the
resulting formula, replacing every $P \rightarrow x$ with $S \rightarrow x$ is
not enough. The same is required for $P' \rightarrow s$ for every $s \in S$.
Common equivalence is achieved only when the replacement is done to all
clauses, not only the ones with head $x$.

A first required lemma is that no tautology is returned. It looks obvious, and
could also be shown as a consequence of a condition on $\mbox{\bf
body\_replace}(F,R,D)$ always returning a subset of $\var(F) \cup R$, but that
would require a recursive proof. Instead, it is obvious when $D=\emptyset$.

\statelemma{no-tautology}{

No run of $\mbox{\bf head\_implicates}(F)$ produces a tautological clause.

}{

\proof The call $\mbox{\bf body\_replace}(F^x,P,\emptyset)$ may only generate
sets of variables contained in $\var(F^x) \cup P$ since the only other sets of
variables involved in the procedure are $D$ and $V$, but $D$ is empty in this
case and $V$ is only used in set subtractions.

Since $F$ does not contain tautologies and $P \rightarrow x \in F$, the
variable $x$ is not in $P$. It is also not in $F^x$ by construction. The first
return value of $\mbox{\bf body\_replace}(F^x,P,\emptyset)$ is a set of
variables $S$ (positive literals) contained in $\var(F^x) \cup P$, which does
not contain $x$. Since $x \not\in S$, the final result $S \rightarrow x$ is not
a tautology.~\qed

}

The following lemma proves that $\mbox{head\_implicates}(F)$ produces the
required clause $S \rightarrow x$. It may also produce other clauses, depending
on the nondeterministic choices.

\statelemma{generate}{

If $F \models P' \rightarrow x$, $P' \subseteq V$ and $x \not\in P'$ then
$\mbox{\bf head\_implicates}(F)$ outputs a clause $S \rightarrow x$ such that
$F^x \cup P' \models S$, provided that $\mbox{\bf body\_replace}()$ always
choses $R'$ as $R \backslash D \backslash V$ for some given set of variables
$V$.

}{

\proof The variable chosen in $\mbox{\bf head\_implicates}(F)$ is $x$. Since $F
\models P' \rightarrow x$ and $x \not\in P'$, Lemma~\ref{set-implies-set}
implies the existence of a clause $P \rightarrow x \in F$ such that $F^x \cup
P' \models P$. This clause is the second nondeterministic choice in $\mbox{\bf
head\_implicates}(F)$. Therefore, $\mbox{\bf body\_replace}(F^x, P, \emptyset)$
is called with $F^x \cup P' \models P$. By Lemma~\ref{entailed}, some
nondeterministic choices of clauses makes the call return with a first return
value that satisfies $F^x \cup P' \models S$ with the given choices of $R'$.
The clause returned by $\mbox{\bf head\_implicates}(F)$ is $S \rightarrow x$,
where $F^x \cup P' \models S$.~\qed

}

The reason why the above is a lemma and not a theorem is that the conclusion
that all common consequences are maintained requires its repeated application:
not only $P \rightarrow x$ is replaced by $S \rightarrow x$, but also $P''
\rightarrow s$ is similarly replaced for every $s \in S \backslash V$.

\statetheorem{head-implicates-theorem}{

With the nondeterministic choices $x \in V$ and $R' = R \backslash D \backslash
V$, $\mbox{\bf head\_implicates}(F)$ returns only clauses $S \rightarrow x$
that are on the alphabet $V$ and are consequences of $F$. If $F \models P'
\rightarrow x$ and $\var(P' \rightarrow x) \subseteq V$ then $P' \rightarrow x$
is entailed by some clauses produced by $\mbox{\bf head\_implicates}(F)$ with
the nondeterministic choices $x \in V$ and $R' = R \backslash D \backslash V$.

}{

\proof Let $S \rightarrow x$ be a clause returned by $\mbox{\bf
head\_implicates}(F)$. Since $x$ is the result of its first nondeterministic
choice, $x \in V$ holds by assumption. If $\mbox{\bf head\_implicates}(F)$
returns, it chose a clause $P \rightarrow x \in F$. By Lemma~\ref{alphabet},
the first return value of
{} $S,E = \mbox{\bf body\_replace}(F^x, P, \emptyset)$
satisfies $S \subseteq V$. This proves that $S \rightarrow x$ only contains
variables of $V$. By the first invariant in
Lemma~\ref{loop-recursion-invariants}, it holds $F^x \cup S \cup \emptyset
\models P \cup E$. Since $P \rightarrow x \in F$, the entailment $F \cup S
\models x$ follows. By the deduction theorem, $F \models S \rightarrow x$.

\

The long part of the proof is to show that if $P' \rightarrow x$ is a
consequence of $F$ then it is also a consequence of some clauses generated by
$\mbox{\bf head\_implicates}(F)$ if $\var(P' \rightarrow x) \subseteq V$.

If $x \in P'$ the clause $P' \rightarrow x$ is tautological and the claim is
proved. Otherwise, Lemma~\ref{generate} tells that $\mbox{\bf
head\_implicates}(F)$ produces a clause $S \rightarrow x$ with $F^x \cup P'
\models S$.

If $S \subseteq P'$ holds, the claim is proved. It is now proved that all $y
\in S \backslash P'$ can be resolved out by some other clauses generated by the
algorithm. Since $F^x \cup P' \models S$, the entailment $F^x \cup P' \models
y$ holds for every $y \in S$, including the variables $y \in S \backslash P'$.
This entailment can be rewritten as $F^x \models P' \rightarrow y$. Since $y
\not\in P'$, Lemma~\ref{generate} tells that $\mbox{\bf head\_implicates}(F^x)$
outputs a clause $S' \rightarrow y$ such that $F^{xy} \cup P' \models S'$. The
clauses $S' \rightarrow y$ and $S \rightarrow x$ resolve on $y \in S$,
producing $(S' \cup (S \backslash \{y\})) \rightarrow x$.

This process can be iterated as long as the antecedent of the clause obtained
by resolution contains some elements that are not in $P'$. In the other way
around, this process terminates only when the body of the clause is a subset of
$P'$.

The process terminates because the formula shrinks at every step: first is $F$,
then $F^x$, then $F^{xy}$. Non-termination would imply a call to $\mbox{\bf
head\_implicates}(\emptyset)$ that produces a clause by Lemma~\ref{generate}.
This clause is a consequence of $\emptyset$ as shown in the first paragraph of
this proof, and is non-tautological by Lemma~\ref{no-tautology}. This is a
contradiction. As a result, the process terminates. As proved in the previous
paragraph, if the process terminates the clause that is the result of
resolution has the form $P'' \rightarrow x$ where $P'' \subseteq P'$. This
clause entails $P' \rightarrow x$ because it is a subset of it.


The final step of the proof accounts for the calls where the formula is not $F$
but $F^x$, $F^{xy}$, etc. Every clause generated by $\mbox{\bf
head\_implicates}(F')$ with $F' \subseteq F$ is also produced by $\mbox{\bf
head\_implicates}(F)$. This is the case because the same variables and clauses
chosen in a run $\mbox{\bf head\_implicates}(F')$ can also be chosen in the run
$\mbox{\bf head\_implicates}(F)$, and the result is the same.~\qed

}

This theorem ensures the correctness of $\mbox{\bf head\_implicates}(F)$ to
forget variables and check common equivalence, as it produces clauses not
containing variables in $V$ but still entailing the same consequences of $F$
among the ones not containing $V$.

What about its efficiency? It works in polynomial space.

\statetheorem{pspace}{

For every set of nondeterministic choices, $\mbox{\bf head\_implicates}(F)$
works in polynomial space.

}{

\proof The statement holds because each individual recursive call only takes
polynomial space, and only a linear number of recursive calls are made in each
nondeterministic branch.

The first claim holds because the data used in each call is the formula $F$, a
clause $P \rightarrow x$ or $P' \rightarrow x$ at a time, a subset $F^x$ or $F^y$
at a time and a constant number of sets of variables. The size of $F$
polynomially bounds all of this.

The second claim holds because of the decreasing size of the formula used in
the recursive sub calls:
{} $\mbox{\bf head\_implicates}(F)$ calls
{} $\mbox{\bf body\_replace}(F^x,\ldots)$,
which calls
{} $\mbox{\bf body\_replace}(F^{xy},\ldots)$,
which calls
{} $\mbox{\bf body\_replace}(F^{xyz},\ldots)$
and so on. Each variable $x,y,z,\ldots$ is the head of a clause of the formula.
All such clauses are removed from the formula before the recursive call.
Therefore, the same variable cannot be chosen again in the same
nondeterministic branch. This implies that the formula is smaller at each
recursive subcall. After at most $|\var(F)|$ calls the formula is empty, and
recursion stops.

This proves that $\mbox{\bf common\_equivalence}(A,B)$ runs in nondeterministic
polynomial space, which is the same as deterministic polynomial
space~\cite{savi-70}.~\qed

}

\subsection{Common equivalence}
\label{section-definite-common}

The algorithm $\mbox{\bf head\_implicates}(F)$ generates a clause for each
sequence of nondeterministic choices. The set of these clauses can be seen as a
CNF formula. Its variables are all in $V$ and it has the same consequences of
$F$ on $V$: it is a way to forget the variables of $\var(F) \backslash V$ from
$F$. Common equivalence could be checked by forgetting all non-shared variables
and then verifying regular equivalence:
{} $\mbox{\bf head\_implicates}(A) \equiv \mbox{\bf head\_implicates}(B)$
with $V = \var(A) \cap \var(B)$. This is correct but may take not only
exponential time but also exponential space because the output of $\mbox{\bf
head\_implicates}()$ may be exponentially larger than $A$.

This is avoided by generating and checking a clause at a time: every clause
produced by $\mbox{\bf head\_implicates}(A)$ is checked against $B$ for
entailment, and the other way around.

\

\hrule
\begingroup
\obeylines
~\newline
\noindent \#\# %
check common equivalence between two formulae
\noindent \# %
input $A$: a formula
\noindent \# %
input $B$: a formula
\noindent \# %
output: true if $A$ is common equivalent to $B$, false otherwise
\endgroup

\begingroup
\noindent
$\mbox{boolean {\bf common\_equivalence}}
(\mbox{formula } A, \mbox{ formula } B)$
\parskip=-10pt

\begin{enumerate}
\parskip=-5pt

\item for each $S \rightarrow x$ generated by
{} $\mbox{\bf head\_implicates}(A)$
with
{} $x \in V$ and $R' = R \backslash D \backslash V$
where 
{} $V = \var(A) \cap \var(B)$

\begin{enumerate}

\item if $B \not\models S \rightarrow x$ return $\false$

\end{enumerate}

\item for each $S \rightarrow x$ generated by
{} $\mbox{\bf head\_implicates}(B)$
with
{} $x \in V$ and $R' = R \backslash D \backslash V$
where 
{} $V = \var(A) \cap \var(B)$

\begin{enumerate}

\item if $A \not\models S \rightarrow x$ return $\false$

\end{enumerate}

\item return $\true$

\end{enumerate}
\endgroup
\hrule

\

Section~\ref{python-common} tells how to generate and process a clause at time
without storing all of them at the same time in the Python programs to forget.
The following theorem proves the correctness of the algorithm.

\statetheorem{return-if-equivalent}{

Algorithm $\mbox{\bf common\_equivalence}(A,B)$ returns whether $A$ and $B$ are
common equivalent.

}{

\proof By Theorem~\ref{head-implicates-theorem},
{} $\mbox{\bf head\_implicates}(A)$
returns some clauses on $V = \var(A) \cap \var(B)$ that are entailed by $A$. If
$B$ does not entail one of them, that clause is a formula on $V$ that is
entailed by $A$ but not by $B$, violating the definition of common equivalence.
The same holds if $A$ does not entail a clause of
{} $\mbox{\bf head\_implicates}(B)$.

The other case is that $A$ entails
{} $\mbox{\bf head\_implicates}(B)$
and $B$ entails
{} $\mbox{\bf head\_implicates}(A)$.
The claim is $A \common B$: if $A \models C$ then $B \models C$ and vice versa
for all formulae $C$ over $V$. Let $C$ be a formula on $V$, and
$C_1,\ldots,C_m$ the clauses in its conjunctive normal form. If $A \models C$
then $A$ entails all clauses $C_i$. By Theorem~\ref{head-implicates-theorem},
{} $\mbox{\bf head\_implicates}(A)$
entails all clauses on $V$ that are entailed by $A$, including all clauses
$C_i$. Therefore, it entails $C$. Since $B$ entails
{} $\mbox{\bf head\_implicates}(A)$,
it entails all $C$. The same argument with $A$ and $B$ swapped proves the
converse.~\qed

}

\subsection{Forget}
\label{section-definite-forget}

Variables are forgotten by iteratively replacing them with other variables.
This is what $\mbox{\bf head\_implicates}()$ does by calling $\mbox{\bf
body\_replace}()$: if the body of a clause $abc \rightarrow d$ contains a
variable to forget $a$, it replaces $a$ with the body of another clause with
$a$ in the head, like $ef \rightarrow a$. Many clauses may contain one or more
variables to forget, and each variable to forget may be the head of multiple
clauses. The choice of which variable to replace in which clause by which body
is chosen nondeterministically. A sequence of choices produces a single clause.
Forgetting is the set of all of them. It is the result of calling $\mbox{\bf
head\_implicates}(F)$ and collecting all clauses it produces in all its
nondeterministic branches.

\

\hrule
\begingroup
\obeylines
~\newline
\noindent \#\# %
forget variables $X$ from formula $F$
\noindent \# %
input $F$: a formula
\noindent \# %
input $X$: a set of variables
\noindent \# %
output: a formula that expresses forgetting $X$ from $F$
\endgroup

\begingroup
\noindent
$\mbox{formula {\bf forget}}(\mbox{formula } F,\mbox{ variables } X)$
\parskip=-10cm

\begin{enumerate}
\parskip=-5pt

\item $G = \emptyset$

\item for each $S \rightarrow x$ generated by $\mbox{\bf head\_implicates}(F)$
with $x \in \var(F) \backslash X$
and $R' = R \backslash D \backslash (\var(F) \backslash X$):

\begin{enumerate}

\item $G = G \cup \{S \rightarrow x\}$

\end{enumerate}

\item return $G$

\end{enumerate}
\endgroup
\hrule

\

Both $\mbox{\bf body\_replace}()$ and $\mbox{\bf head\_implicates}()$ depend on
nondeterministic choices. Those of $x$ and $R'$ are specified in $\mbox{\bf
forget}()$, the others only affect the order of generated clauses, which is
irrelevant to ${\bf forget}()$ as it was to $\mbox{\bf common\_equivalence}()$.

The following lemma proves that $\mbox{\bf forget}(F,X)$ produces the expected
result: a formula that is common equivalent to $F$ and contains only the
variables of $F$ but $X$.

\statetheorem{forget-correct}{

Algorithm $\mbox{\bf forget}(F,X)$ returns a formula on alphabet $\var(F)
\backslash X$ that has the same consequences of $F$ on this alphabet.

}{

\proof By Theorem~\ref{head-implicates-theorem}, the given nondeterministic
choices of $x$ and $R'$ make
{} $\mbox{\bf head\_implicates}(F)$
return only clauses on $V = \var(F) \backslash X$ that are entailed by $F$, and
each clause on $V = \var(F) \backslash X$ that is entailed by $F$ is a
consequence of some clauses returned by
{} $\mbox{\bf head\_implicates}(F)$.
This means that the set of returned clauses is on the alphabet $\var(F)
\backslash X$ and entails every consequence of $F$ on this alphabet.~\qed

}

The definition of forgetting in terms of common equivalence constrains the
variables to be exactly $\var(F) \backslash X$. This is slightly different from
the output of $\mbox{\bf forget}(F,X)$, which may not contain all of them. For
example, when forgetting $d$ from $F = \{a \rightarrow b, c \rightarrow d\}$,
it produces $\{a \rightarrow b\}$, which does not contain $c \in \var(F)
\backslash \{d\}$; this example is in the test file {\tt disappear.py} for the
programs described in Section~\ref{section-python}. This minor difference is
removed by trivially adding tautologies like $\neg c \vee c$.

\

\hrule
\begingroup
\obeylines
~\newline
\noindent \#\# %
forget variables $X$ from formula $F$
\noindent \# %
input $F$: a formula
\noindent \# %
input $X$: a set of variables
\noindent \# %
output: a formula on variables $\var(F) \backslash X$ %
that is common equivalent to $F$
\endgroup

\begingroup
\noindent
$\mbox{formula {\bf forget\_ce}}(\mbox{formula } F, \mbox{ variables } X)$
\parskip=-10cm

\begin{itemize}
\parskip=-5cm

\item[~] return
$\mbox{\bf forget}(F,X) \cup
 \left\{\bigvee \{x, \neg x \mid x \in \var(F) \backslash X\}\right\}$.

\end{itemize}
\endgroup
\hrule

\

The formula returned by this algorithm is equivalent to the one returned by
$\mbox{\bf forget}(F,X)$ since the added clause is a tautology. As a result, it
has the same consequences: the same of $F$ on the alphabet $\var(F) \backslash
X$ by Lemma~\ref{forget-correct}. This alphabet is also the set of their common
variables.

\begin{theorem}
\label{forget-ce-correct}

Algorithm~$\mbox{\bf forget\_ce}(F,X)$ returns a formula that contains exactly
the variables $\var(F) \backslash X$ and is common equivalent to $F$.

\end{theorem}

Since $\mbox{\bf head\_implicates}()$ works in nondeterministic polynomial
space by Theorem~\ref{pspace}, so does $\mbox{\bf forget}()$. As for the common
equivalence algorithm, this requires the clauses to be generated one at a time
rather than all together. Each of them is then output rather than being
accumulated in a set $G$.

While $\mbox{\bf common\_equivalence}()$ is polynomial in space, $\mbox{\bf
forget}()$ is only polynomial in working space; its output may be exponentially
large. In such cases, both algorithms are exponential in time. An example
hitting these upper bounds is the following, shown for $i=3$ in the test file
{\tt exponential.py} for the programs described in
Section~\ref{section-python}.

\[
F =
\{ x_i \rightarrow z_i, y_i \rightarrow z_i \mid 1 \leq i \leq n \} \cup
\{ z_1 \ldots z_n \rightarrow w \}
\]

The call
{} $\mbox{\bf body\_replace}(F^w, \{z_1,\ldots,z_n\}, \emptyset)$
with
{} $V = \{x_1,\ldots,x_n\} \cup \{y_1,\ldots,y_n\} \cup \{w\}$
has two ways to replace $z_1$: by $x_1$ or by $y_1$. For each of these two
nondeterministic choices, other two are generated for replacing $z_2$ by either
$x_2$ or $y_2$. The same for $z_3$ and the other variables up to $z_n$.
Doubling for each $i$ from $1$ to $n$ means exponential in $n$: the final
number of nondeterministic branches is $2^n$. This means exponential time for
$\mbox{\bf common\_equivalence}()$ and exponential output for $\mbox{\bf
forget}()$.

However, the algorithms are not guilty of this. Forgetting $\{z_1,\ldots,z_n\}$
from $F$ always produces an exponential number of clauses. The formula
{} $F' = \{P \rightarrow w \mid
{}         P \in \mbox{\bf body\_replace}(F^w, \{z_1,\ldots,z_n\}, \emptyset)$
is not just a way of forgetting, it is the minimal one. It is a minimal formula
since all its clauses are superredundant~\cite{libe-20}. All formulae
equivalent to it are the same size or larger. Being exponentially large, they
cannot be generated in less than exponential time.

In this sense, $\mbox{\bf forget}()$ does not waste time: exponential output,
exponential running time; nothing better can be done.

Yet, a small variant is backbreaking for $\mbox{\bf forget}()$: removing $x_n$
and $y_n$ from $V$ makes
{} $\mbox{\bf body\_replace}(F^w, \{z_1,\ldots,z_n\}, \emptyset)$
output nothing, but still take exponential time if the variables are replaced
in increasing order of $i$. The same exponential number of nondeterministic
branches are produced, but they all eventually fail because $z_n$ cannot be
replaced.

A simple entailment check would have avoided them. The aim of
{} $\mbox{\bf body\_replace}(F^w, \{z_1,\ldots,z_n\}, \emptyset)$
is to find a set $P' \subseteq V$ such that
{} $F \cup P' \models \{z_1,\ldots,z_n\}$.
The largest possible set $P'$, the one having the most consequences, is $V$
itself. If $F \cup V$ does not entail $\{z_1,\ldots,z_n\}$, no proper subset of
$V$ does. Computation can be cut short when this happens. Since $F$ is Horn,
this check only takes polynomial time:

\[
\mbox{if $F \cup V \not\models P$ then \bf fail}
\]

This check is added right before each call to $\mbox{\bf body\_replace}$, both
the first and the recursive ones.

In the example, $F \cup V$ does not entail $z_n$; therefore, it does not entail
$\{z_1,\ldots,z_n\}$ either. The very first call of $\mbox{\bf body\_replace}$
is skipped. As it should: no subset of $V$ can ever replace
$\{z_1,\ldots,z_n\}$ according to $F$.

Not only this check is useful in avoiding a call that would fail anyway. It
otherwise guarantees its success --- its usefulness. If $F \cup V \models P$
then $\mbox{\bf body\_replace}()$ can replace $P$ with some variables of $V$
entailing it.

All of this is now formally proved: the added check does not harm the
algorithm, but ensures the usefulness of the following call.

The proof of the following lemma requires that the first recursive call is done
with an empty third argument:
{} $\mbox{\bf body\_replace}(F, P, \emptyset)$.
This is what happens anyway, but is specified because the lemma does not work
if the third argument is not empty.

\statelemma{success-implies}{

If $R'$ is always chosen to be $R \backslash D \backslash V$ within a call
{} $\mbox{\bf body\_replace}(F, P, \emptyset)$,
then
{} $F \cup V \models R \cup D \cup E$
holds after every successful subcall
{} $S,E = \mbox{\bf body\_replace}(F', R, D)$.

}{

\proof The claim is proved by top-down induction. The base case is the first
call of recursion, which by assumption is done with formula $F$ and an empty
third argument:
{} $S,E = \mbox{\bf body\_replace}(F, P, \emptyset)$.
Lemma~\ref{loop-recursion-invariants} states
{} $F \cup S \cup \emptyset \models P \cup E$.
Lemma~\ref{alphabet} states $S \subseteq V$. These two facts prove
{} $F \cup V \models P \cup E$,
which implies
{} $F \cup V \models P \cup D \cup E$
since $D = \emptyset$.

The induction case assumes the claim for a recursive call and its return values
and proves it for every one of its recursive subcalls.

The assumption is about an arbitrary recursive call
{} $\mbox{\bf body\_replace}(F, R, D)$,
and states that if its second return value is $E'$ then
{} $F \cup V \models R \cup D \cup E'$

The claim is that after every subcall
{} $S,E = \mbox{\bf body\_replace}(F^y, P, D \cup E')$
it holds
{} $F \cup V \models P \cup (D \cup E') \cup E$.

The first statement of Lemma~\ref{loop-recursion-invariants} is that after
{} $S,E = \mbox{\bf body\_replace}(F^y, P, D \cup E')$
it holds
{} $F^y \cup S \cup (D \cup E') \models P \cup E$.
By Lemma~\ref{alphabet}, also $S \subseteq V$ holds. By definition, $F^y
\subseteq F$. The entailment can therefore be turned into
{} $F \cup V \cup (D \cup E') \models P \cup E$.
The induction assumption is
{} $F \cup V \models R \cup D \cup E'$
for the value of $E'$ at the end of the call. Since $E'$ never loses elements,
its value at the end of the subcall is at most as large as the final one.
Therefore, 
{} $F \cup V \models R \cup D \cup E'$
still holds there. Together with
{} $F \cup V \cup (D \cup E') \models P \cup E$,
this implication proves the claim
{} $F \cup V \models P \cup (D \cup E') \cup E$.~\qed

}

This lemma requires the first recursive call to have an empty third argument,
which is the case when forgetting and checking common equivalence. If so,
adding the instruction
{} ``if $F \cup V \not\models P$ then {\bf fail}''
before the recursive call does not affect the final result.

\statelemma{fail}{

If $F \cup V \not\models P$, the recursive call $\mbox{\bf body\_replace}(F^y,
P, D \cup E')$ fails if $R'$ is always chosen equal to $R \backslash D
\backslash V$.

}{

\proof By contradiction,
{} $\mbox{\bf body\_replace}(F^y, P, D \cup E')$
is assumed successful. Lemma~\ref{success-implies} implies
{} $F \cup V \models P \cup (D \cup E') \cup E$,
which contradicts the assumption
{} $F \cup V \not\models P$.~\qed

}

This lemma proves that the additional check
{} ``if $F \cup V \not\models P$ then {\bf fail}''
before a recursive subcall does not change the result of the algorith, since
the following subcall would fail anyway.

The contrary also holds: if the check succeds, the following call does not
fail.

\statelemma{succed}{

If $F \cup V \models P$ then $\mbox{\bf body\_replace}(F^y, P, D \cup E')$
succeeds if $R'$ is always chosen equal to $R \backslash D \backslash V$.

}{

\proof Lemma~\ref{entailed-arg} states that if $F \cup P' \models P$ and $R'$
is always chosen not to include a variable in $P'$, the recursive call
{} $S,E = \mbox{\bf body\_replace}(F,R,D)$
succeeds for some nondeterministic choices. The preconditions of the lemma hold
for $P' = V$, since $P \cup V \models P$ by assumption and $R' = R \backslash D
\backslash V$. Its consequence therefore follows: the recursive call succeeds
for some nondeterministic choices.~\qed

}

The check $F \cup V \models P$ improves efficiency because it requires
polynomial time (being $F$ Horn) but may save an exponential amount of
recursive calls. It may look like it makes the running time polynomial in the
size of the output~\cite{tarj-73}, since every call succeeds and therefore
produces some clauses. Yet, these are not guaranteed to be different from the
previously generated ones. A counterexample is the following.

\[
F =
\{ a_i \rightarrow x_i \mid 1 \leq i \leq n \} \cup
\{ x_i \rightarrow z_i, y_i \rightarrow z_i \mid 1 \leq i \leq n \} \cup
\{ z_1 \ldots z_n \rightarrow w \}
\]

When replacing $\{z_1,\ldots,z_n\}$ with
{} $V = \{a_i \mid 1 \leq i \leq n\} \cup \{w\}$,
every $z_i$ is replaced by either $x_i$ or $y_i$, leading to exponentially many
combinations. The additional check does not avoid them, since
{} $F \cup V \models \{z_1,\ldots,z_n\}$ holds.
They all produce the same replacement $\{a_1,\ldots,a_n\}$. This is correct
because the only consequence of $F$ on the alphabet of $V$ is $a_1 \ldots a_n
\rightarrow w$. Forgetting takes exponential time to produce a single clause.
This example with $i=3$ is in test file {\tt branches.py} for the programs
described in Section~\ref{section-python}; as expected, the same clause is
produced eight times.

A simple test similar to $F \cup V \models P$ does not suffice. A recursive
call searching for a replacement for $P$ is useful only if it produces a set
which has not been found so far. If these are $A_1,\ldots,A_m$, one needs to
check whether an element of each $A_i$ can be removed from $V$ so that the
result still implies $P$ with $F$. The problem is not that $m$ may be
exponential, since the aim is to bound the running time by a polynomial in $m$.
It is the choice of the elements, since these may be $2^m$ even if each $A_i$
only contains two variables.

\subsection{Single-head}
\label{section-definite-singlehead}

The source of exponentiality in $\mbox{\bf body\_replace}(F,R,D)$ is the
nondeterministic choice of the clause $P \rightarrow y$. Many such clauses may
exist, each spawning a branch of execution. This is not the case if each
variable $y$ is the head of at most a clause $P \rightarrow y$. Formulae with
this property are called {\em single-head}.

This restriction removes nondeterminism in the algorithm. The choice $R'
\subseteq R \backslash D$ is forced to be $R' = R \backslash D \backslash V$
and $x \in \var(F)$ can be turned into a loop over the variables of $F$. The
remaining nondeterministic choices are $P \rightarrow x \in F$ and $P
\rightarrow y \in F$, which becomes deterministic since only a single such
clause may exist for each given $x$ and $y$.

The algorithm simplifies to: for each variable $x \not\in V$, remove the only
clause $P \rightarrow x \in F$ and replace all remaining occurrences of $x$
with $P$. This takes polynomial time because once a variable is replaced, it
disappears from the formula; it is never replaced again. While the
polynommiality of this case looks obvious, it still requires a formal proof.

\statelemma{all-polynomial}{

If $F$ is a single-head definite Horn formula and $V$ a set of variables,
computing all possible return values of
{} $\mbox{\bf head\_implicates}(F)$
with the nondeterministic choices $x \in V$ and $R' = R \backslash D \backslash
V$ only takes time polynomial in the size of $F$.

}{

\proof Since $F$ is single-head, at most one choice of a clause $P \rightarrow
x \in F$ and $P \rightarrow y \in P$ is possible. Since $R'$ is always $R
\backslash D \backslash V$, this nondeterministic choice becomes deterministic.
The only remaining nondeterministic choice is the variable $x \in V$, but only
a linear number of such variables exist. Replacing this nondeterministic choice
with a loop over all variables of $V$ makes the algorithm deterministic.

To prove it polynomial suffices to show that it only performs a polynomial
number of recursive calls, since each call only takes at most a linear amount
of time being a loop over a set of variables.

A recursive subcall $\mbox{\bf body\_replace}(F^y, P, D \cup E')$ is only
performed if $P \rightarrow y \in F$ is the only clause of $F$ having $y$ as
its head. The claim is proved by showing that this may only happen once during
the whole run of the algorithm. In particular, it may not happen again for the
same variable $y$:

\begin{enumerate}

\item in the subcall $\mbox{\bf body\_replace}(F^y, P, D \cup E')$;

\item in the subsequent subcalls of the loop;

\item after the current call returns.

\end{enumerate}

The first argument of the call $\mbox{\bf body\_replace}(F^y, P, D \cup E')$ is
a formula $F^y$ non containing $y$. Therefore, choosing a clause with $y$ as
its head fails.

After this subcall returns, $y$ is added to $E'$. The subsequent subcalls of
the loop $y$ are done with $E'$ as part of the second argument. These subcalls
receive their second argument in $D$; therefore, $y$ can never be in $R' = R
\backslash D \backslash V$; since $D$ is passed to the sub-subcalls, the same
happens there.

Finally, $y$ is not chosen again after the recursive call that calls
{} $\mbox{\bf body\_replace}(F^y, P, D \cup E')$
returns. By assumption,
{} $\mbox{\bf body\_replace}(F^y, P, D \cup E')$
was called because $P \rightarrow y \in F$ and $y$ was a variable of the loop.
This implies that $y$ is added to $E'$ after $\mbox{\bf body\_replace}(F^y, P,
D \cup E')$ returns. After the loop ends, the call returns $E'$ containing $y$
as its second return value. In the caller, the second return value is added to
$E'$ and then returned. This means that $y \in E'$ holds from this point on.
Even if $y \in R'$ at some point, the test $y \in D \cup E'$ succeeds, and the
iteration is cut short before selecting $P \rightarrow y \in F$ again.

This proves that once $P \rightarrow y$ is selected in a call to $\mbox{\bf
body\_replace}()$, it is never selected again in the rest of the run. Since a
recursive call is done only after this selection, and the number of clauses is
linear in the size of the input, the total number of recursive calls is linear
as well. Since each run requires polynomial time, the overall running time is
polynomial.~\qed

}

This proves that both forgetting and checking common equivalence take
polynomial time in the single-head restriction.

\begin{theorem}
\label{single-head-polynomial}

Checking whether $F \common G$ holds can be verified in polynomial time if both
$F$ and $G$ are single-head definite Horn formulae.

\end{theorem}

\section{Non-definite Horn formulae}
\label{section-indefinite}

What about general Horn formulae, where negative clauses like $\neg x \vee \neg
y$ may occur? Such a formula can be turned into a form where all non-definite
clauses are unary. Forgetting can be done by running the algorithm in the
previous section on the definite part of this formula, leaving the non-definite
unary clauses alone. The sequence is:

\begin{itemize}

\item change the formula to make all non-definite clauses unary;

\item run the algorithm in the previous section on the definite clauses;

\item add the non-definite (unary) clauses to the result.

\end{itemize}

Somehow, a non-definite clause $\neg x \vee \neg y$ has a head: it is the same
as $xy \rightarrow \bot$. The head of a negative clause is the truth value of
false. In this form, a negative clause is a definite clause, only with a
special symbol as its head. The algorithm for common equivalence still works
taking $\bot$ as a variable. Or it is a variable, one always forced to be
false.

Technically, a general Horn formula $F$ can be translated into a normal form
where the only negative clauses are unary, that is, they only comprise a single
(negative) literal. This translation is only needed for negative clauses, the
definite ones stay the same.

\begin{definition}

The definite Horn part $def(F)$ of a Horn formula $F$ is the set of definite
clauses of $F$ --- the clauses of $F$ that contain exactly one positive
literal.

\end{definition}

By definition, $def(F) \subseteq F$ holds in general and $def(F) = F$ if $F$ is
definite Horn. The negative clauses $F \backslash def(F)$ can be made definite
in two ways: adding the same positive literals to all of them, or a different
one to each. Both work. The second keeps formulae single-head, but they are
both correct. More importantly, they are both compatible with the common
equivalence algorithm. For this reason, the translation is defined in a general
form; the new heads are taken from a set $Z$, how they are associated to the
negative clauses is not constrained.

\[
Z(F) =
def(F) \cup \{C \vee z \mid C \in F \backslash def(F) ,~ z \in Z\}
\]

An arbitrary set of new variables $Z$ is used as a pool of heads for the
clauses that need one. This definition allows the same variable for all of
them, or one for each. Nothing is said about which $z \in Z$ is chosen for $C
\in F \backslash def(F)$.

Of course, $C$ and $C \vee z$ are not the same; the first is satisfied
regardless of $z$, the second depends on it. Adding $z$ is harmless if $z$ is
guaranteed to be false. For example, $\{C \vee z, \neg z\}$ is common
equivalent to $\{C\}$. At the scale of formulae, $Z(F) \cup \{\neg z \mid z \in
Z\}$ is common equivalent to $F$.

The addition of $\{\neg z \mid z \in Z\}$ may look unnecessary when looking for
common equivalence and the same variable $z$ is the added head of all negative
clauses: if $C \in A$ and $A \common B$, then $B$ implies $C$. Therefore,
$Z(A)$ contains $C \vee z$ and $Z(B)$ implies $C \vee z$. The other direction
is not so easy to prove, and is indeed false. The problem is that $Z(A)$ and
$Z(B)$ may differ on whether they entail a definite clause $P \rightarrow x$,
but this difference is leveled by another clause $P' \rightarrow z$ with $P'
\subseteq P$. When $z$ is not constrained to be false these two clauses are
independent; when it is, the second supersedes the first, making it redundant
and canceling the difference.

Checking formulae for redundancy is not the solution, as the differing clause
may be entailed rather than present in a formula. The following is an example.

\begin{eqnarray*}
A &=& \{\neg a \vee \neg b,
	a \rightarrow a', b \rightarrow b', a'b' \rightarrow c,
	c \rightarrow d\}						\\
B &=& \{\neg a \vee \neg b,
	c \rightarrow d\}
\end{eqnarray*}

When switching from $A$ and $B$ to $Z(A)$ and $Z(B)$, the first formula $Z(A)$
entails $ab \rightarrow c$ while the second $Z(B)$ does not. This is a clause
on their common variables $\var(A) \cap \var(B) = \{a,b,c,d\}$. It proves that
$Z(A)$ and $Z(B)$ are not common equivalent. Yet, $A$ and $B$ are common
equivalent. The differing clause $ab \rightarrow c$ is entailed by $\neg a \vee
\neg b$, which is in both formulae. This is the only possible case where
inequivalence in the definite part of the formulae does not entail equivalence
in the whole: a differing definite clause is superseded by a negative clause that
is entailed by both formulae.

The solution is to make this negative clause stand out from the definite. It
does as soon as its new head is forced to be false: since $Z(B)$ contains $ab
\rightarrow z$ for some $z \in Z$, adding $\neg z$ makes this clause equivalent
to $\neg a \vee \neg b$, which entails $ab \rightarrow z$.

The general solution is to set all new heads $z \in Z$ to false.

The replacement of the non-shared variables is done on the definite Horn
version $Z(A)$ and $Z(B)$ since $\hi()$ only works on definite Horn clauses;
the negation of the new heads is virtually added to the resulting formula. Some
propositional manipulations show that this addition is only necessary to the
other, original formula.

\

\hrule
\begingroup
\obeylines
~\newline
\noindent \#\# %
check common equivalence of two general Horn formulae
\noindent \# %
input $A$: a Horn formula
\noindent \# %
input $B$: a Horn formula
\noindent \# %
output: true if $A \common B$ holds, false otherwise

\endgroup
\noindent
\begingroup
\parskip=-10pt
$\mbox{boolean {\bf common\_equivalence\_horn}}(A,B)$

\begin{enumerate}
\parskip=-5pt

\item for each $P \rightarrow x$ generated by
{} $\mbox{\bf head\_implicates}(Z(A))$
with
{} $x \in V$ and $R' = R \backslash D \backslash V$,
where 
{} $V = \var(A) \cap \var(B)$

\begin{enumerate}

\item if $B \cup \{\neg z \mid z \in Z\} \not\models P \rightarrow x$
return $\false$

\end{enumerate}

\item for each $P \rightarrow x$ generated by
{} $\mbox{\bf head\_implicates}(Z(B))$
with
{} $x \in V$ and $R' = R \backslash D \backslash V$,
where 
{} $V = \var(A) \cap \var(B)$

\begin{enumerate}

\item if $A \cup \{\neg z \mid z \in Z\} \not\models P \rightarrow x$
return $\false$

\end{enumerate}

\item return $\true$

\end{enumerate}
\endgroup
\hrule

\

The following lemma proves that no matter how $Z()$ chooses the heads to attach
to the negative clauses, the result of this algorithm is correct.

\statelemma{common-definite}{

The condition $A \common B$ is equivalent to
{} $B \cup \{\neg z \mid z \in Z\} \models \mbox{\bf head\_implicates}(Z(A))$
and
{} $A \cup \{\neg z \mid z \in Z\} \models \mbox{\bf head\_implicates}(Z(B))$
if $V = Z \cup (\var(A) \cap \var(B))$.

}{

\proof The proof comprises two parts: the first condition implies the other
two, and they imply it.

The first part of the proof begins assuming $A \common B$ and $C \in \hi(Z(A))$
and ends concluding $B \cup \{\neg z \mid z \in Z\} \models C$. By symmetry,
the same holds swapping $A$ and $B$.

By Theorem~\ref{head-implicates-theorem}, all clauses of $\hi(Z(A))$ are
entailed by $Z(A)$. Therefore, $Z(A) \models C$. By monotonicity of entailment,
$Z(A) \cup \{\neg z \mid z \in Z\} \models C$. Since $Z(A) \cup \{\neg z \mid z
\in Z\}$ is equivalent to $A \cup \{\neg z \mid z \in Z\}$, it follows $A \cup
\{\neg z \mid z \in Z\} \models C$. This implies $A \models \{\neg z \mid z \in
Z\} \rightarrow C$, which is the same as $A \models (\vee Z) \vee C$.

Since $(\vee Z) \vee C$ is a consequence of $A$, resolution from $A$ produces a
subset $C' \subseteq (\vee Z) \vee C$. By soundness of resolution, $A \models
C'$. Since $A$ does not contain any variable in $Z$ and resolution does not
introduce literals, this clause $C'$ does not contain any variable in $Z$
either. Therefore, $C' \subseteq C$.

Since $C$ is generated by $\hi(Z(A))$ with $V = Z \cup (\var(A) \cap \var(B))$,
Theorem~\ref{head-implicates-theorem} tells $\var(C) \subseteq Z \cup (\var(A)
\cap \var(B))$. Since $C'$ is a subset of $C$ and does not contain any variable
in $Z$, this containment can be refined as $\var(C') \subseteq \var(A) \cap
\var(B)$. Common equivalence between $A$ and $B$ applies: $A \models C'$
implies $B \models C'$. As a result, $B \models C$ since $C' \subseteq C$. By
monotonicity of entailment, $B \cup \{\neg z \mid z \in Z\} \models C$.

This proves that $C \in \hi(Z(A))$ implies $B \cup \{\neg z \mid z \in Z\}
\models C$ if $A \common B$. By symmetry, the same holds with $A$ in place of
$B$ and vice versa, concluding the first part of the proof.

\

The second part of the proof begins assuming $A \models C$ and
{} $B \cup \{\neg z \mid z \in Z\} \models \mbox{\bf head\_implicates}(Z(A))$
with $\var(C) \subseteq \var(A) \cap \var(B)$ and ends concluding $B \models
C$. Once this is proved, the same holds when swapping $A$ and $B$, proving $A
\common B$.

By monotonicity of entailment, $A \models C$ implies $A \cup \{\neg z \mid z
\in Z\} \models C$. Since $A \cup \{\neg z \mid z \in Z\}$ is equivalent to
$Z(A) \cup \{\neg z \mid z \in Z\}$, the entailment $Z(A) \cup \{\neg z \mid z
\in Z\} \models C$ follows. This is the same as $Z(A) \models \{\neg z \mid z
\in Z\} \rightarrow C$, which can be rewritten as $Z(A) \models (\vee Z) \vee
C$.

Since $\var(C) \subseteq \var(A) \cap \var(B)$, it follows $\var((\vee Z) \vee
C) \subseteq Z \cup (\var(A) \cap \var(B)) = V$.
Theorem~\ref{head-implicates-theorem} states that $\hi(Z(A))$ implies all
consequences of $Z(A)$ on the alphabet $V$. Since $Z(A) \models (\vee Z) \vee
C$ and $\var((\vee Z) \vee C) \subseteq V$, this applies to $(\vee Z) \vee C$:
it is entailed by $\hi(Z(A)$.

By transitivity, $\hi(Z(A)) \models (\vee Z) \vee C$ and the assumption $B \cup
\{\neg z \mid z \in Z\} \models \hi(Z(A))$ imply
{} $B \cup \{\neg z \mid z \in Z\} \models (\vee Z) \vee C$.
This entailment can be rewritten as
{} $B \models \{\neg z \mid z \in Z\} \rightarrow ((\vee Z) \vee C)$,
which is the same as
{} $B \models (\vee Z) \vee ((\vee Z) \vee C)$,
or
{} $B \models (\vee Z) \vee C$.

Since $B \models (\vee Z) \vee C$, a subset of $(\vee Z) \vee C$ is generated
by resolving clauses of $B$. Since $B$ does not contain any variable in $Z$ and
resolution does not introduce literals, this subset $C'$ of $(\vee Z) \vee C$
does not contain any variable in $Z$: it is a subset of $C$. By soundness of
resolution, $B \models C'$. Since $C'$ is a subset of $C$, this implies $B
\models C$, the claim.

By symmetry, the same holds in the other direction, proving that every
consequence of $B$ on the alphabet $V$ is also a consequence of $A$.~\qed

}

The first step of $\mbox{\bf common\_equivalence\_horn}(A,B)$ checks whether
every clause generated by $\hi(Z(A))$ is entailed by $B \cup \{\neg z \mid z
\in Z\}$. This is the same as
{} $B \cup \{\neg z \mid z \in Z\} \models \hi(Z(A))$.
By symmetry, the second step checks
{} $A \cup \{\neg z \mid z \in Z\} \models \hi(Z(B))$.
The lemma proves these two conditions to be equivalent to $A \common B$.

Negative clauses are not a problem. When present, they are turned into definite
clauses before running the replacement algorithm and the newly introduced heads
negated afterwards. If the formula is single-head, it remains so as long as
$Z()$ assigns a different head to each clause.

\

The same mechanism works for forgetting: negative clauses are turned into
definite clauses by adding new variables as heads, the definite Horn algorithm
for forgetting is run and the new variables are replaced with false. The
shorthand $[\bot/Z]$ stands for the set of substitutions $[\bot/z]$ for all $z
\in Z$.

\

\hrule

\begingroup
\obeylines
~\newline
\noindent \#\# %
forget variables $X$ from a general Horn formula $F$
\noindent \# %
input $F$: a Horn formula
\noindent \# %
input $X$: a set of variables
\noindent \# %
output: a formula expressing forgetting $X$ from $F$

\endgroup
\noindent
\begingroup
\parskip=-10pt
$\mbox{formula {\bf forget\_horn}}(\mbox{formula }F,\mbox{ variables } X)$
\begin{enumerate}
\parskip=-5pt

\item $F' = \mbox{\bf forget\_ce}(Z(F),X)$

\item return $F'[\bot/Z]$

\end{enumerate}
\endgroup

\hrule

\

This algorithm is proved correct by the following lemma regardless of how
$Z(F)$ assigns the new heads to the negative clauses.

\statelemma{common-bot}{

For every formula $F$ and set of variables $X$,
if $F' = \mbox{\bf forget\_ce}(Z(F),X)$
then $F'[\bot/Z]$ contains exactly the variables $\var(F) \backslash X$ and
is common equivalent to $F$.

}{

\proof The alphabet of $Z(F)$ is $Z \cup \var(F)$ by construction. Since $Z(F)$
is a definite Horn formula and $\mbox{\bf forget\_ce}()$ is correct on these
formulae by Theorem~\ref{forget-ce-correct}, the variables of $F'$ are
{} $\var(Z(F)) \backslash X = (Z \cup \var(F)) \backslash X$.
Since $Z$ is a set of new variables, it does not intersect $X$. As a result,
the variables of $F'$ are
{} $(Z \cup \var(F)) \backslash X = Z \cup (\var(F) \backslash X)$.
The substitution $F'[\bot/Z]$ removes exactly the variables $Z$, leaving
$\var(F) \backslash X$. This proves the first part of the claim: the return
value of the algorithm is a formula on the alphabet $\var(F) \backslash X$.

The second part of the claim is $F'[\bot/Z] \common F$. The correctness of
$\mbox{\bf forget\_ce}()$ on definite Horn formulae like $Z(F)$ includes $F'
\common Z(F)$. Lemma~\ref{monotonic} tells that $F' \common Z(F)$ still holds
if adding the same formula on the shared variables to both sides of the
equivalence. Since all variables $z \in Z${\plural} are in both formulae, it
implies
{} $F' \cup \{\neg z \mid z \in Z\} \common Z(F) \cup \{\neg z \mid z \in Z\}$.

Lemma~\ref{replace} proves that adding literals is common equivalent to
replacing the same literals with true. This applies to both $Z(F)$ and $F'$,
proving
{} $Z(F)[\bot/Z] \common Z(F) \cup \{\neg z \mid z \in Z\}$
and
{} $F'[\bot/Z] \common F' \cup \{\neg z \mid z \in Z\}$.

The following chain of common equivalences are proved:

\begin{eqnarray*}
Z(F)[\bot/Z]
		& \common &
		Z(F) \cup \{\neg z \mid z \in Z\}	\\
Z(F) \cup \{\neg z \mid z \in Z\}
		& \common &
		F' \cup \{\neg z \mid z \in Z\}		\\
F' \cup \{\neg z \mid z \in Z\}
		& \common &
		F'[\bot/Z]
\end{eqnarray*}

The variables in these formulae are
{} $\var(F)$,
{} $Z \cup \var(F)$,
{} $Z \cup (\var(F) \backslash X)$ and
{} $\var(F) \backslash X$.

Lemma~\ref{transitive} tells that common equivalence is transitive if the
variables shared by the formulae on the sides are all in the formula in the
middle: $A \common B$ and $B \common C$ implies $A \common C$ if $\var(A) \cap
\var(C) \subseteq \var(B)$. This is the case for the first three alphabets:

\[
\var(F) \cap (Z \cup (\var(F) \backslash X)) =
\var(F) \backslash X 
\subseteq
Z \cup \var(F)
\]

Transitivity implies $Z(F)[\bot/Z] \common F' \cup \{\neg z \mid z \in Z\}$.
The alphabets of these two formulae are $\var(F)$ and $Z \cup (\var(F)
\backslash X)$. The same containment among alphabets applies with the last
formula of the chain $F'[\bot/Z]$:

\[
\var(F) \cap (\var(F) \backslash X) = \var(F) \backslash X
\subseteq Z \cup (\var(F) \backslash X)
\]

By transitivity, $Z(F)[\bot/Z] \common F'[\bot/Z]$. Since the first formula is
identical to $F$, its common equivalence with $F'[\bot/Z]$ is proved.~\qed

}

This lemma ensures the correctness of $\mbox{\bf forget\_horn}()$ regardless of
how $Z(F)$ adds the heads to the negative clauses. It may add the same head to
all clauses. Or a different head to each. It works in both cases. The first
involves a slightly simpler formula, but the second keeps the formula
single-head if it was. This proves that forgetting from single-head Horn
formulae is still polynomial in time even when they contain negative clauses.

\section{Single-head minimization}
\label{section-minimize}

The algorithm for forgetting outputs a polynomially large formula since it is
polynomial in time. Is this enough? Polynomial may be linear, or quadratic, or
cubic. Linear is best. Linear may be double the size, ten times the size,
thirty times the size. Double is best. In a best case of a best case,
forgetting doubles the size of the formula. If the goal of forgetting is to
reduce size, doubling size misses it. Even the same size is too large. The aim
is to reduce size, not to increase it; not even to keep it the same.

Not all is lost, however. The algorithm only generates a formula expressing
forgetting, a formula $B$ over the variables $\var(A) \backslash X$ that is
common equivalent to $A$. Other such formulae may exist. Every formula on the
same variables that is equivalent to $B$ fits the definition. The question is
whether such a formula is smaller than $B$. Still better, whether it is small
enough.

The same problem without forgetting has been studied for
decades~\cite{veit-52,mccl-56,rude-sang-87,theo-etal-96}. For example, checking
whether a Horn formula is equivalent to another of a certain size is
\np-complete~\cite{hamm-koga-93}. When the formula is single-head, forgetting
and minimization can be done separately: first forget, in polynomial time; then
minimize. The problem of forgetting within a certain size reduces to the
classic Boolean minimization problem, when the formula is single-head.

Yet, the single-head case has some specificity:

\begin{itemize}

\item the minimally sized formula equivalent to a given single-head one may not
be single-head;

\item therefore, minimizing is actually two distinct problems: find a minimal
formula or a minimal single-head formula;

\item acyclic formulae are easy to minimize, clause by clause;

\item this cannot be done in the cyclic case;

\item a sufficient condition to minimality exists;
it provides directions for minimizing when a formula is not minimal.

\end{itemize}

\

The first point is that minimizing a single-head formula depends on whether the
minimal formula is required to be single-head or not. This is proved by an
example, a single-head formula that is equivalent to a non-single-head one of
size $k$ but to no single-head one of the same or lower size.

It hinges on how equivalences are achieved in single-head formulae: by loops of
implications. If $F$ implies $A \equiv B$ and $B \equiv C$ for three disjoint
sets of variables $A$, $B$ and $C$, the single-head condition requires each set
to imply another in a loop. An example is $A \rightarrow B$, $B \rightarrow C$
and $C \rightarrow A$. This is not the case in general, where the same
equivalences can be realized by entailments centered on one set, for example
$A$, like $A \rightarrow B$, $B \rightarrow A$, $A \rightarrow C$ and $C
\rightarrow A$. Such an alternative may be smaller. An example is the
following.

\[
F = \{
	a \rightarrow b_1,
	a \rightarrow b_2,
	a \rightarrow b_3,
	b_1 b_2 b_3 \rightarrow c_1,
	b_1 b_2 b_3 \rightarrow c_2,
	b_1 b_2 b_3 \rightarrow c_3,
	c_1 c_2 c_3 \rightarrow a
\}
\]

This formula is a specific case of the example above where $A=\{a\}$,
$B=\{b_1,b_2,b_3\}$ and $C=\{c_1,c_2,c_3\}$. It is a single-head formula that
entails the equivalence of $A$, $B$ and $C$ by a loop of entailments:

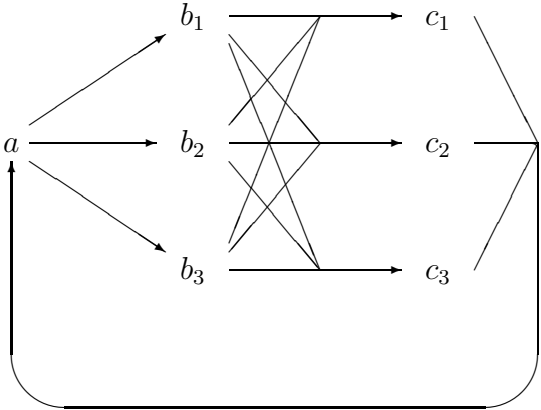
\vbox{
\setlength{\unitlength}{5000sp}%
\begingroup\makeatletter\ifx\SetFigFont\undefined%
\gdef\SetFigFont#1#2#3#4#5{%
  \reset@font\fontsize{#1}{#2pt}%
  \fontfamily{#3}\fontseries{#4}\fontshape{#5}%
  \selectfont}%
\fi\endgroup%
\begin{picture}(2644,2086)(5559,-5342)
\thinlines
{\color[rgb]{0,0,0}\put(5581,-4111){\vector( 0, 1){0}}
\put(6885,-4111){\oval(2608,2444)[bl]}
\put(6885,-4027){\oval(2612,2612)[br]}
\put(8191,-4027){\oval(  0, 12)[tr]}
}%
{\color[rgb]{0,0,0}\put(5671,-3931){\vector( 3, 2){675}}
}%
{\color[rgb]{0,0,0}\put(5671,-4021){\vector( 1, 0){630}}
}%
{\color[rgb]{0,0,0}\put(5671,-4111){\vector( 3,-2){675}}
}%
{\color[rgb]{0,0,0}\put(6661,-3391){\line( 1, 0){450}}
\put(7111,-3391){\line(-2,-5){450}}
}%
{\color[rgb]{0,0,0}\put(7111,-3391){\line(-5,-6){450}}
}%
{\color[rgb]{0,0,0}\put(6661,-3481){\line( 5,-6){450}}
\put(7111,-4021){\line(-5,-6){450}}
}%
{\color[rgb]{0,0,0}\put(6661,-4021){\line( 1, 0){450}}
}%
{\color[rgb]{0,0,0}\put(6661,-3526){\line( 2,-5){450}}
\put(7111,-4651){\line(-1, 0){450}}
}%
{\color[rgb]{0,0,0}\put(6661,-4111){\line( 5,-6){450}}
}%
{\color[rgb]{0,0,0}\put(7111,-3391){\vector( 1, 0){405}}
}%
{\color[rgb]{0,0,0}\put(7111,-4021){\vector( 1, 0){405}}
}%
{\color[rgb]{0,0,0}\put(7111,-4651){\vector( 1, 0){405}}
}%
{\color[rgb]{0,0,0}\put(7876,-3391){\line( 1,-2){315}}
\put(8191,-4021){\line(-1,-2){315}}
}%
{\color[rgb]{0,0,0}\put(7876,-4021){\line( 1, 0){315}}
}%
\put(5581,-4066){\makebox(0,0)[b]{\smash{{\SetFigFont{12}{24.0}
{\rmdefault}{\mddefault}{\updefault}{\color[rgb]{0,0,0}$a$}%
}}}}
\put(6481,-4066){\makebox(0,0)[b]{\smash{{\SetFigFont{12}{24.0}
{\rmdefault}{\mddefault}{\updefault}{\color[rgb]{0,0,0}$b_2$}%
}}}}
\put(6481,-3436){\makebox(0,0)[b]{\smash{{\SetFigFont{12}{24.0}
{\rmdefault}{\mddefault}{\updefault}{\color[rgb]{0,0,0}$b_1$}%
}}}}
\put(6481,-4696){\makebox(0,0)[b]{\smash{{\SetFigFont{12}{24.0}
{\rmdefault}{\mddefault}{\updefault}{\color[rgb]{0,0,0}$b_3$}%
}}}}
\put(7696,-4696){\makebox(0,0)[b]{\smash{{\SetFigFont{12}{24.0}
{\rmdefault}{\mddefault}{\updefault}{\color[rgb]{0,0,0}$c_3$}%
}}}}
\put(7696,-4066){\makebox(0,0)[b]{\smash{{\SetFigFont{12}{24.0}
{\rmdefault}{\mddefault}{\updefault}{\color[rgb]{0,0,0}$c_2$}%
}}}}
\put(7696,-3436){\makebox(0,0)[b]{\smash{{\SetFigFont{12}{24.0}
{\rmdefault}{\mddefault}{\updefault}{\color[rgb]{0,0,0}$c_1$}%
}}}}
\end{picture}%
}
\nop{
++=> a ===> b1 b2 b3 ===> c1 c2 c3 ==++
||                                   ||
++===================================++
}

This formula has $22$ literal occurrences. The three clauses $a \rightarrow
b_i$ have size two each (total: $3 \times 2 = 6$); the three clauses $b_1b_2b_3
\rightarrow c$ have size four each (total: $3 \times 4 = 12$); the single
clause $c_1 c_2 c_3 \rightarrow a$ has size four. The total is
{} $6 + 12 + 4 = 22$.

Every other single-head equivalent formula has the same size: the equivalences
being realized as a loop, every loop of sets of size $1$, $3$ and $3$ is a
closed sequence with a $1-3$ edge, a $3-3$ edge and a $3-1$ edge.

An alternative that is not single-head is $A \rightarrow B$, $B \rightarrow A$,
$A \rightarrow C$ and $C \rightarrow A$:

\[
F' = \{
	a \rightarrow b_1,
	a \rightarrow b_2,
	a \rightarrow b_3,
	b_1 b_2 b_3 \rightarrow a,
	a \rightarrow c_1,
	a \rightarrow c_2,
	a \rightarrow c_3,
	c_1 c_2 c_3 \rightarrow a
\}
\]

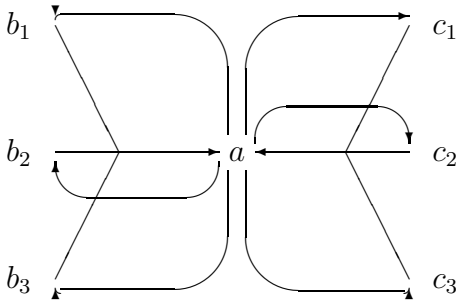
\vbox{
\setlength{\unitlength}{5000sp}%
\begingroup\makeatletter\ifx\SetFigFont\undefined%
\gdef\SetFigFont#1#2#3#4#5{%
  \reset@font\fontsize{#1}{#2pt}%
  \fontfamily{#3}\fontseries{#4}\fontshape{#5}%
  \selectfont}%
\fi\endgroup%
\begin{picture}(2145,1722)(5026,-4888)
\thinlines
{\color[rgb]{0,0,0}\put(5344,-3841){\oval(1464,1190)[tr]}
\put(5344,-3256){\oval(246, 20)[tl]}
\put(5221,-3256){\vector( 0,-1){0}}
}%
{\color[rgb]{0,0,0}\put(5221,-4606){\vector( 0, 1){0}}
\put(5304,-4606){\oval(166,  8)[bl]}
\put(5304,-4021){\oval(1544,1178)[br]}
}%
{\color[rgb]{0,0,0}\put(5221,-3976){\vector( 0, 1){0}}
\put(5626,-3976){\oval(810,364)[bl]}
\put(5626,-3976){\oval(810,364)[br]}
}%
{\color[rgb]{0,0,0}\put(6976,-3256){\vector( 1, 0){0}}
\put(6976,-3841){\oval(1620,1170)[tl]}
}%
{\color[rgb]{0,0,0}\put(6846,-4021){\oval(1360,1194)[bl]}
\put(6846,-4606){\oval(260, 24)[br]}
\put(6976,-4606){\vector( 0, 1){0}}
}%
{\color[rgb]{0,0,0}\put(6976,-3886){\vector( 0,-1){0}}
\put(6594,-3886){\oval(764,362)[tr]}
\put(6594,-3886){\oval(766,362)[tl]}
}%
{\color[rgb]{0,0,0}\put(5221,-3301){\line( 1,-2){315}}
\put(5536,-3931){\line(-1,-2){315}}
}%
{\color[rgb]{0,0,0}\put(5221,-4876){\makebox(0.7938,5.5563){\small.}}
}%
{\color[rgb]{0,0,0}\put(5221,-3931){\vector( 1, 0){810}}
}%
{\color[rgb]{0,0,0}\put(6976,-3301){\line(-1,-2){315}}
\put(6661,-3931){\line( 1,-2){315}}
}%
{\color[rgb]{0,0,0}\put(6976,-3931){\vector(-1, 0){765}}
}%
\put(5041,-3976){\makebox(0,0)[b]{\smash{{\SetFigFont{12}{24.0}
{\rmdefault}{\mddefault}{\updefault}{\color[rgb]{0,0,0}$b_2$}%
}}}}
\put(5041,-3346){\makebox(0,0)[b]{\smash{{\SetFigFont{12}{24.0}
{\rmdefault}{\mddefault}{\updefault}{\color[rgb]{0,0,0}$b_1$}%
}}}}
\put(5041,-4606){\makebox(0,0)[b]{\smash{{\SetFigFont{12}{24.0}
{\rmdefault}{\mddefault}{\updefault}{\color[rgb]{0,0,0}$b_3$}%
}}}}
\put(6121,-3976){\makebox(0,0)[b]{\smash{{\SetFigFont{12}{24.0}
{\rmdefault}{\mddefault}{\updefault}{\color[rgb]{0,0,0}$a$}%
}}}}
\put(7156,-3976){\makebox(0,0)[b]{\smash{{\SetFigFont{12}{24.0}
{\rmdefault}{\mddefault}{\updefault}{\color[rgb]{0,0,0}$c_2$}%
}}}}
\put(7156,-4606){\makebox(0,0)[b]{\smash{{\SetFigFont{12}{24.0}
{\rmdefault}{\mddefault}{\updefault}{\color[rgb]{0,0,0}$c_3$}%
}}}}
\put(7156,-3346){\makebox(0,0)[b]{\smash{{\SetFigFont{12}{24.0}
{\rmdefault}{\mddefault}{\updefault}{\color[rgb]{0,0,0}$c_1$}%
}}}}
\end{picture}%
}
\nop{
b1 b2 b3 <===> a <===> c1 c2 c3
}

This formula is smaller because it centers around the small set $A=\{a\}$,
saving from the costly entailment from a set of size three to another of size
three. The size of $F'$ is the sum of the size the clause $b_1b_2b_2
\rightarrow a$, four, of the three clauses $a \rightarrow b_i$, two each, and
the same for $c_1c_2c_3$. The total is
{} $2 \times (4 + 3 \times 2) = 2 \times 10 = 20$.

This formula is not single-head as it contains two clauses with head $a$. The
answer to the question ``is the single-head formula $F$ equivalent to a formula
of size $20$ or less'' is ``yes''. The answer to the similar question ``...
equivalent to a {\em single-head} formula of size $20$ or less'' is ``no''.

That the second formula is the minimal form of the first is proved by running
{\tt minimize.py} on the first. Running it on the second proves that the second
is not only minimal, but that no other formula of the same size is equivalent
to it. It is the only minimal formula equivalent to itself. This proves that
the only minimal-size formula equivalent to the first formula is single-head.

\begin{verbatim}
minimize.py -f -minimal 'a->b' 'a->c' 'a->d' 'bcd->e' 'bcd->f' 'bcd->g' 'efg->a'
minimize.py -f -minimal 'a->b' 'a->c' 'a->d' 'bcd->a' 'a->e' 'a->f' 'a->g' 'efg->a'
\end{verbatim}

The variables are renamed from $b_1, b_2, b_3, \ldots$ to $b, c, d, \ldots$
because the program only allows for single-letter variables. The {\tt
minimize.py} program is described in a previous article~\cite{libe-20}. It can
currently be retrieved at {\tt https://github.com/paololiberatore/minimize.py}.

\


Regardless of whether the minimal formula has to be single-head or not, finding
it is polynomial if the clauses form no loop, where each clause $P \rightarrow
x$ links every variable in $P$ to $x$~\cite{hamm-koga-95}. Acyclicity is a
subcase of inequivalence
: no two minimally-sized different sets of variables are equivalent.

\begin{condition}
\label{condition-inequivalent}

A formula $F$ is {\em inequivalent} if $F \models A \equiv B$ for two sets of
variables $A$ and $B$ implies $F \models A \equiv A \cap B$.

\end{condition}

The minimal form of an inequivalent single-head formula $F$ can be determined
from the order such that $B \leq_F A$ if $F$ entails $A \rightarrow B$. The
strict part of this order $B <_F A$ is as usual: $B \leq_F A$ but not $A \leq_F
B$. The set $MIN(F)$ is built from this order.

\[
MIN(F) = \{A \rightarrow x \mid
F \models A \rightarrow x \mbox{ and }
\not\exists B ~.~
x \not\in B ,~
F \models B \rightarrow x \mbox{ and }
B <_F A \mbox{ or } B \subset A \}
\]

Every formula equivalent to $F$ implies $MIN(F)$ since the clauses in $MIN(F)$
are all entailed by $F$. In the case of inequivalence, this fact can be pushed
further: every formula equivalent to $F$ contains either a clause of $MIN(F)$
or a superclause of it.

\statelemma{inequivalent-superset}{

If $F$ is inequivalent, it contains a superset of every clause in $MIN(F)$.

}{

\proof Let $A \rightarrow x$ be an arbitrary clause of $MIN(F)$. The definition
of this set includes $F \models A \rightarrow x$. By
Lemma~\ref{set-implies-set}, $F$ contains a clause $B \rightarrow x$ such that
$F \models A \rightarrow B$. This entailment defines $B \leq_F A$. The converse
$A \leq_F B$ may hold or not: either $B <_F A$ or $B \equiv_F A$.

In the first case, $B <_F A$ contradicts the assumption $A \rightarrow x \in
MIN(F)$ since $F$ entails $B \rightarrow x$ with $B <_F A$.

In the second case, $A \equiv_F B$ and the assumption of inequivalence imply $A
\equiv_F A \cap B$. Since $F \models A \rightarrow x$, this implies $F \models
A \cap B \rightarrow x$. The intersection of two sets is always contained in
each, but this containment may be strict or not: either $A \cap B \subset A$ or
$A \cap B = A$. The first case $A \cap B \subset A$ contradicts the assumption
$A \rightarrow x \in MIN(F)$ because $F$ entails $A \cap B \rightarrow x$ with
$A \cap B \subset A$. The other case is $A \cap B = A$. It implies $A \subseteq
B$: the formula $F$ contains a clause $B \rightarrow x$ with $A \subseteq B$.
This holds for every $A \rightarrow x \in MIN(F)$.~\qed

}

Another important property of $MIN(F)$ is that it is equivalent to $F$ if $F$
is single-head~\cite{libe-20-b}.

If $F$ is single-head then $MIN(F)$ is single-head as well. This makes the
problems of the minimal formula and the minimal single-head formula coincide.
Since $MIN(F)$ can be determined in polynomial time if $F$ is inequivalent,
both problems are polynomial in time in the single-head inequivalent case.

\

Does tractability extend from inequivalent single-head formulae to the general
case? A counterexample suggests it does not.

Building $MIN(F)$ is easy because it can be done clause by clause: each $P
\rightarrow x \in F$ is minimized by repeatedly replacing it with $P'
\rightarrow x$ if such a clause is entailed by $F$ and either $P' <_F P$ or $P'
\subset P$. This is not in general possible on formulae containing loops.

As an example, four equivalent sets of variables of size $1$, $1$, $3$ and $3$
are realized by a loop of clauses that makes a one-variable set entail a
three-variable one, that entail the other one-variable set and so on. A
concrete formula for this case is the following.

\begin{equation}
\label{1133}
F = \{
	a \rightarrow b,
	b \rightarrow c_1,
	b \rightarrow c_2,
	b \rightarrow c_3,
	c_1 c_2 c_3 \rightarrow d_1,
	c_1 c_2 c_3 \rightarrow d_2,
	c_1 c_2 c_3 \rightarrow d_3,
	d_1 d_2 d_3 \rightarrow a
\}
\end{equation}

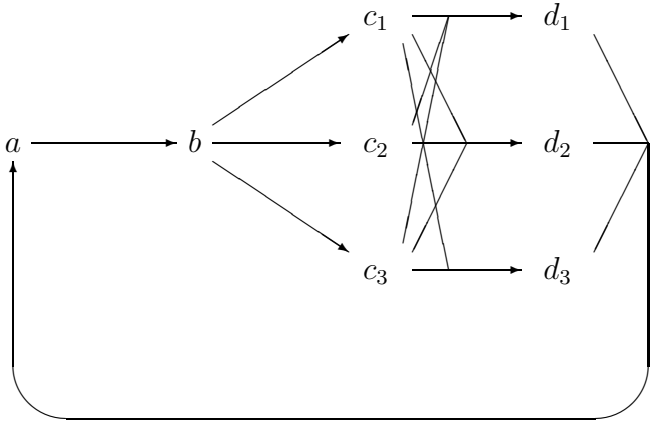
\vbox{
\setlength{\unitlength}{5000sp}%
\begingroup\makeatletter\ifx\SetFigFont\undefined%
\gdef\SetFigFont#1#2#3#4#5{%
  \reset@font\fontsize{#1}{#2pt}%
  \fontfamily{#3}\fontseries{#4}\fontshape{#5}%
  \selectfont}%
\fi\endgroup%
\begin{picture}(3177,2142)(4936,-5308)
\thinlines
{\color[rgb]{0,0,0}\put(4951,-4021){\vector( 0, 1){0}}
\put(6518,-4021){\oval(3134,2556)[bl]}
\put(6518,-3931){\oval(3166,2736)[br]}
}%
{\color[rgb]{0,0,0}\put(5041,-3931){\vector( 1, 0){720}}
}%
{\color[rgb]{0,0,0}\put(5941,-3841){\vector( 3, 2){675}}
}%
{\color[rgb]{0,0,0}\put(5941,-3931){\vector( 1, 0){630}}
}%
{\color[rgb]{0,0,0}\put(5941,-4021){\vector( 3,-2){675}}
}%
{\color[rgb]{0,0,0}\put(6931,-3301){\line( 1, 0){180}}
\put(7111,-3301){\line(-1,-5){225}}
}%
{\color[rgb]{0,0,0}\put(7111,-3301){\line(-1,-3){180}}
}%
{\color[rgb]{0,0,0}\put(6931,-3391){\line( 1,-2){270}}
\put(7201,-3931){\line(-1,-2){270}}
}%
{\color[rgb]{0,0,0}\put(6931,-4561){\line( 1, 0){180}}
\put(7111,-4561){\line(-1, 5){225}}
}%
{\color[rgb]{0,0,0}\put(7111,-4561){\vector( 1, 0){360}}
}%
{\color[rgb]{0,0,0}\put(6931,-3931){\vector( 1, 0){540}}
}%
{\color[rgb]{0,0,0}\put(7111,-3301){\vector( 1, 0){360}}
}%
{\color[rgb]{0,0,0}\put(7831,-3391){\line( 1,-2){270}}
\put(8101,-3931){\line(-1,-2){270}}
}%
{\color[rgb]{0,0,0}\put(7831,-3931){\line( 1, 0){270}}
}%
\put(4951,-3976){\makebox(0,0)[b]{\smash{{\SetFigFont{12}{24.0}
{\rmdefault}{\mddefault}{\updefault}{\color[rgb]{0,0,0}$a$}%
}}}}
\put(5851,-3976){\makebox(0,0)[b]{\smash{{\SetFigFont{12}{24.0}
{\rmdefault}{\mddefault}{\updefault}{\color[rgb]{0,0,0}$b$}%
}}}}
\put(6751,-3976){\makebox(0,0)[b]{\smash{{\SetFigFont{12}{24.0}
{\rmdefault}{\mddefault}{\updefault}{\color[rgb]{0,0,0}$c_2$}%
}}}}
\put(6751,-4606){\makebox(0,0)[b]{\smash{{\SetFigFont{12}{24.0}
{\rmdefault}{\mddefault}{\updefault}{\color[rgb]{0,0,0}$c_3$}%
}}}}
\put(6751,-3346){\makebox(0,0)[b]{\smash{{\SetFigFont{12}{24.0}
{\rmdefault}{\mddefault}{\updefault}{\color[rgb]{0,0,0}$c_1$}%
}}}}
\put(7651,-3346){\makebox(0,0)[b]{\smash{{\SetFigFont{12}{24.0}
{\rmdefault}{\mddefault}{\updefault}{\color[rgb]{0,0,0}$d_1$}%
}}}}
\put(7651,-4606){\makebox(0,0)[b]{\smash{{\SetFigFont{12}{24.0}
{\rmdefault}{\mddefault}{\updefault}{\color[rgb]{0,0,0}$d_3$}%
}}}}
\put(7651,-3976){\makebox(0,0)[b]{\smash{{\SetFigFont{12}{24.0}
{\rmdefault}{\mddefault}{\updefault}{\color[rgb]{0,0,0}$d_2$}%
}}}}
\end{picture}%
}
\nop{
++==> a ===> b ===> c1 c2 c3 ===> d1 d2 d3 ==++
||                                           ||
++===========================================++
}

This formula entails the equivalence of the sets $\{a\}$, $\{b\}$,
$\{c_1,c_2,c_3\}$ and $\{d_1,d_2,d_3\}$ by a loop: $\{a\}$ entails $\{b\}$
which entails $\{c_1,c_2,c_3\}$ which entails $\{d_1,d_2,d_3\}$ which entails
$\{a\}$. Its size is
{} $2 + 2 \times 3 + 4 \times 3 + 4 = 2 + 6 + 12 + 4 = 24$
variable occurrences.

A smaller way to entail the same equivalences is by a loop where $\{a\}$
entails $\{c_1,c_2,c_3\}$, which entails $\{b\}$ which entails
$\{d_1,d_2,d_3\}$ which entails $\{a\}$. The formula is the following.

\begin{equation}
\label{1313}
F' = \{
	a \rightarrow c_1,
	a \rightarrow c_2,
	a \rightarrow c_3,
	c_1 c_2 c_3 \rightarrow b,
	b \rightarrow d_1,
	b \rightarrow d_2,
	b \rightarrow d_3,
	d_1 d_2 d_3 \rightarrow a
\}
\end{equation}

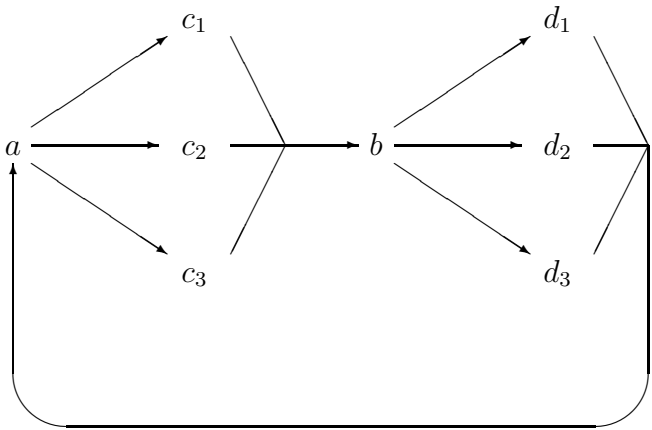
\vbox{
\setlength{\unitlength}{5000sp}%
\begingroup\makeatletter\ifx\SetFigFont\undefined%
\gdef\SetFigFont#1#2#3#4#5{%
  \reset@font\fontsize{#1}{#2pt}%
  \fontfamily{#3}\fontseries{#4}\fontshape{#5}%
  \selectfont}%
\fi\endgroup%
\begin{picture}(3177,2169)(5836,-5335)
\thinlines
{\color[rgb]{0,0,0}\put(5851,-4021){\vector( 0, 1){0}}
\put(7419,-4021){\oval(3136,2610)[bl]}
\put(7419,-3931){\oval(3164,2790)[br]}
}%
{\color[rgb]{0,0,0}\put(5941,-3841){\vector( 3, 2){675}}
}%
{\color[rgb]{0,0,0}\put(5941,-3931){\vector( 1, 0){630}}
}%
{\color[rgb]{0,0,0}\put(5941,-4021){\vector( 3,-2){675}}
}%
{\color[rgb]{0,0,0}\put(6931,-3391){\line( 1,-2){270}}
\put(7201,-3931){\line(-1,-2){270}}
}%
{\color[rgb]{0,0,0}\put(6931,-3931){\vector( 1, 0){630}}
}%
{\color[rgb]{0,0,0}\put(7741,-3841){\vector( 3, 2){675}}
}%
{\color[rgb]{0,0,0}\put(7741,-3931){\vector( 1, 0){630}}
}%
{\color[rgb]{0,0,0}\put(7741,-4021){\vector( 3,-2){675}}
}%
{\color[rgb]{0,0,0}\put(8731,-3391){\line( 1,-2){270}}
\put(9001,-3931){\line(-1,-2){270}}
}%
{\color[rgb]{0,0,0}\put(8731,-3931){\line( 1, 0){270}}
}%
\put(6751,-3976){\makebox(0,0)[b]{\smash{{\SetFigFont{12}{24.0}
{\rmdefault}{\mddefault}{\updefault}{\color[rgb]{0,0,0}$c_2$}%
}}}}
\put(6751,-4606){\makebox(0,0)[b]{\smash{{\SetFigFont{12}{24.0}
{\rmdefault}{\mddefault}{\updefault}{\color[rgb]{0,0,0}$c_3$}%
}}}}
\put(6751,-3346){\makebox(0,0)[b]{\smash{{\SetFigFont{12}{24.0}
{\rmdefault}{\mddefault}{\updefault}{\color[rgb]{0,0,0}$c_1$}%
}}}}
\put(5851,-3976){\makebox(0,0)[b]{\smash{{\SetFigFont{12}{24.0}
{\rmdefault}{\mddefault}{\updefault}{\color[rgb]{0,0,0}$a$}%
}}}}
\put(7651,-3976){\makebox(0,0)[b]{\smash{{\SetFigFont{12}{24.0}
{\rmdefault}{\mddefault}{\updefault}{\color[rgb]{0,0,0}$b$}%
}}}}
\put(8551,-3976){\makebox(0,0)[b]{\smash{{\SetFigFont{12}{24.0}
{\rmdefault}{\mddefault}{\updefault}{\color[rgb]{0,0,0}$d_2$}%
}}}}
\put(8551,-4606){\makebox(0,0)[b]{\smash{{\SetFigFont{12}{24.0}
{\rmdefault}{\mddefault}{\updefault}{\color[rgb]{0,0,0}$d_3$}%
}}}}
\put(8551,-3346){\makebox(0,0)[b]{\smash{{\SetFigFont{12}{24.0}
{\rmdefault}{\mddefault}{\updefault}{\color[rgb]{0,0,0}$d_1$}%
}}}}
\end{picture}%
}
\nop{
++==> a ===> c1 c2 c3 ===> b ===> d1 d2 d3 ==++
||                                           ||
++===========================================++
}

This formula $F'$ is equivalent to $F$ because it entails the same equivalences
between the four sets and nothing else. It contains the same four sets of
variables, each entailing another to form a single loop encompassing all of
them.

Yet, it is smaller:
{} $2 \times 3 + 4 + 2 \times 3 + 4 = 6 + 4 + 6 + 4 = 20$
instead of $24$.

This shows that minimizing cannot be done clause by clause as in the
inequivalent case. Every clause of $F$ is minimal by itself. For example, $a
\rightarrow b$ is minimal; the only sets that are strictly less than $\{a\}$
are $\emptyset$, the proper subsets of $\{c_1,c_2,c_3\}$ and $\{d_1,d_2,d_3\}$
and their union, but none of them entail $b$; the only set that is strictly
contained in $\{a\}$ is $\emptyset$, which again does not entail $b$. While $a
\rightarrow b$ is minimal by itself, it is not in the minimal single-head
formula equivalent to $F$. This formula cannot be minimized by minimizing each
clause it contains; it needs a global restructuring.

\

The question is: how to minimize a single-head formula that is not
inequivalent? The target is a formula that is single-head but minimal: no other
single-head formula is smaller. A simple sufficient condition is based on the
following property.

\statelemma{equivalent-preconditions}{

If $F$ is equivalent to a single-head formula $F'$ that contains the clause $P
\rightarrow x$, then $F$ contains $P' \rightarrow x$ with $F \models P \equiv
P'$.

}{

\proof Since $F'$ contains $P \rightarrow x$ it also entails it: $F' \models P
\rightarrow x$. By the equivalence of the considered formulae, $F$ entails $P
\rightarrow x$. Lemma~\ref{set-implies-set} implies the existence of a set of
variables $P'$ such that $x \not\in P'$, $F \models P \rightarrow P'$ and $P'
\rightarrow x \in F$. The latter condition implies $F \models P' \rightarrow
x$. By equivalence, $F' \models P' \rightarrow x$. Again,
Lemma~\ref{set-implies-set} implies that $F' \models P' \rightarrow P''$ for
some $P'' \rightarrow x \in F'$. Since $F'$ is single-head and contains $P
\rightarrow x$, this is only possible if $P'' = P$. As a result, $F' \models P'
\rightarrow P''$ is the same as $F' \models P' \rightarrow P$. Since $F \models
P \rightarrow P'$ and the two formulae are equivalent, $F \models P \equiv P'$
is proved.~\qed

}

The sufficient condition not only tells when a formula is not minimal, but also
shows how clauses can be replaced.

\statelemma{not-minimal}{

If a single-head definite Horn formula $F$ is not minimal then it contains a
clause $P \rightarrow x$ such that $F \models (BCN(P,F) \backslash \{y\})
\rightarrow y$ where $y \in P$ and $BCN(P,F) = \{x \mid F \cup P\models x\}$.

}{

\proof Let $F$ be a non-minimal single-head definite Horn formula: some other
single-head definite Horn formula is shorter but equivalent. Let $F'$ be the
shortest single-head definite Horn formula equivalent to $F$. Since it is
shorter than $F$, it does not contain a clause of $F$, as otherwise $F
\subseteq F'$. Let $P \rightarrow x \in F$ be that clause.

By Lemma~\ref{equivalent-preconditions}, $F'$ contains a clause $P' \rightarrow
x$ with $F \models P \equiv P'$. If $P=P'$ then $P \rightarrow x \in F'$,
contrary to assumption. If $P \subset P'$ then $F'$ is not minimal: $F \models
P \rightarrow x$ implies $F' \models P \rightarrow x$, which implies $F' \equiv
F' \cup \{P \rightarrow x\}$; in this new formula $P' \rightarrow x$ is
redundant because it is entailed by $P \rightarrow x$; therefore, $F' \cup \{P
\rightarrow x\} \equiv (F' \cup \{P \rightarrow x\}) \backslash \{P'
\rightarrow x\}$. The latter formula is still single-head because it replaces a
clause with another with the same head. It is also equivalent to $F'$ but
smaller than it, contrary to the assumption of minimality of $F'$. This proves
$P \not\subseteq P'$ by contradiction.

This condition $P \not\subseteq P'$ can be rewritten as: $P$ contains an
element $y$ that is not in $P'$. A consequence of $F \models P \equiv P'$ is $F
\cup P \models P'$, which implies $P' \subseteq BCN(P,F)$. Since $y \not\in
P'$, this containment strengthens to $P' \subseteq BCN(P,F) \backslash \{y\}$.
Since $F \models P \equiv P'$ and $P \rightarrow x \in F$, it follows $F
\models P' \rightarrow x$. By monotonicity, $F \models P' \rightarrow x$
implies $F \models (BCN(P,F) \backslash \{y\}) \rightarrow y$.~\qed

}

If $F$ is not minimal, it contains a clause $P \rightarrow x$ such that $F
\models (BCN(P,F) \backslash \{y\}) \rightarrow y$ with $y \in P$. This
condition can be checked easily: for each clause $P \rightarrow x$, determine
the set of all its consequences $BCN(P,F)$ by iteratively adding $x$ to $P$
whenever a clause $P' \rightarrow x$ is in $F$ with $P' \subseteq P$. When no
such clause remains, check
{} $F \models (BCN(P,F) \backslash \{y\}) \rightarrow y$ for each $y \in P$.
Such an entailment suggests that $P \rightarrow x$ could be replaced by a
clause $P' \rightarrow x$ where $P'$ does not contain $y$. In other words, it
not only tells that the formula is not minimal, but it also gives some
directions for minimizing it.

\section{Reducing size by extending the alphabet}
\label{section-extended}

Some formulae cannot be reduced in size by adding new variables, others can.
For example, $F = \{a \vee b\}$ is already minimal: no variable addition
shortens it. On the other hand,
{} $F = \{abcd \rightarrow e, abcd \rightarrow f, abcd \rightarrow g\}$
is common equivalent to the shorter formula
{} $F' = \{abcd \rightarrow n,
{}        n \rightarrow e, n \rightarrow f, n \rightarrow g\}$,
whose size $5+2+2+2 = 11$ is less than $5 + 5 + 5 = 15$.

In this case, size reduction is obtained by summarizing a large body by a
single variable. The new clause $body \rightarrow newvariable$ allows replacing
each occurrence of the body with the single new variable.

Even if a formula contains no whole-body repetition it may still be amenable to
shortening because it contains many repeated body subsets. An example is
{} $F = \{abcd \rightarrow e, abch \rightarrow f, abci \rightarrow g\}$,
which contains no duplicated body, but three of its bodies contain $abc$. A new
clause $abc \rightarrow n$ allows shortening it to
{} $F' = \{abc \rightarrow n,
{}        nd \rightarrow e, nh \rightarrow f, ni \rightarrow g\}$,
reducing size from $5 + 5 + 5 = 15$ to $4 + 3 + 3 + 3 = 13$.

This mechanism is the base of an algorithm that attempts to reduce the size of
a formula employing the addition of new variables. A large set of variables
present in many bodies is summarized by a new variable thanks to an added
clause; all previous occurrences of the set are replaced by the new variable.
This summarization is repeated as long as it shortens the formula. The complete
algorithm is not optimal for two reasons: it is greedy (only works a single
subset at a time) and does not always find the best set of variable to compress
(finding that is \np-hard).

The core of the algorithm is the set/variable replacement: given a formula $F$
and a set of variables $P$, replace every occurrence of $P$ by a new variable
$x$. This is what the $\mbox{newvar}(P,F)$ subroutine does.

\

\hrule

\begingroup
\obeylines
~\newline
\noindent \#\# %
replace $P$ with $x$ in $F$
\noindent \# %
input $P$: a set of variables
\noindent \# %
input $F$: a Horn formula
\noindent \# %
output: formula $F$ with $P$ replaced by $x$

\endgroup
\begingroup
\noindent
$\mbox{formula {\bf newvar}}(\mbox{set }P,\mbox{ formula } F)$
\parskip=-10pt

\begin{enumerate}
\parskip=-5pt

\item create a new variable $x$

\item
$G_1 =
\{((A \backslash P) \cup \{x\}) \rightarrow y
\mid A \rightarrow y \in F \mbox{ and } P \subseteq A\}$

\item
$G_2 =
\{A \rightarrow y
\mid A \rightarrow y \in F \mbox{ and } P \not\subseteq A\}$

\item return $G_1 \cup G_2 \cup \{P \rightarrow x\}$

\end{enumerate}
\endgroup

\hrule

\

No matter which set $P$ is passed to $\mbox{newvar}(P,F)$, the result is
common-equivalent to $F$. This is the basement for building the correctness
proof of the whole algorithm.

\statelemma{newvar-equivalent}{

For every formula $F$ and set of variables $A$,
it holds $F \common newvar(A,F)$.

}{

\proof By Theorem~\ref{equisatisfiable-complete}, common equivalence holds if
and only if $S \cup F$ is equisatisfiable with $S \cup newvar(P,F)$ for every
set $S$ of literals containing all variables the two formulae share. In this
case all variables of $F$ are shared, and the only other variable is $x$, the
new variable.

Let $M$ be the only model of $S$ over variables $\var(F)$. Only one such model
exists because $S$ contains either $y$ or $\neg y$ for every variable $y \in
\var(F)$. Let $M'$ be the model over $\var(newvar(P,F)) = \var(F) \cup \{x\}$
that evaluates every $y \in \var(F)$ as $M$ and $x$ to true if and only if $M$
satisfies $P$.

The claim is proved by showing that $M \models F$ is equivalent to $M' \models
newvar(P,F)$ for all models $M$ over $\var(F)$. This is proved by considering
every kind of clause of $newvar(P,F)$ in turn.

If $M \models P$ then $M' \models P$ and $M' \models x$ by construction.
Therefore, $M' \models P \rightarrow x$. If $M \not\models P$ then $M' \models
P \rightarrow x$. This proves that $M'$ always satisfies $P \rightarrow x$.
This clause of $newvar(P,F)$ can be excluded from the proof of equivalence of
$M \models F$ and $M' \models newvar(P,F)$ since it is always satisfied by
$M'$.

The clauses of $G_2$ are in both $F$ and $newvar(P,F)$ and do not contain $x$.
Therefore, they are satisfied by $M$ if and only if they are satisfied by $M'$,
since the only difference between these two models is the value of $x$.

The only remaining clauses are $A \rightarrow x \in F$ with $P \subseteq A$ and
$((A \backslash P) \cup \{x\}) \rightarrow y \in G_1$. If $M \not\models P$
then $M' \not\models x$; as a result, $M \models A \rightarrow y$ since $P
\subseteq A$ and $M' \models ((A \backslash P) \cup \{x\}) \rightarrow y$ since
$M' \not\models x$. Otherwise, $M \models P$, which implies $M' \models x$. As
a result, $M \models A \rightarrow y$ if and only if $M \models (A \backslash
P) \rightarrow y$ since $P$ is true in $M$. For the same reason, $M' \models (A
\backslash P) \cup \{x\} \rightarrow y$ if and only if $M' \models (A
\backslash P) \rightarrow y$. These conditions coincide, proving the
equivalence of $M \models A \rightarrow y$ and $M' \models ((A \backslash P)
\cup \{x\}) \rightarrow y$.~\qed

}

Calling $\mbox{newvar}(P,F)$ as many times as needed is not a problem, since it
is linear-time. Therefore, it can be called to assess the gain from summarizing
a set of variables $P$ before doing that. A complete algorithm would first call
it over all subsets $P$ and check how large the result is, then replace $F$
with the shortest formula obtained from it and repeat.

The problem is the exponential number of sets of variables. Yet, many can be
excluded based on the formula. Only the sets that are contained in a body need
to be checked. Still better, since the aim is to reduce size, only the sets
that are contained in at least two bodies may be useful. Two bodies are
sometimes enough, for example
{} $F = \{abcdefg \rightarrow i, abcdefh \rightarrow j\}$
is common equivalent to the shorter formula
{} $F' = \{abcdef \rightarrow n, ng \rightarrow i, nh \rightarrow j\}$.

\statelemma{intersection}{

If $|newvar(A,F)| \leq |F|$ for some set of variables $A$ then $|newvar(B,F)|
\leq |newvar(A,F)|$ for some intersection $B$ of the bodies of some clauses of
$F$.

}{

\proof Let
{} $F_1 = \{P \rightarrow y \in F \mid A \subseteq P\}$
be the set of clauses of $F$ whose bodies include $A$. These are the clauses
$newvar(A,F)$ replaces with $((P \backslash A) \cup \{x\}) \rightarrow y$. The
claim is proved for
{} $B = \cap \{P \mid P \rightarrow y \in F_1\}$.
This set contains $A$ because it is the intersection of some sets $P$ that all
contain $A$. If $B$ is equal to $A$, the claim trivially holds. Therefore, only
the case of strict containment $A \subset B$ is considered.

If $F_1$ is empty then no clause of $F$ contains $A$ in its body. Therefore,
$newvar(A,F) = \{A \rightarrow x\} \cup F$. This implies $|newvar(A,F)| = |A| +
1 + |F| > |F|$, contrary to the assumption $|newvar(A,F) \leq |F|$. As a
result, $F_1$ contains at least a clause: $|F_1| \geq 1$.

Both $newvar(A,F)$ and $newvar(B,F)$ contain $F \backslash F_1$. They however
differ in two other ways: they respectively contain
{} $A \rightarrow x$ and
{} $B \rightarrow x$,
and they respectively contain
{} $((P \backslash A) \cup \{x\}) \rightarrow y$ and
{} $((P \backslash B) \cup \{x\}) \rightarrow y$
for each $P \rightarrow y \in F_1$.

Since $A \subset B$, the first clause $A \rightarrow x$ is smaller than $B
\rightarrow x$. The difference in size is $|B| - |A|$.

The opposite difference exists between $((P \backslash A) \cup \{x\})
\rightarrow y$ and $P \backslash (B \cup \{x\}) \rightarrow y$, where $P
\rightarrow y \in F_1$. Since both $A$ and $B$ are contained in $P$ and $x$ is
not (because it is a new variable), the size of these clauses can be computed
by simple additions and subtractions: $|P| - |A| + 1 + 1$ and $|P| - |B| + 1 +
1$. The difference is negative: $|A| - |B|$.

The overall size difference between $newvar(A,F)$ and $newvar(B,F)$ is $|B| -
|A|$ for the first clause and $|A| - |B|$ for each clause of $F_1$, and is
therefore 
{} $|B| - |A| + |F_1| * (|A| - |B|) =
{}  (|F_1| - 1) * (|A| - |B|)$.
Since $|F_1| \geq 1$, this amount is negative, meaning that $|newvar(B,F)|$ is
smaller than $|newvar(A,F)|$.~\qed

}

This lemma gives directions to the algorithm by restricting to sets $P$ that
are intersections of bodies of the formula.

This is not enough for polynomiality because the number of intersections is
still exponential. A further consideration reduces search still more: size
tends to decrease more for large sets $P$ than for small, and the intersection
of few sets tends to be larger than the intersection of many. Following this
principle, the algorithm searches for the best intersection of two bodies and
then tries to improve it by intersecting it with a third body and so on.

This procedure is always correct because it repeatedly applies
$\mbox{newvar}(P,F)$, which is proved common-equivalent to $F$ by
Lemma~\ref{intersection}. Since it only performs a replacement with a shorter
formula, it is guaranteed to reduce size.

\

\hrule

\begingroup
\obeylines
~\newline
\noindent \#\# %
try to reduce size by introducing new variables
\noindent \# %
input $F$: a Horn formula
\noindent \# %
output: a formula that is common equivalent to $F$

\endgroup

\begingroup
\noindent
$\mbox{formula {\bf minimize}}(\mbox{formula } F)$
\parskip=-10pt

\begin{enumerate}
\parskip=-5pt


\item $F' = F$

\item for each $A \rightarrow x, B \rightarrow y \in F$

\begin{enumerate}
\item if $|newvar(A \cap B,F)| < |F'|$
\begin{enumerate}
\item $N = A \cap B$
\item $F' = newvar(N,F)$
\end{enumerate}
\end{enumerate}

\item if $F' = F$ return $F$
\label{greedy-extend-return}


\item $F'' = F$
\label{addbody}

\item for each $C \rightarrow x \in F$

\begin{enumerate}
\item if $|newvar(N \cap C,F)| < |F''|$
\begin{enumerate}
\item $M = N \cap C$
\item $F'' = newvar(M,F)$
\label{greedy-extend-smaller}
\end{enumerate}
\end{enumerate}

\item if $F'' \not= F'$

\begin{enumerate}

\item $N = M$

\item goto Step~\ref{addbody}

\end{enumerate}

\item $F = F''$
\label{greedy-extend-replace}

\item goto 1

\end{enumerate}
\endgroup

\hrule

\

This algorithm is implemented by the {\tt newvar.py} program. It is proved
correct by the following lemma. It does not always reduce the size of the input
formula, which may already be minimal. Yet, it never increases size. That it is
optimal is later disproved by a counterexample.

\statelemma{minimize-return}{

The formula returned by $\mbox{\bf minimize}(F)$ is common-equivalent to $F$
and not larger than it.

}{

\proof The only return instruction in the algorithm is
Step~\ref{greedy-extend-return}, which returns $F$. This variable $F$ initially
contains the input formula and is only changed in
Step~\ref{greedy-extend-replace}, which copies formula $F''$ into it. In turn,
$F''$ is initialized to $F$ and is only changed in
Step~\ref{greedy-extend-smaller} to $\mbox{newvar}(M,F)$. Therefore, the return
value is the same as the last of a sequence of instructions
{} $F = \mbox{newvar}(M,F)$
for some sets of variables $M$.

Every formula produces by such a sequence is proved by induction to be common
equivalent to the first. The claim trivially holds if the sequence is empty.
This is the base case. In the inductive case, the initial formula is assumed
common equivalent to the current; it is proved common equivalent to the next.
These three formulae are called $F_i$, $F_c$ and $F_n$. The assumption is that
the initial formula $F_i$ is common equivalent to the current $F_c$. The next
is $F_n = \mbox{newvar}(M,F_c)$. It is common equivalent to $F_c$ by
Lemma~\ref{newvar-equivalent}. Its variables are the same as these of the
initial formula $F_i$ and some new ones. Therefore, The variables shared
between $F_n$ and the initial formula $F_i$ are the variables of $F_i$. These
are also in the current formula $F_c$. Lemma~\ref{transitive} shows that common
equivalence is transitive in this case: the new formula $F_n$ is common
equivalent to the initial $F_i$. This proves the claim in the inductive case.

Since $M = N \cap C$, Step~\ref{greedy-extend-smaller}, replaces $F''$ with
$newvar(N \cap C,F)$. Since this instruction falls within the scope of the
conditional $|newvar(N \cap C,F)| < |F''|$, this replacement may only decrease
the size of $F''$. Since $F''$ is initialized to $F$, and $F$ is then assigned
$F''$, the size of $F$ monotonically decreases or stays the same during the
execution of the algorithm. This proves the second part of the claim: the
return value is a formula that is smaller than or equal to the input
formula.~\qed

}

This lemma proves that $\mbox{\bf minimize}(F)$ meets the minimal requirements
for correctness and usefulness: it outputs a formula that is the same as the
input apart from the new variables and is shorter or the same size. Hopefully,
it is shorter. Ideally, it is as short as possible. This is not always the
case, as the following counterexample shows.

\begin{eqnarray*}
F &=&
   \hbox to 6pt{\hfill\{} x_1 x_2 x_3 x_4 x_5 x_6 \rightarrow z_1,	\\
&& \hbox to 6pt{} x_1 x_2 x_3 x_4 x_5 x_6 \rightarrow z_2,		\\
&& \hbox to 6pt{} x_1 x_2 x_3 \rightarrow z_3,				\\
&& \hbox to 6pt{} x_4 x_5 x_6 \rightarrow z_4 \}
\end{eqnarray*}

The intersections of the bodies of this formula are
{} $A = \{x_1,x_2,x_3,x_4,x_5,x_6\}$,
{} $B = \{x_1, x_2, x_3\}$ and
{} $C = \{x_4, x_5, x_6\}$.
The first step of the algorithm takes is determining the minimum between
$\mbox{newvar}(A,F)$, $\mbox{newvar}(B,F)$ and $\mbox{newvar}(C,F)$.

The result of $\mbox{newvar}(A,F)$ is the following formula,
of size $7 + 2 + 2 + 4 + 4 = 19$.

\begin{eqnarray*}
\mbox{newvar}(A,F) &=&
   \hbox to 6pt{\hfill\{} x_1 x_2 x_3 x_4 x_5 x_6 \rightarrow y,	\\
&& \hbox to 6pt{} y \rightarrow z_1,					\\
&& \hbox to 6pt{} y \rightarrow z_2,					\\
&& \hbox to 6pt{} x_1 x_2 x_3 \rightarrow z_3,				\\
&& \hbox to 6pt{} x_4 x_5 x_6 \rightarrow z_4 \}
\end{eqnarray*}

The result of $\mbox{newvar}(B,F)$ is the following formula, of size $4 + 5 + 5
+ 2 + 4 = 20$. By symmetry, this is also the size of $\mbox{newvar}(C,F)$.

\begin{eqnarray*}
\mbox{newvar}(B,F) &=&
   \hbox to 6pt{\hfill\{} x_1 x_2 x_3 \rightarrow y,			\\
&& \hbox to 6pt{} y x_4 x_5 x_6 \rightarrow z_1,			\\
&& \hbox to 6pt{} y x_4 x_5 x_6 \rightarrow z_2,			\\
&& \hbox to 6pt{} y \rightarrow z_3,					\\
&& \hbox to 6pt{} x_4 x_5 x_6 \rightarrow z_4 \}
\end{eqnarray*}

Since $\mbox{newvar}(A,F)$ has size $19$ while $\mbox{newvar}(B,F)$ and
$\mbox{newvar}(C,F)$ have size $20$ each, the algorithm chooses $A$. It then
checks whether intersecting
{} $A = \{x_1,x_2,x_3,x_4,x_5,x_6\}$
with another body further reduces size. These intersections are equal to $B$
and $C$, which have been proved to produce a larger formula instead. Therefore,
the first step ends with $\mbox{newvar}(A,F)$ replacing $F$.

The second step tries to further reduce the size of $\mbox{newvar}(A,F)$ by an
intersection of its bodies. The intersections are the same as
{} $B = \{x_1, x_2, x_3\}$ and
{} $C = \{x_4, x_5, x_6\}$.
By symmetry, only $\mbox{newvar}(B,\mbox{newvar}(A,F))$ is analyzed. Its size
is $4 + 5 + 2 + 2 + 2 + 4 = 19$, the same as $\mbox{newvar}(A,F)$.

\begin{eqnarray*}
\mbox{newvar}(B,\mbox{newvar}(A,F)) &=&
   \hbox to 6pt{\hfill\{} x_1 x_2 x_3 \rightarrow y',			\\
&& \hbox to 6pt{} y' x_4 x_5 x_6 \rightarrow y,				\\
&& \hbox to 6pt{} y \rightarrow z_1,					\\
&& \hbox to 6pt{} y \rightarrow z_2,					\\
&& \hbox to 6pt{} y' \rightarrow z_3,					\\
&& \hbox to 6pt{} x_4 x_5 x_6 \rightarrow z_4 \}
\end{eqnarray*}

This formula is not shorter than the current. Therefore, the algorithm returns
the current: $\mbox{newvar}(A,F)$. This is a formula of size $19$.

A shorter common equivalent formula exists, and can even be found by a sequence
of $\mbox{newvar}()$ calls. Using $B$ and $C$ on $F$ produces the following
formula, of size $4 + 4 + 3 + 3 + 2 + 2 = 18$.

\begin{eqnarray*}
newvar(B,newvar(C,F)) &=&
   \hbox to 6pt{\hfill\{} x_1 x_2 x_3 \rightarrow y,			\\
&& \hbox to 6pt{} x_4 x_5 x_6 \rightarrow y',				\\
&& \hbox to 6pt{} y y' \rightarrow z_1,					\\
&& \hbox to 6pt{} y y' \rightarrow z_2,					\\
&& \hbox to 6pt{} y \rightarrow z_3,					\\
&& \hbox to 6pt{} y' \rightarrow z_4 \}
\end{eqnarray*}

In summary: the algorithm first finds the single best intersection $A$ of two
bodies; it then tries to intersect it with another body, but none further
reduces size; it therefore replaces $F$ with $\mbox{newvar}(A,F)$; on this new
formula, it searches all intersections of two bodies but none reduces size.
Yet, a smaller formula exists, and is obtainable by two non-minimal
intersections.

The problem is with the greedy procedure: the bird in the hand $A$ is not worth
the two $B$ and $C$ in the bush. Every algorithm driven by the best immediate
gain falls in this trap.

The counterexample is single-head and acyclic (inequivalent). Even in this very
restrictive case, the greedy algorithm may fail to find the shortest
common-equivalent formula. In fairness, this is not the fault of the algorithm:
the problem itself is \np-complete.

Membership to \np\  is relatively easy to prove.

\statelemma{extended-exists}{

Given a single-head formula $F$ and an integer $l$, deciding the existence of a
common-equivalent single-head formula $G$ such that $\var(F) \subseteq \var(G)$
and $|G| \leq l$ is in \np.

}{

\proof If $l \geq |F|$ the answer is trivial: yes, $G = F$. Otherwise, $G$ is
found by guessing a formula that contains at most $l$ variables which include
$\var(F)$ and then checking $F \common G$. Since $l < |F|$, such a guessing can
be done in nondeterministic polynomial time.
Theorem~\ref{single-head-polynomial} proves that checking $F \common G$ is also
polynomial in time since both $F$ and $G$ are single-head. As a result, the
whole problem can be solved in nondeterministic polynomial time.~\qed

}

Hardness holds even in the subcase of single-head acyclic definite Horn
formulae. It does not require the output formula to be single-head. In other
words, whether a single-head acyclic definite Horn has a common-equivalent
formula of a certain size is \np-hard, regardless of whether such an equivalent
formula is required be single-head or not.

\statelemma{extended-hard}{

Given a single-head minimal-size acyclic definite Horn formula $A$ and an
integer $m$, deciding the existence of a common equivalent definite Horn
formula $B$ such that $\var(A) \subseteq \var(B)$ and the size of $B$ is
bounded by $m$ is \np-hard. The same holds if $B$ is constrained to be acyclic
or single-head.

}{

\proof The claim is proved by reduction from the vertex cover problem. A vertex
cover of a graph $(V,E)$, where $V=\{v_1,\ldots,v_n\}$ and
$E=\{e_1,\ldots,e_m\}$, is a subset $C \subseteq V$ such that either $v_i \in
C$ or $v_j \in C$ for every $e_l = (v_i,v_j) \in E$. The vertex cover problem
is to establish whether a given graph has a vertex cover of size $k$ or less.

Given a graph, the corresponding formula and integer are as follows. For each
node $v_i$ of the graph the formula contains five variables $v_i$, $r_i$,
$r_i'$, $s_i$ and $s_i'$; for each edge $e_l$ it contains two variables $e_l$
and $e_l'$; finally, it contains a single other variable $w$.

\ttytex{
\begin{eqnarray*}
A &=&
\{
v_i w v_j \rightarrow e_l ,~
v_i w v_j \rightarrow e_l'
\mid e_l = (v_i,v_j) \in E
\}
\cup \\
&&
\{
v_i w r_i \rightarrow s_i,
v_i w r_i' \rightarrow s_i'
\mid v_i \in V
\}						\\
m &=& 6 \times |E| + 8 \times |V| + k
\end{eqnarray*}
}{
A = { vi w vj -> el,
      vi w vj -> el' | for each edge el = (vi,vj) in E } u
    { vi w ri -> si,
      vi w ri' -> si' | for each node vi in V }
m = 6 x |E| + 8 x |V| + k
}

This formula is acyclic and single-head. It can be shortened by introducing new
variables. In particular, the clause $v_i w \rightarrow y_i$ allows for the
following replacements:

\begin{eqnarray*}
v_i w v_j \rightarrow e_l & ~~~ \Rightarrow ~~~ & y_i v_j \rightarrow e_l \\
v_i w v_j \rightarrow e_l' & ~~~ \Rightarrow ~~~ & y_i v_j \rightarrow e_l' \\
v_i w r_i \rightarrow s_i & ~~~ \Rightarrow ~~~ & y_i r_i \rightarrow s_i \\
v_i w r_i' \rightarrow s_i' & ~~~ \Rightarrow ~~~ & y_i r_i' \rightarrow s_i'
\end{eqnarray*}

The new clause $v_i w \rightarrow y_i$ contains three literals. The clauses
$v_i w r_i \rightarrow s_i$ and $v_i w r_i' \rightarrow s_i'$ are shortened by
one literal each, leaving the balance at plus one. Therefore, the new
implication is only convenient it shortens some clauses $v_i w v_j \rightarrow
e_l$ and $v_i w v_j \rightarrow e_l'$. It can unless they have already been
shortened to $v_i y_j \rightarrow e_l$ and $v_i y_j \rightarrow e_l'$ by the
other clause $v_j w \rightarrow y_j$.

For each edge $e_l = (v_i,v_j)$ in the graph, introducing either $v_iw
\rightarrow y_i$ or $v_jw \rightarrow y_j$ is always convenient because it
allows shortening the clauses $v_i w v_j \rightarrow e_l$ and $v_i w v_j
\rightarrow e_l'$ of $e_l$ by one literal each, overmatching the increase of
one due to the new clause. The variables $v_i$ and $v_j$ of the new clauses of
the minimal formula that is common-equivalent to $A$ form a vertex cover of the
graph.

\

The formal proof comprises three acts:

\begin{enumerate}

\item $A$ is a minimal formula; it cannot be shortened without introducing new
variables;

\item if a formula is common-equivalent to $A$, it can be put in a certain form
without a size increase while maintaining common equivalence; this form is what
used in the proof sketch above: new clauses $v_i w \rightarrow y_i$ are
introduced to shorten other clauses by replacing $v_i w$ with $y_i$; it is
acyclic and single-head, proving that at least an acyclic single-head formula is
minimal;

\item such formulae correspond to vertex covers.

\end{enumerate}

\


The first part of the proof shows that $A$ is minimal. The variables in $A$
occur either only negative (like $v_i$) or only positive (like $e_l$). As a
result, the clauses of $A$ do not resolve. Since no clause is contained in
another, $A$ is equal to the set of its prime implicates, and none of them is a
consequence of the others. Therefore, $A$ is minimal. Reducing its size is not
possible without introducing new variables.

\


The second part of the proof shows that if $B$ is common-equivalent to $A$
and $\var(A) \subseteq \var(B)$, then $B$ is at least as large as a formula
obtained from $A$ by introducing some clauses $v_i w \rightarrow y_i$ and
replacing some occurrences of $v_i w$ with $y_i$.

Since $\var(A) \subseteq \var(B)$, Lemma~\ref{contained} tells that $B \models
A$, that is, $B$ entails every clause of $A$. For example, $B \models v_i w v_j
\rightarrow e_l$. If some clauses of $B$ are not involved in any such an
entailment, they are unnecessary and can be removed from $B$ without affecting
common equivalence.

This is a consequence of Theorem~\ref{contained-equisat}, which tells that if
$\var(A) \subseteq \var(B)$, then $A \common B$ is the same as $B \models A$
and the satisfiability of $A \cup S$ entails that of $B \cup S$ for every
consistent set of literals over the common alphabet that contains all its
variables. If a clause $c$ of $B$ is not involved in any implication of a
clause of $A$ from $B$, then $c$ can be removed from $B$ without affecting
common equivalence. Indeed, $(B \backslash \{c\}) \models A$ still holds
because $c$ was not involved in $B \models A$, and the satisfiability of $A
\cup S$ still entails that of $(B \backslash \{c\}) \cup S$ since it entails
that of $B \cup S$.

The entailment $B \models v_i w v_j \rightarrow e_l$ is equivalent to $B \cup
\{v_i,w,v_j\} \models e_l$. Since $B$ is definite Horn and contains no unit
clauses, $e_l$ is obtained by propagating $\{v_i,w,v_j\}$ on the clauses of
$B$.

Let $n$ be the first variable derived in this propagation. Since it is derived,
a clause $S \rightarrow n$ is in $B$. Since it is the first derived from
$\{v_i,w,v_j\}$, the premise of this clause may only contain these three
variables: $S \subseteq \{v_i,w,v_j\}$.

The clause $S \rightarrow n$ is used for deriving $e_l$ from $\{v_i,w,v_j\}$,
but may be used in other derivations as well. However, since $S \subseteq
\{v_i,w,v_j\}$, it can only be used in derivations where the premises contain
$S$, since otherwise it would be possible to derive a variable of $S \subseteq
\{v_i,w,v_j\}$ from other variables of $A$, which is not possible since $A$
does not contain any positive occurrences of these variables.

What has been said for $v_i w v_j \rightarrow e_l$ happens for the other
clauses of $A$ as well. These will be mentioned only when they significantly
differ from $v_i w v_j \rightarrow e_l$.

The possible cases are analyzed by the size of $S$.

\begin{description}

\item[$|S|=0$]: the clause $S \rightarrow n$ is $n$ alone; since $A$ does not
contain unit clauses, $n$ is not a variable of it: $n \in \var(B) \backslash
\var(A)$; removing the clause $n$ and each negative occurrence of $n$ does not
affect common-equivalence while reducing size, disproving $B$ minimal;

\item[$|S|=1$]: the clause $S \rightarrow n$ is binary; since $A$ does not
contain binary clauses, $n$ cannot be one of its variables: $n \in \var(B)
\backslash \var(A)$; removing this clause and replacing each negative
occurrence of $n$ by the only variable of $S$ reduces size without affecting
common equivalence, disproving $B$ minimal;

\item[$|S|=3$]: the only subset of $\{v_i,w,v_j\}$ having size three is
$S=\{v_i,w,v_j\}$; the clause is therefore $v_i,w,v_j \rightarrow n$; this is
the only case where $n$ may be a variable of $A$; it may not as well;

\begin{itemize}

\item $n$ is a new variable: $n \in \var(B) \backslash \var(A)$; as shown
above, such a clause can only be used in derivations from sets of literals that
contain $\{v_i,w,v_j\}$; only the derivations of $e_l$ and $e_l'$ have such
premises; since $n$ is a new variable, it is neither $e_l$ nor $e_l'$; as a
result, at least two other clauses are necessary to derive $e_l$ and $e_l'$;
none of them can be unary, since otherwise $A$ would imply either $e_l$ or
$e_l'$; as a result, the derivations of $e_l$ and $e_l'$ require at least two
binary clauses, in addition to $v_i w v_j \rightarrow n$; their total size is
$4+2+2$; replacing them with the two original clauses $v_i w v_j \rightarrow
e_l$ and $v_i w v_j \rightarrow e_l'$ does not change size nor affect common
equivalence;

\item $n$ is a variable of $A$; the clause $v_i w v_j \rightarrow n$ only
comprises variables of $A$; as a result, it is entailed by $A$; the only two
clauses of $A$ with these premises are $v_i w v_j \rightarrow e_l$ and $v_i w
v_j \rightarrow e_l'$; if $B$ contains both then the derivations of $e_l$ and
$e_l'$ require two clauses of four literals each, like in $A$;

otherwise, $B$ contains only one of them, for example $v_i w v_j \rightarrow
e_l'$; this clause alone allows deriving $e_l'$ from $\{v_i,w,v_j\}$; as shown
above, it can also be used in other derivations, but only those having premises
that contain $\{v_i,w,v_j\}$; the only other one is the derivation of $e_l$
from $\{v_i,w,v_j\}$;

if this derivation uses $v_i,w,v_j \rightarrow e_l'$, then $e_l$ is obtained by
propagating $\{v_i,w,v_j,e_l'\}$ on the clauses of $B \backslash \{v_i w v_j
\rightarrow e_l'\}$; if $v_i$ is not used in this propagation, then $B \models
w v_j e_l' \rightarrow e_l$; since this clause only contains variables of $A$,
it is entailed by it; this is not the case; the same applies to $w$ and $v_j$,
and proves that the derivation of $e_l$ from $\{v_i,w,v_j,e_l'\}$ uses all
three variables $\{v_i,w,v_j\}$; it also contains $e_l$, meaning that the size
of the clauses involved in it is at least four;

the derivations of $e_l$ and $e_l'$ require $v_i w v_j \rightarrow e_l'$ and
other clauses of size at least four; this is the same as in $A$, which contains
$v_i w v_j \rightarrow e_l$ and $v_i w v_j \rightarrow e_l'$

\end{itemize}

this shows that no clause $v_i w v_j \rightarrow n$ allows decreasing size; $B$
may contain other clauses of size four: $v_i w r_i \rightarrow n$ and $v_i w
r_i' \rightarrow n$; the analysis is the same, except that the first clause may
only be used in the derivation of $s_i$; therefore, either $n=s_i$ and then $B$
contains the original clause $v_i w r_i \rightarrow s_i$ of $A$ or $n$ is a new
variable and then $B$ contains another clause comprising $s_i$; in both cases,
the clauses used in this derivation have at least the same size of the original
clause of $A$; the same for the derivation of $s_i'$;

\item [$|S|=2$]: the clause $S \rightarrow n$ is ternary; since $A$ does not
contain ternary clauses, $n$ is not one of its variables: $n \in \var(B)
\backslash \var(A)$; since $S \subseteq \{v_i,w,v_j\}$, only three cases are
possible: $S$ is either $\{v_i,w\}$, $\{v_i,v_j\}$ or $\{w,v_j\}$; the third
case is analogous to the first, so it is not considered;

a step back is necessary: this is the analysis of the first clause $S
\rightarrow n$ of the derivation of $e_l$ from $v_iwv_j$ in $B$, which is
necessary since $A$ contains $v_iwv_j \rightarrow e_l$; the cases where $|S|$
is not $2$ are the same for the other clauses of $A$; this one is not; the
derivation of $s_i$ from $v_iwr_i$ involves $S=\{v_i,r_i\}$ and $S=\{w,r_i\}$,
and the same for $s_i'$;

\begin{itemize}

\item $w r_i \rightarrow n$ and $v_i r_i \rightarrow n$; the only derivation
whose premises contain either $\{w,r_i\}$ or $\{v_i,w\}$ is that of $s_i$ from
$\{v_i,w,r_i\}$; since $n$ is a new variable, it is not $s_i$; as a result,
another clause comprising $s_i$ is necessary; this clause cannot be unary, as
otherwise $B$ would entail $s_i$ while $A$ does not; therefore, the derivation
requires a clause of size two in addition to $S \rightarrow n$, of size three;
the total size is $3+2=5$, while the same derivation can be done in size four
with the original clause $v_i w r_i \rightarrow s_i$;

\item $v_i v_j \rightarrow n$; the only derivations having $\{v_i,v_j\}$ in
their premises are those of $e_l$ and $e_l'$; since $n$ is a new variable it is
neither $e_l$ nor $e_l'$; as a result, other two clauses are necessary to
entail them: $S' \rightarrow e_l$ and $S'' \rightarrow e_l'$; none of these two
clauses may be unary since otherwise $B$ would entail either $e_l$ or $e_l'$
while $A$ does not; if either $|S'| > 1$ or $|S''| > 1$, the total size is at
least $3+3+2 = 8$, the same as the original clauses of $A$;

otherwise, the size of both $S'$ and $S''$ is one: the two clauses have the
form $m \rightarrow e_l$ and $k \rightarrow e_l'$; by symmetry, only the first
needs to be analyzed; $m$ cannot be one of the original variables since $A$
does not entail any binary clauses; it cannot be $n$ as otherwise $B$ would
imply $v_iv_j \rightarrow e_l$ while $A$ does not; as a result, $B$ contains
another clause $S''' \rightarrow m$ as otherwise $m$ would not be involved in
the derivation of $e_l$;

if the size of $S'''$ is one, the same argument applies to its only variable;
it has size two or more, the total size for $e_l$ alone is at least $3 + 3 + 2
= 8$, regardless of how $e_l'$ is generated; this is the same size as the
clauses of $A$ for deriving both;

\item $v_i w \rightarrow n$; changing the name of $n$ to $y_i$, this is the
clause $v_i w \rightarrow y_i$

\end{itemize}

\end{description}

All of this shows that every formula that is common-equivalent to $A$ can be
transformed without a size increase into a formula $B$ where the first clause
used in each derivation $B \cup \{v_i,w,v_j\} \models e_l$ is either the
original clause $v_i w v_j \rightarrow e_l$ of $A$ or $v_i w \rightarrow y_i$
or $v_j w \rightarrow y_j$. The derivations of $s_i$ and $s_i'$ are similar but
with the first two choices only.

What is required to complete the proof is to pinpoint the other clauses
involved in the derivations. If $B$ contains $v_i w v_j \rightarrow e_l$, no
other clause is necessary to derive $e_l$ from $\{v_i,w,v_j\}$ in $B$. If it
contains either $v_i w \rightarrow y_i$ or $v_j w \rightarrow y_j$, another
clause with head $e_l$ is required.

This clause may be
{} $y_i v_j \rightarrow e_l$ if $v_i w \rightarrow y_i \in B$,
{} $v_i y_j \rightarrow e_l$ if $v_j w \rightarrow y_j \in B$, and
{} $y_i y_j \rightarrow e_l$ if both clauses are in $B$.
In the third case, $y_i y_j \rightarrow e_l$ can be replaced by $y_i v_j
\rightarrow e_l$ with no increase in size. All these clauses have size three.

Shorter clauses $P \rightarrow e_l$ do not work. If the size of $P \rightarrow
e_l$ is one then $B$ implies $e_l$ while $A$ does not.

If the size of $P \rightarrow e_l$ is two, it is $n \rightarrow e_l$ for some
variable $n$. Since $A$ does not contain binary clauses, $n$ is not a variable
of $A$. Since $n \rightarrow e_l$ is the last clause used in the derivation of
$e_l$ from $\{v_i,w,v_j\}$, its precondition $n$ is entailed exactly by
$\{v_i,w,v_j\}$. It is not entailed by any other group of variables of $A$,
since otherwise that group would entail $e_l$ in $A$. Therefore, $n$ may only
be used in the derivations from $\{v_i,w,v_j\}$, which are those producing
$e_l$ and $e_l'$. However, $n$ is neither $y_i$ nor $y_j$, as otherwise $B$
would imply $v_i w \rightarrow e_l$ or $v_j w \rightarrow e_l'$. Therefore, $n$
is derived by another clause $P' \rightarrow n$. The precondition $P'$ of this
clause may not be empty, as otherwise $B$ would imply $e_l$ alone while $A$
does not. Therefore, this clause has at least a precondition: $m \rightarrow
n$. If it is used in the derivation of $e_l$ alone, the two clauses $m
\rightarrow n$ and $n \rightarrow e_l$ can be replaced by $m \rightarrow e_l$.
Otherwise, the three clauses $m \rightarrow n$, $n \rightarrow e_l$ and $n
\rightarrow e_l'$ have the same size (6) of
{} $v_i y_i \rightarrow e_l$ and $v_i y_i \rightarrow e_l'$
and the same size of
{} $v_j y_j \rightarrow e_l$ and $v_j y_j \rightarrow e_l'$.

The conclusion is that if a formula that is common equivalent to $A$ has a
certain size, another exists that implies $v_i w v_j \rightarrow e_l$ through
either
{} $\{v_i w v_j \rightarrow e_l\}$ or
{} $\{v_i w \rightarrow y_i, y_i v_j \rightarrow e_l\}$ or 
{} $\{v_j w \rightarrow y_j, y_j v_j \rightarrow e_l\}$,
and similarly for the derivations of the clauses with head $e_l'$, $s_i$ or
$s_i'$. If this formula contains other clauses, these are not necessary for
common equivalence and can be removed. All remaining clauses of $B$ are
redundant since they do not contribute to entail any clause of $A$.

The first case
{} $v_i w v_j \rightarrow e_l \in B$
is now excluded for minimal formulae $B$. This only applies to the clauses
having $e_l$ or $e_l'$ as their head, not $s_i$ or $s_i'$. If $B$ already
contains $v_i w \rightarrow y_i$, the clause $v_i w v_j \rightarrow e_l$ can be
replaced by $y_i w \rightarrow e_l$, reducing size by one. The same if $B$
contains $v_j w \rightarrow y_i$. If $B$ contains neither, then it contains
$v_i w v_j \rightarrow e_l$, $v_i w v_j \rightarrow e_l'$, $v_i w r_i
\rightarrow s_i$ and $v_i w r_i' \rightarrow s_i'$. These clauses have total
size 16, but are common equivalent to $v_i w \rightarrow y_i$, $y_i v_i
\rightarrow e_l$, $y_i v_i \rightarrow e_l'$, $y_i r_i \rightarrow s_i$, and
$y_i r_i' \rightarrow s_i'$, which have total size 15.

This bans the original clauses
{} $v_i w v_j \rightarrow e_l$
from being in $B$. This is instead possible for the clauses
{} $v_i w r_i \rightarrow s_i$.
However, such a clause cannot be in a minimal $B$ if this formula also contains
$v_i w \rightarrow y_i$, since $v_i w r_i \rightarrow s_i$ could be replaced by
the shorter clause $y_i r_i \rightarrow s_i$. The same applies to the clause
containing $s_i'$.

In summary, the formulae $B$ common equivalent to $A$ built over a superset of
$\var(A)$ are either:

\begin{enumerate}

\item formulae that contain

\begin{itemize}

\item either
{} $\{v_i w \rightarrow y_i, y_i v_j \rightarrow e_l\}$ or
{} $\{v_j w \rightarrow y_j, y_j v_j \rightarrow e_l\}$
for each clause $v_i w v_j \rightarrow e_l \in A$, and similarly for the
clauses that contain $e_l'$;

\item
{} $v_i w \rightarrow y_i$ if $y_i r_i \rightarrow s_i \in B$
{} and $v_i w r_i \rightarrow s_i$ otherwise,
and similarly for clauses that contain $s_i'$;

\end{itemize}

\item formulae that are as large as the formulae of the first kind or larger.

\end{enumerate}

The formulae of the first kind are single-head because their clauses have
either the same heads as the original single-head formula or heads $y_i$. Since
no $y_i$ is in the original formula and only one clause is introduced with each
head $y_i$, no two clauses have the same head. The formula is also acyclic
because all its edges are from the variables $V$, $R$ and $\{w\}$ to $Y$, and
from all of these to $E$ and $S$; no edge back means acyclic.

\


The third part of the proof shows that the minimal formulae $B$ have size
linear in the size of the minimal vertex covers of the graph.

According to the conclusion of the previous part of the proof, every minimal
formula that is common equivalent to $A$ can be rewritten without size increase
as a formula $B$ that contains either
{} $\{v_i w \rightarrow y_i, y_i v_j \rightarrow e_l\}$ or
{} $\{v_j w \rightarrow y_j, y_j v_j \rightarrow e_l\}$
for each clause $v_i w v_i \rightarrow e_l \in A$. A consequence of this
condition is that $B$ contains either $v_i w \rightarrow y_i$ or $v_j w
\rightarrow y_j$ for each $e_l \in E$. Therefore,
{} $C = \{v_i \mid v_i w \rightarrow y_i \in B\}$
is a vertex cover of the graph $G$.

The size of the formula and the cover are related as follows:

\begin{itemize}

\item for each clause $v_i w v_j \rightarrow e_l$ of $A$ a clause $y_i v_j
\rightarrow e_l$ or $y_j v_j \rightarrow e_l$ is in $B$, and the same for the
clause containing $e_l'$; their literal occurrence count is $3+3$; since a pair
of such clauses is in $A$ for every $e_l \in E$, this part of $B$ has size $|E|
\times 6$;

\item $B$ contains a clause $v_i w \rightarrow y_i$ for each $v_i \in C$; total
size is $|C| \times 3$;

\item $B$ contains $v_i w r_i \rightarrow s_i$ if $v_i w \rightarrow y_i
\not\in B$, which is the same as $v_i \not\in C$; otherwise, it contains $y_i
r_i \rightarrow s_i$, if $v_i \in C$; total size is
{} $|C| \times 3 + (|V| - |C|) \times 4 =
{}  |C| \times 3 + |V| \times 4 - |V| \times 4 =
{}  |V| \times 4 - |C| \times 2$.
Adding the clauses with $s_i'$ doubles this size to
{} $|V| \times 8 - |C| \times 2$.

\end{itemize}

The overall size is therefore
{} $|E| \times 6 + |C| \times 3 + 8 \times |V| - |C| \times 2 =
{}  |E| \times 6 + |C| + |V| \times 8$.
The part $|E| \times 6 + |V| \times 8$ depends only on the graph and not on the
cover. The size of the formula and the cover are therefore linearly related:
the formula contains as many literals as the nodes in the cover, apart a
constant.

Having proved that each formula corresponds to a cover, what is missing is that
every cover corresponds to a formula $B$ of the considered kind.

Let $C$ be a vertex cover for $(V,E)$ such that $|C| = k$. A formula $B$ of
size bounded by $m$ is proved to be common equivalent to $A$ with $\var(A)
\subseteq \var(B)$:

\begin{eqnarray}
\nonumber
B &=&
\{
	y_i v_j \rightarrow e_l,
	y_i v_j \rightarrow e_l'
		\mid v_i \in C \mbox{ and } (v_j \not\in C \mbox{ or } i < j)
\} \cup							\\
\nonumber
&&
\{
	v_i y_j \rightarrow e_l,
	v_i y_j \rightarrow e_l'
		\mid v_j \in C \mbox{ and } (v_i \not\in C \mbox{ or } j < i)
\} \cup							\\
\nonumber
&&
\{
	v_i w \rightarrow y_i,
	y_i r_i \rightarrow s_i,
	y_i r_i' \rightarrow s_i'
		\mid v_i \in C
\} \cup							\\
&&
\label{form-of-b}
\{
	v_i w r_i \rightarrow s_i,
	v_i w r_i' \rightarrow s_i'
		\mid v_i \not\in C
\}
\end{eqnarray}

This formula is common equivalent to $A$ because resolving out all new
variables $y_i$ produces $A$. For example, $y_i v_j \rightarrow e_l$ is in $B$
only if $v_i \in C$, which implies that $B$ also contains $v_i w \rightarrow
y_i$; resolving these two clauses produces the original $v_i w v_j \rightarrow
e_l$.

The number of occurrences of literals in the formula is now determined. For
each edge $e_l \in E$, the formula contains either $y_i v_j \rightarrow e_l$ or
$v_i y_j \rightarrow e_l$, and the same for $e_l'$; the total size of these
clauses is $6$. For each node $v_i \in C$, the formula contains three
implications of three literals each. For each node $v_i \not\in C$, the formula
contains two implications of size four each. The total is therefore:

\[
6 \times |E| + 9 \times |C| + 8 \times (|V| - |C|) =
6 \times |E| + 8 \times |V| + |C| =
6 \times |E| + 8 \times |V| + k
\]

This proves the last part of the claim: if the graph has a vertex cover of size
$k$, then $A$ has a common-equivalent formula of size $m = 6 \times |E| + 8
\times |V| + k$.~\qed

}

Minimizing a formula by introducing new variables is \np-complete in the
single-head case, and remains hard in the acyclic case even when releasing the
constraint on the minimal formula to be acyclic or single-head. The problem
hardness excuses the greedy mechanism of $\mbox{\bf minimize}(F)$ for not being
optimal. Being polynomial and correct, it cannot also be optimal unless
$\p=\np$.

\section{Implementations}
\label{section-python}

The algorithms presented in this article have been implemented in
Python~\cite{vanr-drak-11} and Bash~\cite{bash-07}: forgetting, common
equivalence verification and minimization by variable introduction. They can
currently be retrieved from
{\tt https://github.com/paololiberatore/forget}.

Formulae are sets of clauses; each clause is written {\tt abc->d}. Implications
and equivalences between sets of variables like {\tt ab->cd} and {\tt ab=cd}
are allowed, and are internally translated into clauses. In order to allow this
simple way of writing clauses, variables can only be single characters. This
limits their number to the character range in Python, currently about a
million.

Most formulae used in the article are made into test files for these programs.
They can be checked by passing the name of the file to the programs as their
only commandline argument.

\subsection{Forget}
\label{python-forget}

Forgetting hinges around
{} $\mbox{\bf body\_replace()}$ and
{} $\mbox{\bf head\_implicates()}$.
The main issue in their implementation is how to realize their nondeterministic
choices: the variable $x$ in $\mbox{\bf head\_implicates()}$ and the choice of
a clause in both $\mbox{\bf body\_replace()}$ and $\mbox{\bf
head\_implicates()}$.

The two choices in $\mbox{\bf head\_implicates()}$ can be realized by a single
loop over the clauses of $F$: the clauses $P \rightarrow x \in F$ with $x
\not\in V$ are disregarded, the others are changed by calling $\mbox{\bf
body\_replace()}$ on their bodies to replace the variables to be forgotten in
all possible ways.

Variables may be forgotten in many ways. For example, $a$ in $a \rightarrow x$
may be replaced by $b$, $cd$ and $e$ if the formula contains $b \rightarrow a$,
$cd \rightarrow a$ and $e \rightarrow a$. If some of the replacing variables
$b$, $c$, $d$ and $e$ are to be forgotten as well, they are also recursively
replaced in all possible ways. This is what nondeterminism does in $\mbox{\bf
body\_replace()}$: each nondeterministic choice is a possible way of replacing
a variable.

The first implementation of nondeterminism restricts to formulae that do not
need it: the single-head ones. The second iterates over all nondeterministic
choices, collecting the results in a set. The third exploits operating systems
multitasking primitives. These three implementations are in three separate
programs:

\begin{description}

\item[{\tt forget-singlehead.py}:] when the input formula is single-head, the
choice of a clause $P \rightarrow y \in P$ with head $y$ becomes deterministic
since only zero or one such clause may exist; it is still realized as a loop,
but only the first clause with head $y$ is used; this is correct as long as the
formula is single-head; the algorithm is simple as it does not implement any
special mechanism for nondeterminism, but is only correct on single-head
formulae;

\item[{\tt forget-set.py}:] the function $\mbox{\bf bo{}dy\_replace()}$ returns
a set of pairs instead of a pair as in the algorithm; this set contains the
possible return values of all nondeterministic choices; a recursive call to
$\mbox{\bf body\_replace()}$ no longer returns a pair $S,E$ but a set of such
pairs; these are the possible ways to replace the variables to forget in the
set $P$; each such replacement is further pursued by the caller, possibly
leading to other replacements; these are collected in a set of pairs, which is
returned;

the increased complication of the code was to be expected, but this
implementation suffers from a more serious drawback: since the nondeterministic
choices are realized by sets, these may grow exponentially even if the output
is not; an extreme example is in the test file {\tt branches.py}, where large
sets are built to produce the single clause $klm \rightarrow j$ when forgetting
$V$ in $F$.

\begin{eqnarray*}
F &=& \{
	k \rightarrow a, k \rightarrow b,
	l \rightarrow d, l \rightarrow e,
	m \rightarrow g, m \rightarrow h,
\\
&&
	a \rightarrow c, b \rightarrow c,
	d \rightarrow f, e \rightarrow f,
	g \rightarrow i, h \rightarrow i,
	cfi \rightarrow j
\}
\\
V &=& \{k, l, m, j\}
\end{eqnarray*}

\item[{\tt forget-fork.py}:] this implementation exploits multitasking; each
nondeterministic possible choice is made into a branch of execution; the step
``choose $P \rightarrow y \in F$'' is turned into a loop over the clauses of
$F$ with head $y$; for each such clause, execution is forked; the first branch
continues as if the clause were the only possible choice, the other waits for
the first to terminate; only when the original call to $\mbox{\bf
head\_implicates()}$ ends the other branch continues the loop; if other clauses
$P \rightarrow y \in F$ exists, it repeats on one of them;

this solution may look complicated, but its implementation is quite simple: in
the $\mbox{\bf bo{}dy\_replace()}$ function, the instruction ``choose $P
\rightarrow y \in F$'' is implemented as {\tt c = choose(h)} where {\tt h} is
the set of clauses with head $y$ and {\tt choose()} is as follows;

\begin{verbatim}
def choose(s):
    sys.stdout.flush()
    for i,c in enumerate(s):
        if i == len(s) - 1 or os.fork() == 0:
            return c
        else:
            os.wait()
    else:
        fail()

def fail():
    sys.stdout.flush()
    os._exit(0)
\end{verbatim}

the {\tt choose()} function receives an arbitrary collection, not only a set of
clauses; if for example it is called on the list {\tt [1,2,3]}, it iterates
over its elements {\tt 1}, {\tt 2} and {\tt 3}; the first iteration is on {\tt
1}; execution branches into a parent and child processes; the child cuts the
loop short by immediately returning {\tt 1}, the parents waits for the child to
terminate; only when the whole child process terminates the parent continues
with the second iteration of the loop, over the element {\tt 2}; the same
happens again: the child returns {\tt 2}, the parent waits for it to finish;
overall, {\tt choose([1,2,3])} returns {\tt 1} in a branch of execution; when
that finishes it returns {\tt 2} in another branch of execution; when that
finishes it returns {\tt 3}; in general, it creates a branch of execution for
each element of the set passed as its first argument;

the {\tt nondeterministic.py} program shows a similar example: {\tt choose()}
is called two times in a row; in the first it is passed the list of strings
{\tt ['I', 'you', 'they']}, in the second the list {\tt ['go', 'wait', 'jump',
'sleep']}; what follows the two calls is executed in a separate branch for each
string of the first list and each of the second; for example, a branch runs on
{\tt 'I'} and {\tt 'sleep'}, another on {\tt 'they'} and {\tt 'wait'} and so on
for all possible combinations;

each branch of execution terminates when the program ends, not when the calling
function ends; this is required: a nondeterminstic branch ends when the
computation ends;

while it works as required, this creates a problem: everything executed after
{} $\mbox{\bf head\_implicates()}$
runs in all successful branches of execution; for example, calling
{} $\mbox{\bf head\_implicates()}$
on a second formula processes the second formula once for each clause generated
on the first; the solution is to begin
{} $\mbox{\bf head\_implicates()}$
by forking, returning only when the child ends;

\begin{verbatim}
def forget(f, v):
    a = choose(['run', 'wait'])
    if a == 'wait':
         return
    # run and print result
    fail()
\end{verbatim}

currently, the program just outputs the result of forgetting; using it in the
same program is not obvious, as each clause is output in a separate branch of
execution; collecting them in the parent introduces the slight complication of
managing a memory area shared by processes;

the run/wait split saves from the danger of exponentially growing branches: at
each nondeterministic fork, only one choice is pursued; only when done the next
road is taken;

while this solution may superficially appear to be also realizable by the
setjmp/longjmp Linux system calls, it is not; neither by Posix threads; these
two mechanisms only allow branches to remain separated until they return to the
branching point; nondeterministic choices last until the end of the main call
to
{} $\mbox{\bf head\_implicates()}$;

the cost of forking is small in operating systems implementing the
copy-on-write mechanism~\cite{smit-magu-88}, such as Linux: only when memory
pages start differing between the child and the parent processes they are
duplicated; the creation of the child process is cheap by itself;

the number of processes in simultaneous execution is also small: each
nondeterministic choice generates a waiting parent and a running child; since
the parent is waiting it does not spawn any other process until the child
terminates; overall, only one process is running while a polynomial number of
other processes are waiting; polynomiality is guaranteed by the second
parameter of $\mbox{\bf body\_replace()}$, which forbids replacing the same
variable twice; so no, this is not a fork bomb.


\end{description}

\subsection{Common equivalence}
\label{python-common}

The {\tt commonequivalent} bash scripts takes two formulae and tells whether
they are common equivalent. It first computes their shared variables, then runs
the {\tt forget-fork.py} program on the first formula and checks whether each
produced clause is entailed by the second formula by calling {\tt entail.py},
then does the same with reversed formulae.

While the number of generated clauses may be exponential, {\tt forget-fork.py}
runs in polynomial space. Collecting all these clauses in a formula and then
calling {\tt entail.py} to check for equivalence or entailment may require
exponential space. Instead, each clause produced by {\tt forget-fork.py} is
immediately piped to {\tt entail.py} as soon as it is generated. As soon as an
execution branch of {\tt forget-fork.py} successfully ends, the clause it
outputs is passed to {\tt entail.py} and immediately discarded. Since each
branch of execution takes polynomial space and {\tt entail.py} does the same,
the whole mechanism only requires polynomial space.

\begin{verbatim}
forget-fork.py $V $A | \
grep -v '[= ]' | \
{
	while read F;
	do
		echo -n "$B |= $F"
		entail.py $F $B && echo " no" && exit 1
		echo " yes"
	done
	exit 0;
}

[ $? = 1 ] && echo "no" && exit
\end{verbatim}

\def\returnvalue{{\tt \$?}}

This is the check in the first direction: each clause obtained by forgetting
$V$ from $A$ is entailed by $B$. The two {\tt exit} instructions do not
terminate the script but only the pipeline. They are needed to provide it a
return value that can then be retrieved in the variable \returnvalue.

A single Python program for checking common equivalence can be made by
extending {\tt forget-fork.py}. It requires some mechanism for terminating
execution as soon as a generated clause is not entailed by the other formula.

\subsection{Reducing size by variables addition}

The {\tt newvar.py} Python program implements the $\mbox{newvar}(P,F)$
subroutine of Section~\ref{section-extended}, the greedy algorithm for reducing
formula size by introducing new variables. It works in the definite Horn case,
not necessarily single-head or acyclic, but is not optimal even in the most
restrictive case. The counterexample showing that the algorithm is not optimal
is in the testing file {\tt greedy.py}.

\section{Conclusions}
\label{section-conclusions}

Some comments are in order about the results in this article:

\begin{itemize}


\item the increase in size due to variable forgetting may be contrasted by the
introduction of new variables; common equivalence is the semantical foundation
of this process; in turns, common equivalence can be expressed in terms of
forgetting; however, the computational properties of the two concepts are
different: while both can be performed with only a polynomial amount of memory,
forgetting may produce an exponentially large output while checking common
equivalence does not; checking common equivalence by forgetting is still
possible but requires some care to avoid an unnecessarily large consumption of
memory; in addition, some properties of common equivalence such as its limited
transitivity and its insensitivity to new variables are not obvious to express
in terms of forgetting;


\item common equivalence is \S{2}-complete even in the case where it expresses
forgetting and \conp-complete for Horn formulae; unsurprisingly, complexity is
one level higher than entailment and satisfiability; this is a common trait of
many forms of complex reasoning ~\cite{baum-gott-99,eite-gott-91-b}; efficient
solvers for problems of this complexity exist~\cite{baly-etal-17,puli-seid-19},
but the increase in hardness still affects speed;

forgetting may exponentially increase size even when formula minimization is
allowed; this affects forgetting usage; when it is necessary (e.g., for
privacy), it has to be done anyway; when it is done for efficiency, it may be
counterproductive; in such cases, giving up forgetting some variables may be
the best course of action;


\item an algorithm for Horn formulae running in polynomial space for forgetting
and checking common equivalence is presented; it is equivalent to resolving out
or unfolding variable occurrences in a certain order; it does not process all
occurrences of each variable at time as previous algorithms
do~\cite{delg-wass-13,delg-17,wang-etal-05}, which may exponentially enlarge
the formula; a side effect of this result is that while forgetting a set of
variables is semantically equivalent to forgetting each variable at time, it is
computationally different; the algorithm presented in this article shows how to
forget a set of variables in polynomial space, which may not be possible when
forgetting variable by variable;


\item the algorithm runs in polynomial time, in addition to polynomial space,
when each variable is the head of one clause at most; this subcase formalizes
situations where each fact only obtains as the result of a single set of
premises; it excludes situations where something results from two or more
different causes; still, a fact may be equivalent to another, like when $a$
implies $b$ which implies $a$; forbidding such loops on the top of the
single-head restriction guarantees that not only forgetting can be performed in
polynomial time, but the result can also be minimized in polynomial time; this
additional constraint can be satisfied in simple cases like $a$ equivalent to
$b$ by replacing all occurrences of $b$ with $a$;


\item an algorithm for minimizing a formula when new variables can be
introduced is shown; it is not optimal, but the problem itself is \np-hard even
in the single-head acyclic case; under the common assumptions in complexity
theory, an algorithm for this problem would be either incomplete or exponential
in time; the one presented in this article is incomplete: it does not always
find a minimal formula; at the same time, it is guaranteed never to increase
the size of the original formula.

\end{itemize}

An open question is whether minimization is polynomial-time or \np-hard in the
single-head cyclic case with no variable introduction. It is polynomial-time
for acyclic formulae, but is \np-hard with variable introduction. The open case
is in the middle: not acyclic, but no new variables either. Some preliminary
investigation hint it is \np-hard.

All of this applies to propositional logic, mostly in the Horn case. An open
question is what extends to other logics where forgetting is applied:
{} first-order logic~\cite{lin-reit-94,zhou-zhan-11},
{} description logics~\cite{eite-06,zhan-16}
and
{} modal logics~\cite{zhan-zhou-09,vand-etal-09},
where forgetting is often referred to as its dual concept of uniform
interpolant, and also
{} temporal logics~\cite{feng-etal-20},
{} logics for reasoning about actions~\cite{erde-ferr-07,raja-etal-14},
{} circumscription~\cite{wang-etal-15}
{} defeasible logics~\cite{anto-etal-12} and
{} abstract argumentation systems~\cite{baum-etal-20}.
Answer set programming~\cite{gonc-etal-16} is very close to Horn logics, as
clauses like $abc \rightarrow d$ resemble positive rules like {\tt d <- a,b,c}.
Yet, answer set programming is significantly different from propositional logic
due to the semantics of negation as failure. For example, a method for
forgetting $b$ from the program comprising the rules {\tt a <- b} and {\tt b <-
c} produces the empty program~\cite{eite-wang-06} instead of {\tt a <- c} like
forgetting in propositional logic does. The original program has the empty set
as its only answer set: since no rule forces a variable to be true, they are
all false. Removing $b$ from this answer set leaves it empty, and the empty set
is also the only answer set of the empty program for the same reason. This way
of forgetting is therefore correct if forgetting is defined as removing
variables from the answer sets. Other ways of forgetting instead produce {\tt a
<- c}, similarly to propositional
logic~\cite{bert-etal-19,knor-alfe-14,zhan-foo-06}. The algorithms for
forgetting this way include replacing atoms with bodies; therefore, they suffer
from size increase when forgetting variable by variable. The solution is to
nondeterministically forget each occurrence of a variable instead of all
occurrences of same variable at once applies, as done by the algorithm in
Section~\ref{section-definite} for propositional Horn clauses.



In the other direction, different restrictions of the propositional formulae
may simplify the problem. Binary clauses is a common subclass with good
computational properties. Other subclasses are in Post's
lattice~\cite{chap-etal-07}.

A different kind of solution is to cap the size of the formulae produced by
forgetting while maintaining the consequences of the original formula as much
as possible. If the size bound is a hard constraint, something which cannot be
overcome, this may be the only possible solution when no formula expressing
forgetting is sufficiently small. The problem is changed by this variant,
similar to how approximation changes formula
minimization~\cite{hamm-koga-93,bhat-etal-10}, bounding the number of
quantifiers changes first-order forgetting~\cite{zhou-zhan-11} and limiting
size changes several \pspace\  problems~\cite{libe-05}.

Given that single-head formulae allow for a simple algorithm for forgetting,
another question is whether a formula that is not single-head can be turned in
this form. For example,
{} $\{a \rightarrow b, b \rightarrow a, b \rightarrow c, c \rightarrow b\}$
is not single-head, but is equivalent to the single-head formula
{} $\{a \rightarrow b, b \rightarrow c, c \rightarrow a\}$.
The problem of identifying such formula is not trivial as it may
look~\cite{libe-20-b}.

\appendix
\section{Proofs}
\label{section-proofs}

\subsection{Proof of Section~\ref{section-common}}

\re{theorem}{equisatisfiable}

\re{theorem}{equisatisfiable-complete}

\re{lemma}{contained}

\re{lemma}{transitive}

\re{lemma}{new-variables}

\re{lemma}{replace}

\re{lemma}{monotonic}

\re{theorem}{common-hard}

\re{theorem}{horn-conp-complete}

\re{theorem}{definite-horn-conp-complete}

\re{theorem}{horn-common-horn}

\subsection{Proofs of Section~\ref{section-size}}

\re{lemma}{no-smaller-cnf}

\re{lemma}{a-common-b}

\re{lemma}{exponential}

\subsection{Proofs of Section~\ref{section-definite}}

\re{lemma}{set-implies-set}

\re{lemma}{r-prime-empty}

\re{lemma}{loop-recursion-invariants}

\re{lemma}{entailed-arg}

\re{lemma}{alphabet}

\re{lemma}{entailed}

\re{lemma}{no-tautology}

\re{lemma}{generate}

\re{theorem}{head-implicates-theorem}

\re{theorem}{pspace}

\re{theorem}{return-if-equivalent}

\re{theorem}{forget-correct}

\re{lemma}{success-implies}

\re{lemma}{fail}

\re{lemma}{succed}

\re{lemma}{all-polynomial}

\subsection{Proofs of Section~\ref{section-indefinite}}

\re{lemma}{common-definite}

\re{lemma}{common-bot}

\subsection{Proofs of Section~\ref{section-minimize}}

\re{lemma}{inequivalent-superset}

\re{lemma}{equivalent-preconditions}

\re{lemma}{not-minimal}

\subsection{Proofs of Section~\ref{section-extended}}

\re{lemma}{newvar-equivalent}

\re{lemma}{intersection}

\re{lemma}{minimize-return}

\re{lemma}{extended-exists}

\re{lemma}{extended-hard}

\let\c=\cedilla
\bibliographystyle{alpha}
\newcommand{\etalchar}[1]{$^{#1}$}

\end{document}